\documentclass[twocolumn,longauthor]{aastex701}

\def\lessim{\mathrel{\hbox{\rlap{\hbox{\lower4pt\hbox{$\sim$}}}\hbox{$<$}}}}
\def\grtsim{\mathrel{\hbox{\rlap{\hbox{\lower4pt\hbox{$\sim$}}}\hbox{$>$}}}}

\def\lessim{\mathrel{\hbox{\rlap{\hbox{\lower4pt\hbox{$\sim$}}}\hbox{$<$}}}}
\def\grtsim{\mathrel{\hbox{\rlap{\hbox{\lower4pt\hbox{$\sim$}}}\hbox{$>$}}}}

\begin{document}

\title{A Century of Novae in the Large Magellanic Cloud}

\correspondingauthor{A. W. Shafter}

\author[0000-0002-1276-1486]{Allen W. Shafter}
\affiliation{Department of Astronomy and Mount Laguna Observatory, San Diego State University, San Diego, CA 92182, USA}
\email[show]{ashafter@sdsu.edu}

\begin{abstract}

A comprehensive study of novae in the Large Magellanic Cloud (LMC) is presented.
A total of 66 nova
eruptions have been reported in the LMC over the past century (1926--2025).
Of these, a total of 48 eruptions from 39 unique progenitor systems
(nine eruptions are recurrences of four known recurrent novae) have sufficient
photometric data to permit the maximum magnitudes and rates of decline to
be reliably estimated. These data confirm earlier studies showing that
LMC novae are, on average, slightly more luminous and faster evolving
compared with novae in the Andromeda galaxy (M31) or the Milky Way.
For the first time,
the nova models of Yaron et al. have been used in conjunction with the
light-curve data to estimate the fundamental properties of the LMC nova
population. The models suggest that LMC novae are characterized
by generally higher white dwarf (WD) masses and higher expansion velocities compared with
M31 novae. The higher average WD masses have resulted in a higher percentage
of recurrent nova eruptions in the LMC ($\sim20.6$\%) compared with that seen in M31
($\sim6.40$\%). Future observations made possible by the Vera Rubin
Observatory promise to revolutionize our understanding of nova
populations in the LMC and in galaxies beyond the Local Group.

\end{abstract}



\section{Introduction} 

Nova eruptions result from a thermonuclear runaway (TNR)
on the surface of an accreting WD in a close
binary system.
The resulting outbursts can reach luminosities of $M_V\simeq-10$,
making them easily visible in Local Group galaxies
with moderate-sized telescopes.
The observed properties of a nova (its peak luminosity,
decline rate, ejecta mass and velocity and recurrence time) depend
sensitively on the properties of the underlying binary system; in particular,
the WD mass and its rate of accretion.
Novae that have relatively short intervals between successive
eruptions (typically less than $\sim$100~yr),
and have been observed to have more than one eruption, are
referred to as recurrent novae (RNe).
Observations of extragalactic novae offer an unparalleled opportunity
to study the properties of large and equidistant samples of novae.

Novae in Local Group galaxies have been studied extensively over the
past century going all the way back to the seminal work at the
Mount Wilson Observatory by Hubble and collaborators
on M33 \citep{Hubble1926} and the Andromeda galaxy, M31 \citep{Hubble1929}.
During the century that followed more than 60 novae have been recorded in M33,
and more than 1360 in M31. The extensive
observations of the latter galaxy, in particular, have allowed
the properties of the nova population
(e.g., peak luminosities, rates of decline, spectroscopic class)
of the nova population
to be thoroughly explored
\citep[e.g.,][]{Arp1956,Rosino1973,Ciardullo1987,Capaccioli1989,Shafter2001,
Darnley2006,Shafter2011,Rector2022,Clark2024,Shafter2026a}.

Observations of novae in our nearest neighbor galaxies, the Large and
Small Magellanic Clouds (LMC and SMC),
lagged somewhat in the early 20th century as a result of the limited 
observing facilities available in the Southern Hemisphere at that time.
Interestingly, the first nova to be seen in the Magellanic Clouds
was discovered in the SMC. The object, HV 849, was discovered by H. S. Leavitt
on plates taken at the Boyden Station in Arequipa, Peru, in late 1897,
and originally classified as a variable star \citep{Leavitt1908}. Subsequent
analysis by \cite{Nail1951} strongly suggested the object was a nova, making it
the first such eruption to be recorded outside the confines
of the Milky Way.

\begin{figure}
\plotone{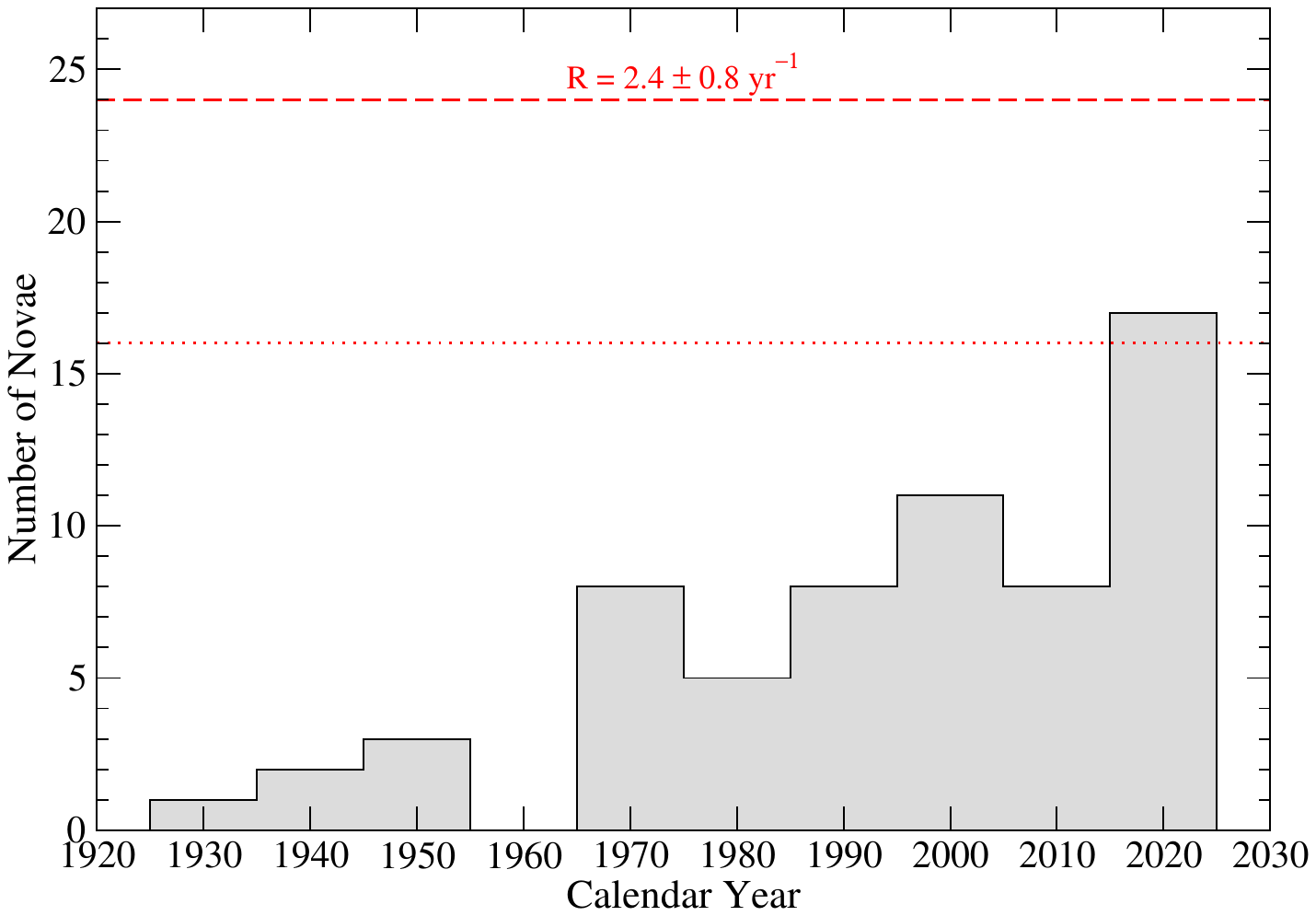}
\caption{The history of nova candidate discoveries in the LMC plotted in decade-wide
bins from 1926 to 2025. The dashed red line shows the number of novae
expected to erupt each decade (and its $1\sigma$ lower limit: dotted line)
based on an estimated annual nova rate of $2.4\pm0.8$~yr$^{-1}$
\citep{Mroz2016a}.
It has not been until the past decade that the number of
novae observed has exceeded $\grtsim50$\% of the expected total.
}
\label{fig1}
\end{figure}

A quarter century would elapse before the first nova was detected
in the LMC. Once again, the nova was registered
on Harvard Observatory plates taken at the Arequipa Station.
The discovery was made by \cite{Luyten1927}
when blinking images taken on 1926 September 28~UT.
Over the century that has passed
since that initial discovery a total of 66 nova
candidates have now been logged in the LMC.
The history of nova discoveries in the LMC is illustrated in Figure~\ref{fig1}.

In this paper, I compile and examine all available LMC nova observations
with the goal of determining the fundamental
properties of the progenitor binaries. Of the 66 known nova candidates,
a total of 39 systems were judged to have
sufficient photometric coverage to determine the peak luminosities
and subsequent rates of decline. These measurements are used to
construct a Maximum Magnitude -- Rate of Decline (MMRD) relation
for the LMC, which is compared with that determined by
\cite{Clark2024} for M31. The light-curve parameters are then analyzed
using the nova models of \cite{Yaron2005}, following the procedure
outlined by \cite{Shafter2026a} in their study of
the fundamental properties of novae in M31. I conclude by comparing
the resulting LMC nova parameters
(WD masses, accretion rates, maximum ejection velocities, and
recurrence times between successive eruptions) with their counterparts in M31.

\section{The LMC Nova Population}

\cite{Shafter2013} (hereafter Paper~I) compiled data for the 43 known and suspected LMC nova eruptions prior to 2013.
At the time, three RNe were recognized in the LMC (LMCN 1937-11a, 1968-12a, and 1971-08a).
The main driver for that study was to characterize the photometric and spectroscopic
properties of the LMC nova population and to compare it with that seen in M31 and the Galaxy.
Light curve properties (peak luminosities and rates of decline) were determined for 29 LMC nova eruptions (26 systems).
A particular emphasis was placed on determining the spectroscopic class, as defined by \cite{Williams1992}, 
where it was found that of the 22 eruptions with available spectroscopic data, half (11) belonged to the \ion{Fe}{2} class
with the remaining 11 belonging to the He/N or Hybrid\footnote{Hybrid novae are sometimes referred to
as broad-lined \ion{Fe}{2} (\ion{Fe}{2}b) novae.} classes. The percentage of He/N and Hybrid novae
is significantly higher than that seen in M31 or the galaxy ($\sim20$\%). All three of the known RNe were members of the
He/N class.

Over the dozen years that have elapsed since the publication of Paper~I
the available data for LMC novae has increased markedly. An additional
23 nova eruptions have been recorded, including 4 eruptions
from prior to 2013 that have been only
recently recognized (LMCN 2001-08a, 2002-10a, 2010-11a, 2011-08a).
Two of these (2002-10a and 2010-10a)
were previously unrecorded eruptions of the known recurrent nova (RN)
LMCRN 1968-12a.

\begin{figure*}
\plotone{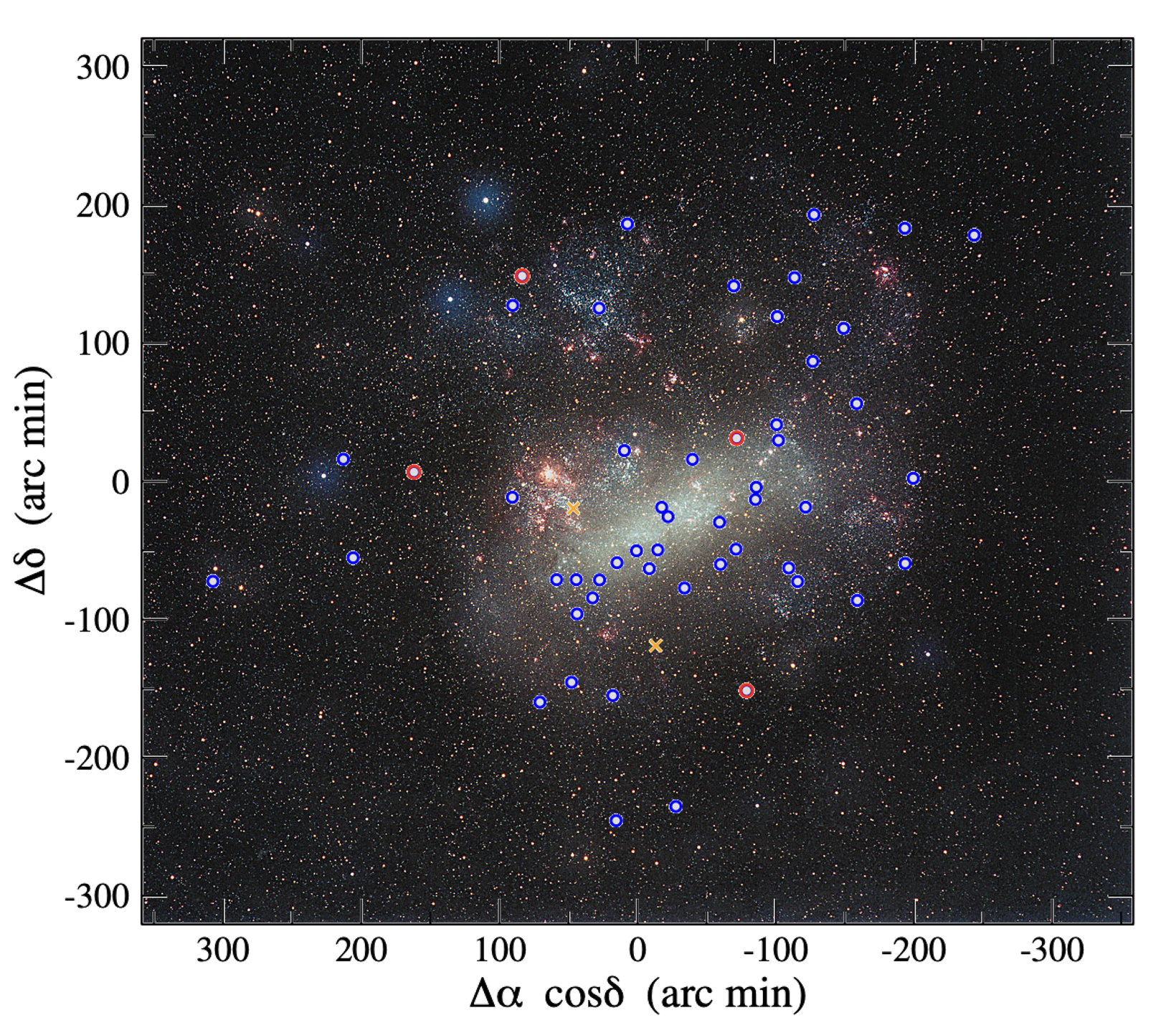}
\caption{The positions of the LMC nova candidates. The offsets were
computed assuming the center of the LMC lies at
R.A. = 5h 26m 41.19s; Decl. = $-69^{\circ}$ $11'$ $9.15''$.
The blue circled points represent known novae, the red circled points
represent the 4 known RNe, and the two orange X symbols represent
doubtful novae (LMCN 1998-12a and 2001-06a).
Two additional novae,
LMCN 2013-10a and 2015-03a, were observed in the outskirts of the LMC
(at positions $123'$~E, $352'$~S and $116'$~E, $367'$~S, respectively)
and fell outside the image.
The photograph of the LMC covers $717.64' \times 641.19'$ and
was taken by Eckhard Slawik (ESA/Hubble).
}
\label{fig2}
\end{figure*}

In all, a total of 66 candidate
nova eruptions, from 57 unique progenitors, have been
recorded in the LMC over the past century.
Their spatial positions within the LMC are shown in Figure~\ref{fig2}.
In a few cases the observations are
limited, making it difficult to establish conclusively that the transients
were actually novae. Examples include, LMCN 1935-09a, which was almost
certainly a supernova in NGC~1511 \citep{vandenbergh1988a},
LMCN~1952 and LMC~1966 \citep{Shida2004} were possibly novae, but
very little is known about these objects. More recently, two additional
nova candidates LMCN 1998-12a \citep{Becker1999} and LMCN 2002-06a
\citep{Liller2002a,Liller2003a} were recorded,
but are also unlikely to be novae.
Of the remaining eruptions, the vast majority can
be confidently associated with novae. However,
not all of these eruptions have photometric coverage
sufficient to characterize their light-curve properties and permit
further analysis.

\subsection{The Nova Candidates}

All 66 candidate nova
eruptions are summarized in Table~\ref{tab1}, and the
available photometric and spectroscopic
properties are briefly discussed individually below.
Twenty-two of these novae were discussed in Paper~I, but for completeness,
we have included them as well, updating their information as necessary
based on the availability of new data.

\subsubsection{LMCN 1926-09a}
LMCN 1926-09a (RY~Dor) was discovered by \cite{Luyten1927} and was
the first nova to be recorded in the LMC. The photometric data
are scanty, but available data suggest that the nova likely
reached a peak brightness of $V\simeq12.0$ \citep{Henize1954}, and evolved
slowly thereafter, fading by $\sim$3 magnitudes over roughly the
next 200~d \cite{Buscombe1955}. For our analysis
we adopt $m_{pg}\mathrm{(max)}=12.0\pm0.4$, and $t_2=106\pm19$~d.

\subsubsection{LMCN 1935-09a}
LMCN 1935-09a, also known as Nova Hydri 1935 and HV~11970, was observed in the outskirts of
the LMC, and near the position of NGC~1511. \cite{Boyce1943} considered it possible that
the object is a supernova in the latter galaxy. Later analysis by \cite{vandenbergh1988a} has
made a convincing case that the object was indeed a supernova in NGC~1511, and it has
since been designated SN 1935C \citep{vandenbergh1988b}. Given these revelations,
we have omitted this object from our analysis.

\subsubsection{LMCN 1936-02a}
LMCN 1936-02a (Nova Doradus 1935) was a relatively fast nova that was estimated to have
reached $pg\simeq10.5$ \citep{Henize1954}.
The rate of decline can be parameterized by $t_3\simeq31.6$ \citep{Buscombe1955}.
For our analysis, we adopt $m_{pg}\mathrm{(max)}=10.5\pm0.5$ and estimate $t_2=16\pm3$~d.

\subsubsection{LMCRN 1937-11a}
LMCRN 1937-11a (Nova Doradus 1937; YY Dor) is a RN. It was first seen to erupt
in November of 1937, reaching
$pg\simeq10.7$ \citep{Henize1954}. The subsequent
rate of decline was rapid, corresponding to $t_3\simeq20$ as estimated by \cite{Buscombe1955}.
Based on the decline relations from \citet{Shafter2026b}, we estimate $t_2=10\pm2$~d.
The nova was seen to erupt again in 2004 October (LMCN 2004-10a).

\subsubsection{LMCN 1948-12a}
LMCN 1848-12a (Nova Doradus 1948) was an unusual object with a very slow rise to maximum light, which is
estimated by \cite{Henize1954} to have reached $pg\simeq13.0$. Given that the
maximum appears to have been relatively well covered, we assume an uncertainty
of only 0.2~mag. \cite{Buscombe1955} estimate
$t_3\simeq100$~d. From the relations in \cite{Shafter2026a}, we estimate $t_2=53\pm9$.

\subsubsection{LMCN 1951-01a}
LMCN 1951-01a (Nova Mensae 1951) was discovered on H$\alpha$ objective prism plates taken at the
Lamont-Hussey Observatory \citep{Henize1954}. The nova reached $m=11.95$ as
measured by the H$\alpha$ plates.
The nova faded quickly, with
\cite{Buscombe1955} estimating $t_3\simeq6.3$~d and noting that the nova
possibly erupted in a region with heavy extinction. Our best estimate for the
light-curve parameters are
$R\mathrm{max}=11.9\pm0.2$ and $t_2=3.2\pm0.4$~d. However, given the 
likelihood of heavy extinction and the possibility that maximum light
was missed, we have not included this nova in our analysis
to determine the system's fundamental properties.

\subsubsection{LMCN 1952}
LMCN 1952 is a poorly known object reaching $V\lessim11.4$ with
very limited information \citep{Shida2004}. No
light-curve or spectroscopic observations have been published.

\subsubsection{LMCN 1966}
Like LMCN 1952, LMCN 1966 is also a poorly known object
claimed by \cite{Shida2004} to have reached $V\lessim11.1$. Again,
no light-curve or spectroscopic data have been published.

\subsubsection{LMCRN 1968-12a}
LMCRN 1968-12a (Nova mensae 1968) is a RN, and the first such object
to be discovered outside the Milky Way. The peak magnitude of the 1968 eruption was missed,
but \cite{Capaccioli1990a} estimated the nova to have reached $V\simeq10.4$. \cite{Sievers1970}
claimed that the nova evolved quickly after discovery at $m=10.9$, dropping 2 mag in just 4 days. Thus, the 1968
eruption suggests $t_2=4$~d; however, the initial decline may have been even faster.
The nova has now been seen to erupt 6 more times (1990, 2002, 2010, 2016, 2020, 2024).
None of the other three RN in the LMC has been observed to have more than one subsequent eruption.

\subsubsection{LMCN 1970-03a}
LMCN 1970-03a was discovered on an objective
prism plate taken on 1970 March 8.2~UT \citep{MacConnell1970}.
The nova was discovered at $V\sim12$, 
with no light-curve information available.
The spectrum shows bright, broad and flat-topped Balmer emission
along with similarly broad but fainter \ion{Fe}{2} emission features 
at $\lambda$4924 and $\lambda$5018.
Although, line widths were not specified,
it appears that LMCN 1970-03a is most likely
a broad-lined \ion{Fe}{2} or Hybrid nova.

\subsubsection{LMCN 1970-11a}
LMCN 1970-11a was discovered while already in decline.
Observations showed that the nova faded rapidly,
with \cite{Graham1971a} estimating a decline rate of $\sim0.25$ mag~d$^{-1}$.
Spectra, which were obtained by \cite{Havlen1972}
starting a little more than a week after the
estimated date of maximum light (1970 October 30~UT)
reveals the nova to be a member of the \ion{Fe}{2}
spectroscopic class. Unfortunately, no reliable estimates
for the peak magnitude or rate of decline are available.

\begin{figure}
\epsscale{1.3}
\plotone{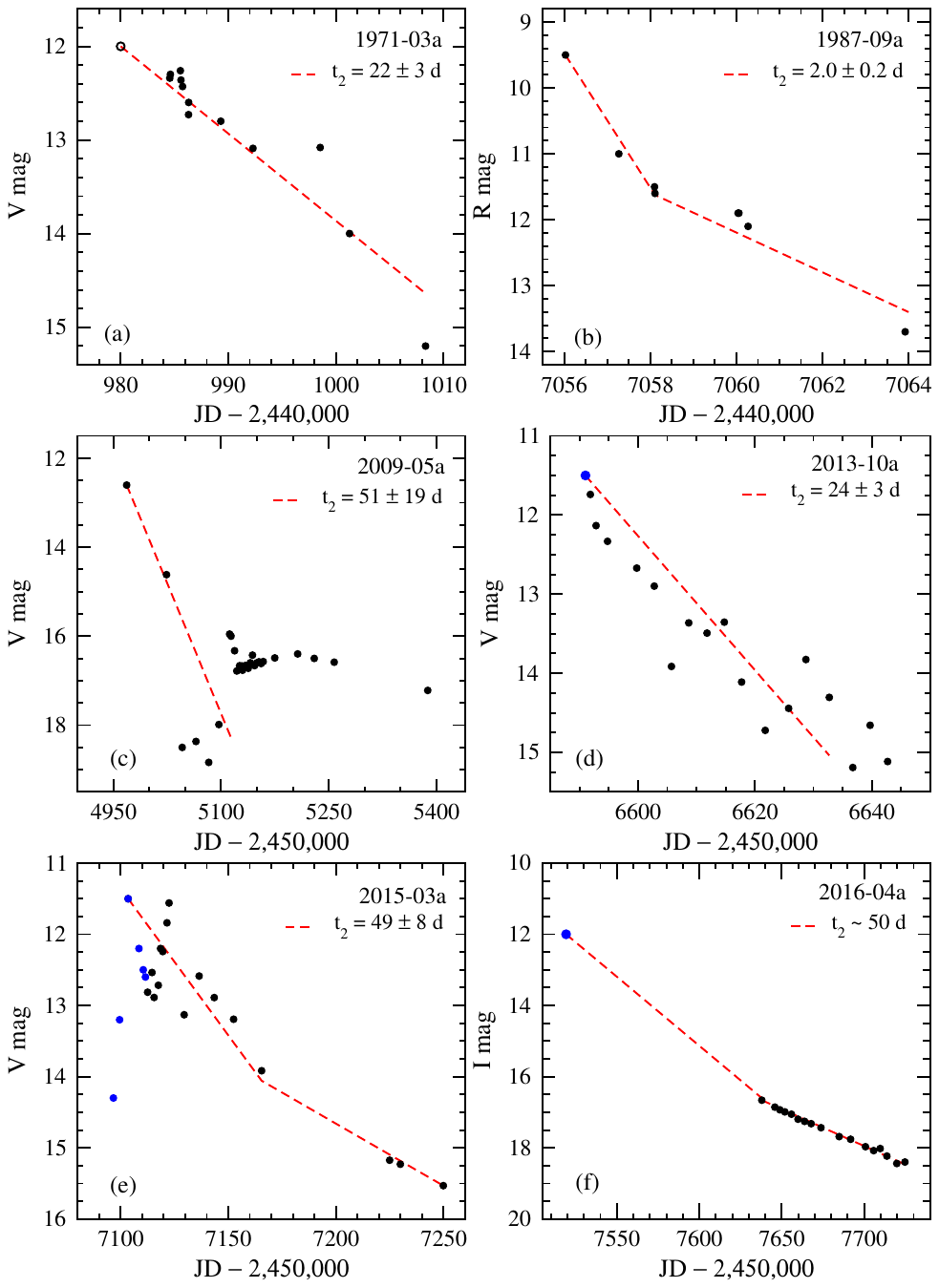}
\caption{Light curves for selected novae used for the determination of $t_2$ times.
(a) LMCN 1971-03a: $V$-band light-curve from Table~II of \cite{Graham1971a}. From these data we estimate a decline rate of
    0.091 mag~d$^{-1}$, or $t_2=22\pm2$~d.
(b) LMCN 1987-09a: $R$-band data from \cite{McNaught1987} yield an initial decline rate of 1 mag~d$^{-1}$ giving $t_2 = 2.0\pm0.2$~d.
(c) LMCN 2009-05a: $V$-band data from SMARTS. A fit to the first 8 points gives a decline rate
of 0.39~mag~d$^{-1}$, with $t_2=51\pm9$~d.
(d) LMCN 2013-10a: $V$-band data from SMARTS (black points).
    The blue point is estimated from OGLE data \citep[see][]{Mroz2016a,Hachisu2018}.
    The rate of decline is 0.085~mag~d$^{-1}$, corresponding to $t_2=24\pm3$.
(e) LMCN 2015-03a: $V$-band photometry from ASAS-SN (blue points) and  SMARTS (black points).
    A linear fit to the data yields a decline rate of 0.042~mag~d$^{-1}$ and $t_2=49\pm8$~d.
(f) LMCN 2016-04a: Peak magnitude from MASTER \citep{Gorbovskoy2016}, and the remaining data is
    $I$-band photometry from OGLE (black points). Assuming a linear decline of 0.04~mag~d$^{-1}$,
    yields $t_2\simeq50\pm10$~d.
}
\label{fig3}
\end{figure}

\subsubsection{LMCN 1971-03a}
LMCN 1971-03a was first recorded at $V=12.34$ on 1971 February 1.1~UT as part of the University of Michigan
Cerro Tololo survey \citep{Graham1971a}. The nova was not recognized for more than a month (IAU Circ. 2305, 2307),
hence the March designation.
\cite{Graham1971a} estimate that the nova had likely reached maximum light $\sim4$ days earlier,
at a magnitude $V\simeq12.0$ based on a previous nondetection on January 24.

For our analysis,
we assume $V_\mathrm{max}=12.0\pm0.3$. From an analysis of the light-curve
data presented in their Taable~II (see Figure~\ref{fig3}), we estimate that the nova was moderately fast with
$t_2\simeq22\pm3$~d and $t_3=30\pm3$~d.
These values are consistent with $t_3=28.3$ given in \cite{Subramaniam2002}.

\subsubsection{LMCRN 1971-08a}
LMCRN 1971-08a is a RN.
It was initially discovered by \cite{Graham1971b} at $m\sim13$ on 1971 August 16.4~UT, likely well after maximum light.
No other information regarding the 1971 eruption is available. The nova erupted again in February of 2009 (LMCN 2009-02a) providing
important light-curve data (see below).

\subsubsection{LMCN 1972-08a}
LMCN 1972-08a was discovered by J. A. Graham at $m\simeq13$
on 1972 August 22~UT at $m\sim13$ \citep{Bateson1972}. Prediscovery images of the region obtained
by B. Ward revealed that the nova reached $m=11$ on August 6.319~UT. No light-curve or spectroscopic
information is available.

\subsubsection{LMCN 1973-09a}
LMCN 1973-09a was discovered by \cite{Graham1973}. The nova was seen to
reach $m_{pg}=11.6$ \citep{Bateson1974}, but no light-curve or spectroscopic
follow-up observations have been published.

\subsubsection{LMCN 1977-02a}
LMCN 1977-02a was discovered by \cite{Graham1977a}.
Photometry by \cite{Lewis1977} shows that the nova reached $V=12.65$, but
no light-curve or spectroscopic information is available.

\subsubsection{LMCN 1977-03a}
LMCN 1977-03a was discovered by \cite{Graham1977b}. It was
a moderately fast nova that was well covered photometrically
\citep{Canterna1977}. The rise to maximum light was observed, allowing the peak brightness
to be accurately measured at $V=10.67\pm0.04$. The decline was well covered, yielding
$t_2=11.2$~d and $t_3=20.7$~d \citep{Canterna1981}, where we assume an uncertainty of $\pm1$~d
for $t_2$ and $\pm2$~d for $t_3$.
Spectra obtained by \cite{Canterna1981} approximately two weeks after peak
were dominated by moderately broad Balmer and \ion{Fe}{2} emission lines consistent with the
identification of the object as a member of the \ion{Fe}{2} spectroscopic class.

\subsubsection{LMCN 1978-03a}
LMCN 1978-03a was discovered on 1978 March 29~UT by \cite{Graham1978} at $m=12$.
An objective prism plate taken on the same day is described by \cite{Graham1979}.
The discovery spectrum shows broad ($\sim2000$~km~s$^{-1}$) Balmer plus Bowen blend ($4640, 4650$~\AA) emission features
suggesting that the nova might belong to the Hybrid class, but a definitive classification cannot be made.
At discovery, the nova was at $V\simeq12$ and declining rapidly. Maximum light was clearly missed, but it was speculated that
if the evolution was typical of fast novae, the object likely reached a peak magnitude of $V=9.5-10.0$ between
March 18 and 20. Given that the data are so sparse, we have opted to exclude this nova from further analysis.

\subsubsection{LMCN 1978-11a}
LMCN 1978-11a was discovered well after maximum light at $V\simeq16$
\citep{Pesch1978}. Consequently, no useful light-curve information
is available. A late-time spectrum was acquired showing blended [O~III]
emission, which strengthens the case that the object was indeed a nova.
No spectroscopic classification is possible.

\subsubsection{LMCN 1981-09a}
LMCN 1981-09a was discovered by \cite{Maza1981} at seen to reach $pg=11.8$
by A. A. Page on 1981 October 7.64~UT. Unfortunately, photometric coverage
near maximum light was very sporadic, and no reliable estimates
of the peak magnitude or subsequent decline rate are available. A spectrum
obtained by H. W. Duerbeck on October 7.32~UT reveals strong and broad
emission lines He~I, N~II, and \ion{Fe}{2} along with the Balmer series
(FWHM H$\alpha\simeq3800$~km~s$^{-1}$) consistent with a
Hybrid spectroscopic classification. The spectrum was said to
closely resemble that of V1500~Cyg and of CP~Pup 4.5 mag below maximum light
\citep{Maza1981,Andrillat1983}.

\subsubsection{LMCN 1987-09a}
LMCN 1987-09a was reported by \cite{McNaught1987} to have reached $V=9.6$ on 1987 September 17.53~UT. The maximum
was well covered as was the subsequent and rapid initial decline \citep{Capaccioli1990a}. We assume the peak
magnitude is accurate to within $\pm0.1$~mag. Based on available data (see Figure~\ref{fig3} we
estimate $t_2=2.0\pm0.2$~d, followed by a slower decline that resulted in $t_3=5.0\pm0.2$.
These values are consistent with the findings of \cite{Subramaniam2002} and \cite{Capaccioli1990a}.

\subsubsection{LMCN 1988-03a}
LMCN 1988-03a was discovered by \cite{Garradd1988} at $V=11.4$ on 1988 March 21.484~UT. \cite{Schwarz1998} have presented an extensive
multiwavelength study of the nova that suggests maximum light on March 23 at $V\simeq11.0$. The subsequent moderately fast decline
resulted in $t_2=22\pm2$~d and $t_3=40\pm2$~d. Adopting the brightest observed magnitude of
$V=11.2$ on March 23.43 UT \citep{McNaught1988a}, fits to the available photometry by \cite{Hearnshaw2004} have
resulted in the following parameters: $V_\mathrm{max}=11.2\pm0.3$, with
$t_2=22.5\pm4.0$~d and $t_3=38.4_{-5.0}^{+5.9}$~d. For the purposes of our analysis,
we adopt the more conservative estimate for $V_\mathrm{max}=11.2\pm0.2$, along with the $t_2$ time given by Hearnshaw.
Spectroscopic observations during the month following eruption \citep{Williams1991,Schwarz1998} revealed
relatively narrow Balmer and \ion{Fe}{2} emission lines consistent with the \ion{Fe}{2} spectroscopic class.

\subsubsection{LMCN 1988-10a}
LMCN 1988-10a was discovered by \cite{McNaught1988b} on 1988 October 12.48~UT at $V=11.3$,
reaching $V=10.3$ on October 13.75~UT \citep{Martin1988}. A subsequent spectroscopic
study by \cite{Sekiguchi1989} revealed prominent Balmer, He, and N
emission lines (FWZI $\sim6000$~km~s$^{-1}$) typical of the He/N novae. These authors
also compiled the available photometry and concluded that $t_2\simeq5$~d and $t_3\simeq10$~d,
establishing the object to be a very fast nova.
The maximum of this nova was well covered ($\pm0.2$ mag), and the decline rates accurately measured ($\pm0.5$~d), making it ideal for our study.

\subsubsection{LMCN 1990-01a}
LMCN 1990-01a was discovered by \cite{McNaught1990} on 1990 January 16.47~UT
at $V=11.5$. Maximum light was missed, but \cite{Liller2005} have suggested
that an extrapolation of the light-curve indicates that the nova may have reached $V\simeq9.7$.
Here, we take a more conservative approach and assume
$V_\mathrm{max}=10.0\pm0.5$ to better reflect
the uncertainty associated with this nova.
Available data show that the nova faded rapidly at a rate of 0.59 mag per day, which suggests that
$t_2\simeq3.4$ and $t_3\simeq5.1$. \cite{Vanlandingham1999} give $t_2=4.5$~d (and $t_3=9$~d) for this nova.
For the purposes of our study, we take an average of these estimates, adopting $t_2=4.0\pm0.5$~d and $t_3=7\pm2$.

A series of spectra obtained by \cite{Dopita1990} between one and two weeks postdiscovery
initially exhibit broad (FWHM $\sim5600$~km~s$^{-1}$),
flat-topped Balmer and He~I emission lines with
He~II and [Ne~III] lines becoming more prominent by January 30. These data establish
that LMCN 1990-01a was an ONe nova, and a member of the He/N spectroscopic class.

\subsubsection{LMCN 1990-02a = LMCRN 1968-12a}
LMCN 1990-02a was the second recorded eruption of LMCN 1968-12a was discovered
by W. Liller on 1990 February 14.1~UT at $V=11.2$ \citep{Williams1990}.
Spectroscopic observations
obtained approximately a week postdiscovery
established that the nova was a likely member of the He/N
spectroscopic class \citep{Sekiguchi1990,Shore1991}.
\cite{Sekiguchi1990} noted that the nova faded rapidly, dropping
an estimated 4.5 mag in the week following discovery, while
\cite{Liller2005} speculated that the nova may have reached $V\simeq10.2$ and
faded at a rate of $\sim$0.59 mag~d$^{-1}$.

\subsubsection{LMCN 1991-04a}
LMCN 1991-04a was discovered by \cite{Liller1991} on 1991 April 18~UT during its rise to maximum light.
On 1991 April 24
the nova reached $V\simeq9.0\pm0.2$ based on the average of three visual estimates between
April 23.76 and April 24.77 \cite{Gilmore1991},
making it the brightest nova ever observed in the LMC.
Post-maximum, the nova faded relatively rapidly,
with $t_2(V)$ and $t_3(V)$ of $6\pm1$ and $8\pm1$ days, respectively \citep{Schwarz2001}.
A spectrum obtained by \cite{DellaValle1991} more than 2 weeks after maximum light
showed relatively broad
Balmer emission lines (FWHM H$\alpha$ = 2500~km~s$^{-1}$), along
with a prominent CIII/NIII blend near 4645\AA\ consistent with an
\ion{Fe}{2}b (Hybrid) classification.

\subsubsection{LMCN 1992-11a}
LMCN 1992-11a was discovered at $R=10.7$ on 1992 November 11.21~UT,
reaching maximum at $V=10.2\pm0.2$ approximately a day later \citep{Liller1992}.
The nova declined relatively quickly with $t_2(V)=6.9\pm1.1$~d and $t_3(V)=13.7\pm1.6$~d
as estimated by \cite{Hearnshaw2004}.
Spectroscopic observations by \cite{DellaValle1992} and \citet{Duerbeck1992}
obtained near maximum light suggest
that the nova is a member of the \ion{Fe}{2} spectroscopic class.

\subsubsection{LMCN 1995-02a}
LMCN 1995-02a was discovered by \citet{Liller1995} at $m\simeq10.7$
on 1995 March 2.11~UT before reaching a peak magnitude of $V\simeq10.2$.
An analysis of the light-curve by \citet{Hearnshaw2004} showed that the nova
reached $V=10.35\pm0.5$ before declining with
$t_2=11\pm3$~d and $t_3=19.6\pm3.2$~d.
The nova was detected as a supersoft X-ray source (SSS) three years
after outburst \citep{Orio1999}. In our analysis we
assume $V_\mathrm{max}=10.3\pm0.2$. Spectroscopic
observations obtained near maximum light by \cite{DellaValle1995} suggest that the nova
was likely a member of the \ion{Fe}{2} class.

\subsubsection{LMCRN 1996-11a}
LMCRN 1996-11a was discovered between 1996 November 10.66 and November 12.67~UT
during the course of the MACHO survey \citep{Shida2004}\footnote{https://ogle.astrouw.edu.pl/cont/4$\_$main/var/OGLE-2018-NOVA-01.pdf}.
\cite{Liller2004b} estimate a decline rate of 0.22 mag~d$^{-1}$, corresponding to $t_2\sim9$~d and $t_3\sim14$~d.
Very little attention was paid to the nova until it erupted again in 2018 (LMCN 2018-02a, discussed below).

\subsubsection{LMCN 1997-06a}
LMCN 1997-06a was discovered by the MACHO project
at $V\sim13.5$ \citep{Alcock1997}. Available observations constrain the
time of maximum to be between 1997 June 3.398~UT and June 16.395~UT;
however, no useful light-curve or spectroscopic
data are available.

\subsubsection{LMCN 1998-12a}
LMCN 1998-12a was also discovered by the MACHO project as a variable source
that started brightening from a quiescent level ($R=20.5$; $V=21.0$) in
late 1998 December, reaching $R=19.9$, $V=17.1$ on 1999 January~14.059~UT
\citep{Becker1999}. Spectroscopic observation taken on January 14.059~UT
show two broad absorption troughs near 441.8~nm and 513.7~nm attributed
to the transient, along with nebular emission features.
The nova nature of the object is questionable.

\subsubsection{LMCN 1999-09a}
LMCN 1999-09a was discovered by the OGLE experiment on 1999 September 10~UT at a magnitude of $I=12.6$ \citep{Mroz2016a}.
Peak brightness has been estimated at $V=12.5$ by \cite{Shida2004}, but maximum light was missed and the nova may have gotten
even brighter. Decline rates of $t_2=16$~d and $t_3=19$~d have been estimated by \cite{Mroz2016a}.
In our analysis, we adopt $V_\mathrm{max}=12.5\pm0.5$ and $t_2=15\pm2$~d in the visual band.
No spectroscopic data are available to determine the spectroscopic class.

\subsubsection{LMCN 2000-07a}
LMCN 2000-07a was discovered by \citet{Liller2000}
on 2000 July 12.4~UT at $m=11.2$.
Although maximum light was not covered, \cite{Shida2004} speculates the nova reached $V=10.1$, while
\cite{Hearnshaw2004} estimate that the nova reached $V=10.7\pm0.5$ and that it
subsequently faded with $t_2$ and $t_3$ times of $8_{-3.5}^{+4.5}$~d and $20_{-7.0}^{+12.0}$~d,
respectively. \cite{Greiner2003} arrive at similar values, finding that $V_\mathrm{max}\simeq10.5$, along with
$t_2\simeq9\pm2$~d and $t_3\simeq22\pm2$~d. For the purpose of our study, we conservatively adopt $V_\mathrm{max}=10.4\pm0.4$,
along with a decline characterized by $t_2=9\pm3$ and $t_3=21\pm5$.
Spectroscopy obtained approximately three days after discovery
by \cite{Duerbeck2000} established the nova to be a member of the \ion{Fe}{2} class.

\subsubsection{LMCN 2001-06a}
LMCN 2001-06a is an unusual variable object (LMC V2434)
normally in a quiescent state
near $m=13$ that briefly reached $R\simeq9.7$ on 2001 June 6.96~UT
\citep{Liller2002a,Liller2003a}.
The object is almost certainly not a nova, but likely another
type of variable, possibly a flare star.

\subsubsection{LMCN 2001-08a}
LMCN 2001-08a was discovered in archival OGLE data \citep{Mroz2016a}. The light-curve
was peculiar, rising slowly to a peak of $I=13.3$ before an initial rapid decline
($t_2=4.8$~d; $t_3=7.2$~d).
\cite{Mroz2016a} speculated that the object may be a rare
C-type nova \citep[see][for a definition]{Strope2010}.
This classification seems plausible given that the apparent
maximum was of relatively low luminosity with the true maximum likely being missed.
No spectroscopic data are available for a spectroscopic classification.
Given that reliable estimates for peak brightness and rate of decline are not
available, we are omitting this nova from further analysis.

\subsubsection{LMCN 2002-02a}
LMCN 2002-02a was discovered by \cite{Liller2002b}, and reached
$V=10.5$ on 2002 March 3.066~UT. Subsequently,
\citet{Liller2005} estimated that the nova eventually reached $V=10.1$ at maximum light
before fading at $\sim0.1$ mag~d$^{-1}$.
Observations by \cite{Mason2005}, which were
obtained about a week after peak brightness
established that LMCN 2002-02a was an \ion{Fe}{2} nova, and that $t_2=12$~d.
\citet{Subramaniam2002} estimate that $t_3=23$~d.
For the purposes of our study, we adopt $V_\mathrm{max}=10.3\pm0.3$
and $t_2=12\pm2$~d.

\subsubsection{LMCN 2002-10a = LMCRN 1968-12a}
A 2002 eruption of the LMCRN 1968-12a was apparently detected in All Sky Automated Survey \citep[ASAS;][]{Pojmanski2002},
The nova was not visible on 2002 October 9.304~UT, but it was detected on October 11.256~UT at $V=11.15\pm0.02$ and at $V=14.17\pm0.13$ on October 15.222~UT.
From these measurements we can conclude that the nova declined by $\sim0.76$~mag~d$^{-1}$, corresponding to $t_2\simeq2.6$~d.

\subsubsection{LMCN 2003-06a}
LMCN 2003-06a was discovered by \citet{Liller2003b} at $V=11.0$ on 2003
June 19~UT.
\citet{Liller2005} estimate a decline rate of 0.25 mag per day, resulting
in $t_2$ and $t_3$ times of $\sim8$ and $\sim12$ days, respectively.
The light-curve is not well sampled, so we assume relatively large uncertainties,
adopting $V_\mathrm{max}=11.0\pm0.5$ and $t_2=8\pm2$ in our analysis.
As described in Paper~I available spectroscopic
data point toward a He/N (or possibly a Hybrid) classification.

\subsubsection{LMCN 2004-10a = LMCRN 1937-11a}
LMCN 2004-10a was discovered by \cite{Liller2004a} on 2004 October 20.193~UT near the
position of LMCN 1937-11a, establishing that the latter nova is recurrent with $P_\mathrm{rec}=66.9$~yr.
Given the relatively long interval between eruptions at a time when the LMC was not
always frequently monitored, it is possible that the true recurrence time is shorter than observed.

Photometric observations by \citet{Liller2005} suggest that the nova may have reached
$V\sim10.9$ before fading at $\sim$0.17 mag per day.
From these data we can roughly estimate $t_2$ and $t_3$ times of 12 and 18 days for the 2004
eruption, respectively. Fortunately,
the nova was also captured in the OGLE survey, which established $I$-band decline times
of $t_2=11$~d and $t_3=24$~d \citep{Mroz2016a}.
Spectroscopy carried out soon after discovery by \citet{Bond2004} and by \citet{Mason2004}
clearly establish the nova as belonging to the He/N class,
as expected for a RN.
For the purposes of our study, we have taken the mean of the peak magnitudes for two eruptions
(noting that $V-pg\simeq0.1$ for 1937-11a) yielding
$V_\mathrm{max}=10.9\pm0.3$, and adopted $t_2=11\pm2$~d and $t_3\simeq21\pm3$~d.

\subsubsection{LMCN 2005-09a}
LMCN 2005-09a was identified on 2006 July 18~UT
via its X-ray emission $\sim$10 months
after eruption \citep{Read2009}. Analysis
of archival photometry from ASAS \citep{Pojmanski2002}
showed that the nova erupted between 2005 September 18 (last nondetection) and
September 30 when it was detected at $V\sim12$. The limited ASAS
data suggested that the nova faded relatively rapidly, with $t_2$ roughly
estimated to be $\sim8\pm2$~d. The nova was also detected in the OGLE survey at $I=11.4$
on September 30 \cite{Mroz2016a}. The subsequent $I$-band decline was characterized by $t_2$ and $t_3$
times of 16 and 29 days, respectively. Taking into account the
better coverage of the OGLE data, but noting that the decline rate of novae tends to be more rapid
in $V$ compared with $I$ \citep{Hachisu2015}, we adopt $I=11.4\pm0.2$ and $t_2=14\pm4$ as
reasonable estimates for the peak brightness and rate of decline.
Unfortunately, no spectroscopic data near the time
of eruption are available to firmly establish the spectroscopic class. Based on
observations from roughly a year after eruption, \cite{Read2009}
have speculated that the nova was most likely an \ion{Fe}{2} system.

\subsubsection{LMCN 2005-11a}
LMCN 2005-11a was discovered initially by \citet{Liller2005b} on 2005 November 22.065~UT
at $m\sim12.8$. A subsequent analysis of the light-curve
showed that the nova likely reached $V=11.5$ and then
faded very slowly with $t_2$ and $t_3$ times of 63 and 94 days, respectively \citep{Liller2007}.
The OGLE survey covered the light-curve well, establishing $t_2$ and $t_3$ times of 79 and 112 days,
respectively. Again, taking into account the expected slower decline in the $I$ band, we adopt
$V_\mathrm{max}=11.5\pm0.2$ and $t_2=70\pm7$ as reasonable estimates of the maximum magnitude and rate of decline
for use in our analysis.
Spectroscopic observations by \citet{Walter2012}
starting shortly after discovery show the object to be a member of the \ion{Fe}{2} spectroscopic class.

\subsubsection{LMCN 2009-02a = LMCRN 1971-08a}
LMCN 2009-02a was observed
by \cite{Liller2009a} on 2009 February 5.067~UT at $m\sim10.6$
to erupt coincident with the position of LMCN 1971-08a,
establishing the latter nova as the 3rd known RN in the LMC. Comprehensive multiwavelength observations of LMCN 2009-02a by \cite{Bode2016}
established $V_\mathrm{max}=10.6$, with $t_2$ and $t_3$ times of 5 and 10.4 days, respectively. For our analysis, we
adopt 20\% uncertainties in each parameter. Spectroscopic observations by the same
authors showing broad Balmer emission lines (FWHM of H$\beta \simeq 3900$~km~s$^{-1}$) along with the absence of significant \ion{Fe}{2} emission
establish that the nova belongs to the He/N class.
Assuming no eruptions were missed going back to 1971, the recurrence time of LMCRN 1971-08a, $P_\mathrm{rec}=37.7$~yr.

\subsubsection{LMCN 2009-05a}
LMCN 2009-05a was discovered by \cite{Liller2009b} at $m\sim12.1$
on 2009 May 4.994~UT. We assume an uncertainty of 0.5~mag
in the peak brightness. Photometric and spectroscopic
observations by \citet{Walter2012} starting $\sim3$ days after discovery
show that the object was
a slowly fading \ion{Fe}{2} nova. Based on the $V$-band
photometry of \cite{Walter2012}, we estimate $t_2\sim51\pm19$~d
(see Figure~\ref{fig3}).
The nova was also observed by the OGLE survey, but long after maximum
so no useful data for determining decline rate parameters.

\subsubsection{LMCN 2010-11a = LMCRN 1968-12a}
Another eruption of RN 1968-12a occurred within a day of 2010 November 20.2~UT.
According to \cite{Mroz2014}, the nova reached a peak magnitude of $V=11.7\pm0.3$
before subsequently fading at a rate consistent with $t_2\simeq3.5$~d and
$t_3\simeq5.0$~d. These observations provide arguably the best coverage of
any recurrence of LMCRN 1968-12a, although previous eruptions were estimated to have reached
peak magnitudes that were $\sim0.5$mag brighter.

The light-curve of LMCN 2010-11a shows a periodicity at $P=1.26432(8)$ day,
which likely represents the orbital period of the system.
If so, the secondary star in LMCN 2010-11a must be slightly evolved.

\subsubsection{LMCN 2011-08a}
LMCN 2011-08a was discovered in archival data from
the OGLE survey \citep{Mroz2016a}. The nova was at $I\sim13.1$ and fading
rapidly at the time of discovery with characteristic decline times estimated to be
$t_2\lessim9$~d and $t_3\lessim14$~d. Unfortunately,
maximum light was missed, so we are unable to include this nova in our analysis.

\subsubsection{LMCN 2012-03a}
LMCN 2012-03a was discovered by \cite{Seach2012} on 2012 March 26.397~UT
at $m=10.7$. The nova declined rapidly after discovery,
with $t_2\sim1.1$~d and $t_3\sim2.1$~d as
estimated by \cite{Walter2012}.
Unfortunately, available data do not
tightly constrain the time of maximum light.
For the present analysis, we assume $V_\mathrm{max}=10.5\pm0.5$,
and take $t_2=1.1\pm0.2$~d.
Extensive multiwavelength observations were reported by \cite{Schwarz2015}.
Spectroscopic observations by \cite{Prieto2012}
on March 27 and less than a day after discovery established
the nova as a member of the He/N class.

\subsubsection{LMCN 2012-10a}
LMCN 2012-10a was first recognized at $I=13.9$ by \cite{Wyrzykowski2012}
on 2012 November 04.358~UT
during the course of the OGLE survey. Available
data show that the nova reached maximum light between
2012 October 22.374 and 25.310~UT. The October 25.310~UT image showed
the nova saturated, with an $I$-band magnitude between 11 and 12.
The $t_2(I)$ and $t_3(I)$ times were initially estimated to be $\sim10$~d and $\sim15$~d,
and later refined by \cite{Mroz2016a} to be 12 and 22 days, respectively.
For the subsequent analysis, we adopt $I_\mathrm{max}=11.5\pm0.5$, along with $t_2=10\pm2$~d.
Spectroscopy by \cite{Walter2012} shows
the nova to be a likely member of the He/N class.

\subsubsection{LMCN 2013-10a}
LMCN 2013-10a was discovered on the rise by 
the OGLE survey at $I\simeq13.2$ \citep{Wyrzykowski2013}.
The maximum appears to be fairly well covered with \cite{Mroz2016a}
reporting $t_2$ and $t_3$ times of 52 and 67 days, respectively.
The nova was also observed by the SMARTS program in multiple bandpasses \citep{Walter2012}.
Figure~\ref{fig3} shows the $V$-band light-curve based on the SMARTS data.
A best-fit to these data suggests that $t_2=24\pm3$~d. \cite{Hachisu2018} have analyzed
the same data and concluded that the nova reached $V=11.5$ before declining
with $t_2\simeq21$~d and $t_3\simeq47$~d. For the present analysis, we
adopt $V_\mathrm{max}=11.5\pm0.3$ and $t_2=24\pm3$~d as reasonable estimates
for these parameters.
The spectroscopic class of LMCN 2013-10a is unknown.

\subsubsection{LMCN 2015-03a}
LMCN 2015-03a (ASASSN-15fd) was discovered by \cite{Danilet2015} as part of the All-Sky Automated Survey for SuperNovae (ASAS-SN) survey \citep{Kochanek2017}
on 2015 March 15.12~UT at $V=14.3$, rising to $V=13.2$ on March 17.99. Continuing
observations by ASAS-SN showed that the nova reached $V=11.5$ on March 21.99 \citep{Stanek2015}.
The nova was also covered by SMARTS, and Figure~\ref{fig3} shows the resulting light-curve. A fit to below
2 mag from peak reveals that $t_2\simeq49\pm8$~d. The light-curve is a bit jittery,
and we estimate an uncertainty in the magnitude at maximum to be of order 0.3~mag.
Spectroscopic observations by \cite{Walter2015}
establish the nova to be a member of the \ion{Fe}{2} class.

\subsubsection{LMCN 2016-01a = LMCRN 1968-12a}
LMCN 2016-01a was the 4th recorded eruption of LMCRN 1968-12a.
The 2016 eruption was discovered by the OGLE survey
on 2016 January 21.2~UT at $I=11.5$ \citep{Mroz2016b} and followed up by \cite{Munari2016} who found
decline rates consistent with those found for the 2010 eruption:
$t_2=4.5$~d and $t_3=6.7$~d.
Spectroscopic observations by \cite{DiMille2016} revealed broad Balmer (FWZI=10000~km~s$^{-1}$) and He~I
lines consistent with a He/N classification.
The most comprehensive study of LMCN 2016-01a was conducted by \cite{Kuin2020} a few months
before the 2020 eruption in May of that year. These authors compiled available data from
previous eruptions and settled on characteristic values of $t_2=4.6\pm0.5$~d and $t_3=7.0\pm1.0$~d.

\subsubsection{LMCN 2016-04a}
LMCN 2016-04a was discovered by \cite{Gorbovskoy2016} on 2016 May 10.728~UT at $m=12$
as part of the MASTER program \citep{Lipunov2010}.
Unfortunately, no further observations were available until the OGLE survey picked up the
nova below the 16th magnitude. We assume an uncertainty of 0.5 mag in the peak brightness.
Figure~\ref{fig3} shows the available photometry. Assuming that
the nova declined linearly (in mag) between the discovery and the first OGLE detection,
we estimate $t_2\simeq50\pm10$~d, where the uncertainty has been assumed to be of order 20\%.
Given the sparsity of data, the derived $t_2$ value is best considered an upper limit.

The progenitor system, which is visible in OGLE data at $I\simeq20.64$, appears to
show eclipse-like variability with a period of $P=2.6508\pm0.0002$~d.
Assuming this is the orbital period, the secondary star in LMCN 2016-04a must be somewhat evolved.
A spectrum obtained by \cite{Bohlsen2016} shows strong and broad Balmer emission
lines (FWHM H$\alpha \simeq3000$~km~s$^{-1}$, with similarly broad, but
weaker He and N emission features consistent with a He/N (or possibly Hybrid) classification.

\subsubsection{LMCN 2017-11a}
LMCN 2017-11a was discovered by \cite{Chomiuk2018a} as part of the ASAS-SN survey. The nova
evolved slowly reaching a maximum of $V=11.8$ on 2017 December 7~UT. It then declined
very slowly over the next several months as described in \cite{Aydi2019a}. We estimate the uncertainty
in the peak magnitude to be $\sim0.1$~mag, with that in $t_2$ to be of order 20~d. The
spectrum is that of a very slow \citep[J class in the][classification system]{Strope2010} \ion{Fe}{2} nova with
an estimated $t_2\grtsim100$~d.

\begin{figure}
\epsscale{1.3}
\plotone{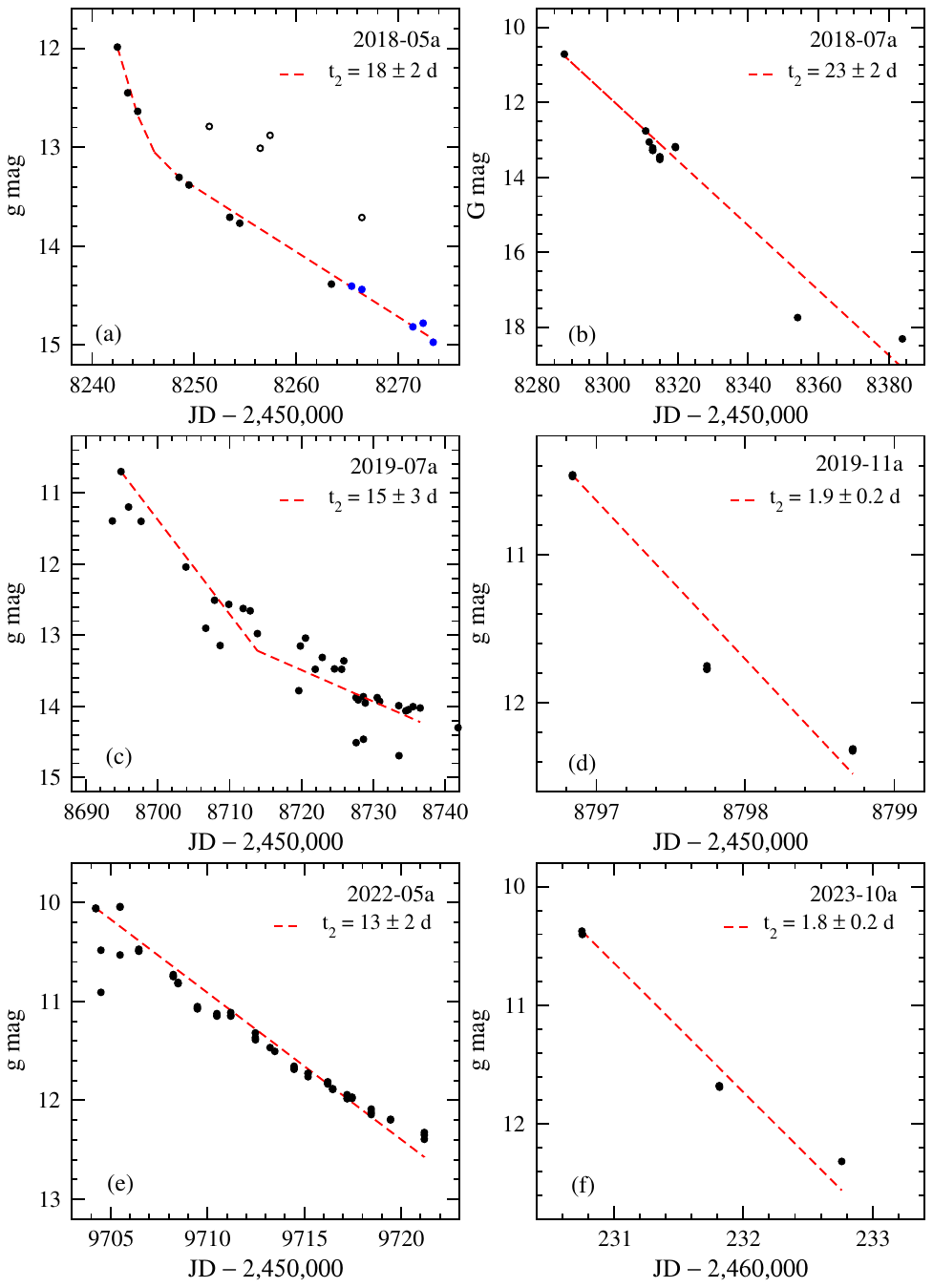}
\caption{Light curves for selected novae used for the determination of $t_2$ times.
(a) LMCN 2018-05a: $R$-band photometry from SMARTS (black points); $I$-band GAIA data (blue points). A two-component fit to data
    brighter than $m=15$ yields an estimated $t_2=18\pm2$~d.
(b) LMCN 2018-07a: A linear fit to GAIA photometry shows a 0.087~mag~d$^{-1}$ decline rate, corresponding to an estimated $t_2=23\pm2$~d.
(c) LMCN 2019-07a: An initial decline rate of $0.132\pm0.016$~mag~d$^{-1}$ yields $t_2=15\pm3$.
(d) LMCN 2019-11a: A linear fit to the limited ASAS-SN photometry suggests $t_2=1.9\pm0.2$~d.
(e) LMCN 2022-05a: $g$-band light-curve based on ASAS-SN data. The light declines at a rate of 0.15 mag~d$^{-1}$,
corresponding to $t_2=13\pm2$~d.
(f) LMCN 2023-10a: $g$-band photometry from the ASAS-SN survey showed the nova declined rapidly
    at a rate of 1.1 mag~d$^{-1}$, with $t_2=1.8\pm0.2$~d.
}
\label{fig4}
\end{figure}

\begin{figure}
\epsscale{1.3}
\plotone{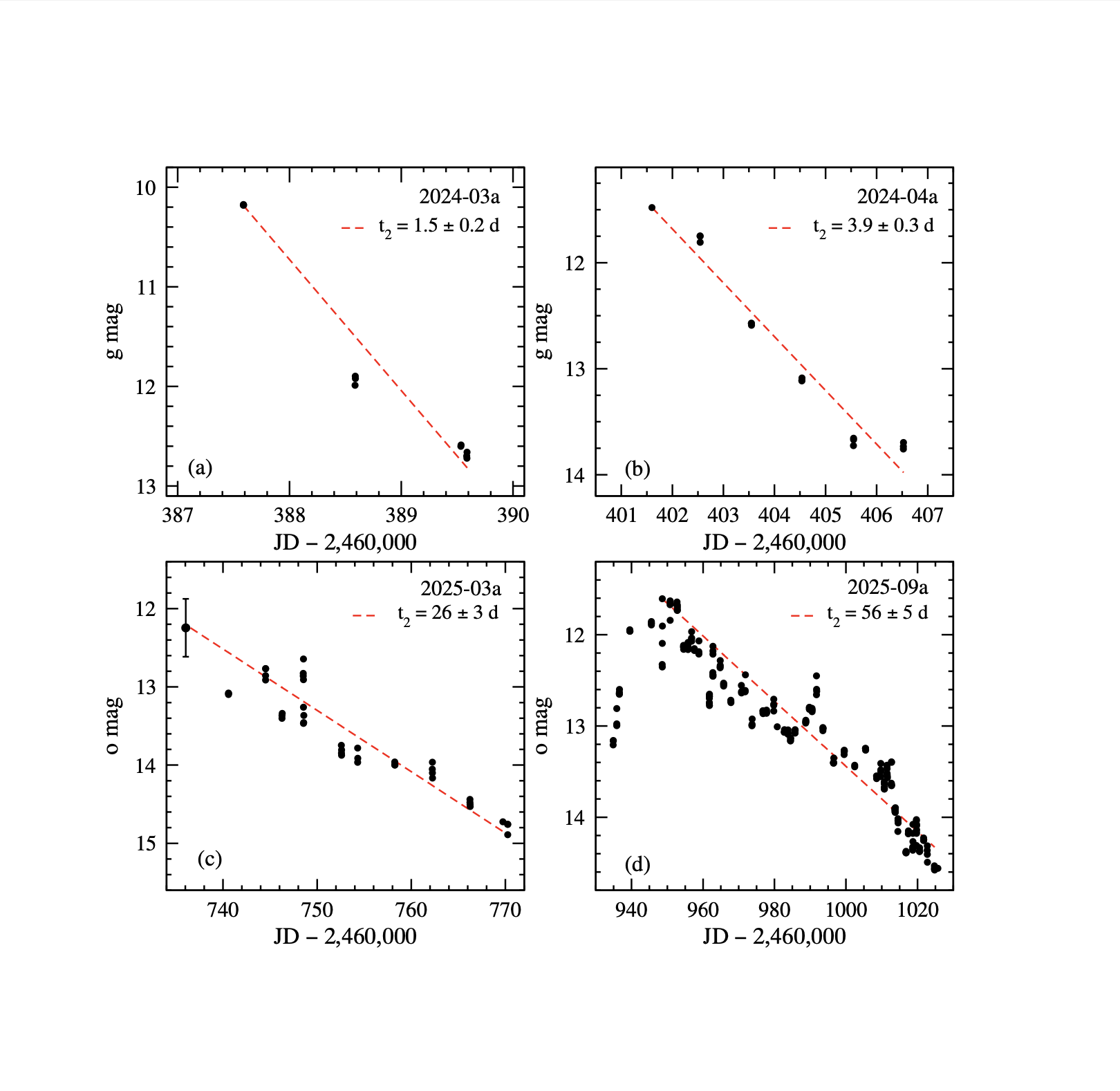}
\caption{Light curves for selected novae used for the determination of $t_2$ times.
(a) LMCN 2024-03a: $g$-band photometry from ASAS-SN shows a very rapid decline (1.3 mag~d$^{-1}$) characterized by $t_2=1.5\pm0.2$.
(b) LMCN 2024-04a: ASAS-SN $g$-band photometry showing a decline rate of 0.51~mag~d$^{-1}$, corresponding to $t_2=3.9\pm0.3$~d.
(c) LMCN 2025-03a: $o$-band light-curve based on ATLAS data. The light declines at a rate of 0.15 mag~d$^{-1}$,
    corresponding to $t_2=26\pm3$~d.
(d) LMCN 2025-09a: $o$-band photometry from the ATLAS survey showed the nova declined slowly
    with an estimated $t_2=56\pm5$~d.
}
\label{fig5}
\end{figure}

\subsubsection{LMCN 2018-02a = LMCRN 1996-11a}
LMCN 2018-02a was observed to be spatially coincident with a relatively
unstudied nova, LMCN 1996-11a, establishing the object as
the 4th known RN in the LMC and providing the opportunity to determine
the nova's peak luminosity and rate of decline. The 2018 eruption
was well covered by the ASAS-SN survey \citep{Chomiuk2018b}, An examination of the ASAS-SN light-curve
suggests $V_\mathrm{max}=11.0\pm0.2$, with $t_2\simeq=4\pm1$ and $t_3\simeq8\pm1$.
The progenitor was detected in the OGLE images with a mean brightness of $I=19.73$ mag along with
eclipse-like variability with a period of $2.84995\pm0.00003$~d \citep{Mroz2018}. Assuming this is the
orbital period of the system, the secondary star must be somewhat evolved.
Spectroscopic observations of the 2018 eruption establishes the nova as a member of the He/N class \citep{Walter2018}.
If no recent eruptions have been missed, the recurrence time of LMCRN 1996-11a, $P_\mathrm{rec}=21.3$~yr.

\subsubsection{LMCN 2018-05a}
LMCN 2018-05a was discovered on 2018 May 3 23:02~UT at $g=12.1$ by the ASAS-SN survey \citep{Chomiuk2018c}.
The ASAS-SN light-curve supplemented by SMARTS $V$-band data at late times
is shown in Figure~\ref{fig4}.
From these data, we estimate $t_2\sim18\pm2$~d and $R_\mathrm{max}=12.0\pm0.2$.
We note that the nova was very red ($V-R\simeq1.0$) based on the SMARTS data,
and that the decline from outburst in GAIA data was significantly
slower ($t_2\simeq45$~d) than seen in the ASAS-SN and SMARTS data.
Spectroscopy reported by \citep{Chomiuk2018c}, which was taken
within 2 days of peak brightness,
is suggests a possible \ion{Fe}{2} classification.

\subsubsection{LMCN 2018-07a}
LMCN 2018-07a was first reported
by \cite{Stanek2018} during the course of the ASAS-SN survey.
A search of archival GAIA data reveals that the nova was first detected
at $G=10.7$ on 2018 June 18~UT. The available photometry\footnote{\tt https://gsaweb.ast.cam.ac.uk/alerts/alert/Gaia18bvg/}
yields the light-curve shown in Figure~\ref{fig4}. The maximum was not well covered,
but a fit to the data gives $t_2=23\pm2$~d. For the purposes of our analysis,
we conservatively estimate $V_\mathrm{max}=10.5\pm0.5$, and take $t_2=23\pm2$.

\subsubsection{LMCN 2019-07a}
LMCN 2019-07a (AT 2019lvm)
was discovered by \cite{Jacques2019} on 2019 July 27.3496~UT
at $m=12.6$ as part of the Brazilian Transient Search (BraTS) program.
The nova was subsequently seen to
rise to $m\sim10.7$ (unfiltered) by August 1.
The eruption was covered by ASAS-SN, and the available photometry
is shown in Figure~\ref{fig4}, supplemented by the BraTS magnitude
at maximum light. A fit to the light-curve yields a fade rate of
0.133 mag~d$^{-1}$, corresponding to $t_2=15\pm3$~d. This value
is somewhat shorter, but consistent within the uncertainties, to a value
of $t_2\simeq21$~d found by T. A. Napoleao from AAVSO data.
A spectrum taken on 2019 July 29.38~UT by \cite{Aydi2019b}
is consistent with a nova at or near maximum light, but precludes
the determination of the spectroscopic class.

\subsubsection{LMCN 2019-11a}
LMCN 2019-11a (AT 2019uni) was discovered by the BraTS
\citep{Pimentel2019} on 2019 November 11.2263~UT at $m=12.0$.
Subsequent observations by ASAS-SN
showed that the nova reached $g=10.46$ before declining at a rate of
$\sim1.07$ mag~d$^{-1}$ (see Figure~\ref{fig4}), which corresponds to $t_2=1.9\pm0.2$~d.
High-resolution spectroscopy by \cite{Aydi2019c} established that the
nova is a member of the He/N class.

\subsubsection{LMCN 2020-05a = LMCRN 1968-12a}
LMCN 2020-05a was the sixth eruption of LMCRN 1968-11a. An ASAS-SN light-curve is available for
the 2020 eruption, but the data are sparse. The partial ASAS-SN light-curve
shows a decline rate
suggesting $t_2\sim6.5$~d, but otherwise no useful estimates
of the rate of decline are available.

\subsubsection{LMCN 2020-11a}
LMCN 2020-11a was discovered on 2020 November 17.32~UT at $g\simeq11.2$
during the course of the ASAS-SN survey \citep{Stanek2020}. Spectroscopic
observations by \cite{Aydi2020} reveal broad Balmer emission
(FWZI$\sim9000$~lm~s$^{-1}$) along with \ion{Fe}{2} features. It appears that
the nova is likely a member of the broad-lined \ion{Fe}{2} (Hybrid)
spectroscopic class. Unfortunately, it does not appear
that follow-up photometry suitable
for determining the rate of decline has been published.

\subsubsection{LMCN 2022-05a}
LMCN 2022-05a was discovered on 2022 May 21.8916~UT by the BraTS team
Observations taken as part of the
ASAS-SN survey showed that the eruption actually began approximately 3 weeks earlier on 2022 April 30.766~UT. The light-curve
was well covered by ASAS-SN rising to a peak of $g=10.05$ on
May 5.9~UT \citep{Aydi2022}. The ASAS-SN light-curve is reproduced
in Figure~\ref{fig4},
from which we have determined $g_\mathrm{max}=10.1\pm0.1$ and $t_2=13\pm2$~d.
Spectroscopic observations reported by \cite{Aydi2022} noted only that the spectrum displayed strong emission lines of \ion{H}{1}, \ion{Fe}{2},
and \ion{O}{1}.
From their description, it is likely that the nova belonged either to the \ion{Fe}{2} or Hybrid spectroscopic class,
but a definitive determination cannot be made.

\subsubsection{LMCN 2023-10a}
LMCN 2023-10a (ASASSN-23hd) was discovered on 2023 October 13.9~UT in the ASAS-SN survey at $g=10.37$ \citep{Strader2023}.
A prediscovery nondetection ($g>18$) was recorded approximately a day earlier,
confirming that the maximum was likely covered. The nova faded rapidly as shown by subsequent ASAS-SN data (see Figure~\ref{fig4}). From
these data we deduce $g_\mathrm{max}=10.4\pm0.1$ and that $t_2=1.8\pm0.2$~d. Spectroscopic observations reported by \cite{Strader2023}
revealed strong emission lines of H I, He I, N II, and N III. The Balmer lines were characterized by extremely broad wings
(extending up to $\pm6000$~km~s$^{-1}$ in H$\beta$ and up to $\pm10000$ in H$\alpha$)
consistent with a He/N classification.

\subsubsection{LMCN 2024-03a}
LMCN 2024-03a (ASASSN-24by, AT 2024epj) was discovered on 2024 March 18.08~UT at $g=10.2$, also by ASAS-SN. The peak brightness is reasonably secure
given that the last nondetection showed $g>17.1$ on the previous day, so we assume an uncertainty of 0.1~mag. Follow-up photometry by ASAS-SN and by \cite{Perez-Fournon2024} showed that
the nova faded very quickly after reaching maximum light. Figure~\ref{fig5}
shows the resulting light-curve, from which we derive $t_2=1.5\pm0.2$~d.
Spectroscopic observations made $\sim$2 days postmaximum by \cite{Merc2024} show H$\alpha$ (FWHM $\simeq 6600$~km~s$^{-1}$, along
with possible He I and N III emission. The novae appears to be a member of the He/N class.

\subsubsection{LMCN 2024-04a}
LMCN 2024-04a (ASASSN-24ck, AT 2024fjh) was discovered on 2024 April 1.1~UT at $g=11.48$, once again by ASAS-SN \citep{Aydi2024b}.
Peak brightness is again well constrained ($\pm$0.1 mag assumed)
with the last nondetection $\sim1$ day prior ($g>18.3$).
The ASAS-SN light-curve,
which is reproduced in Figure~\ref{fig5}, shows that the nova declined rapidly at a rate characterized by $t_2=3.9\pm0.3$~d.
Spectroscopic observations by \cite{Aydi2024b} and \cite{Shore2024a} show broad Balmer, He~II and N~III emission lines (FWHM H$\alpha \sim 6000$~km~s$^{-1}$)
consistent with a He/N classification.

\subsubsection{LMCN 2024-08a = LMCRN 1968-12a}
LMCN 2024-08a is the most recent (and 7th) recorded eruption of
LMCRN 1968-12a that was detected on 2024 August 1.8~UT with the Swift satellite \citep{Darnley2024}.
Multiwavelength observations reported by \cite{Basu2025} suggest a fade rate of $\sim0.8$~mag~d$^{-1}$.
Spectroscopic observations by \citet{Shore2024b} reveal broad (FWZI $\sim10000$~km~s$^{-1}$) Balmer and He~II emission.

\begin{figure}
\plotone{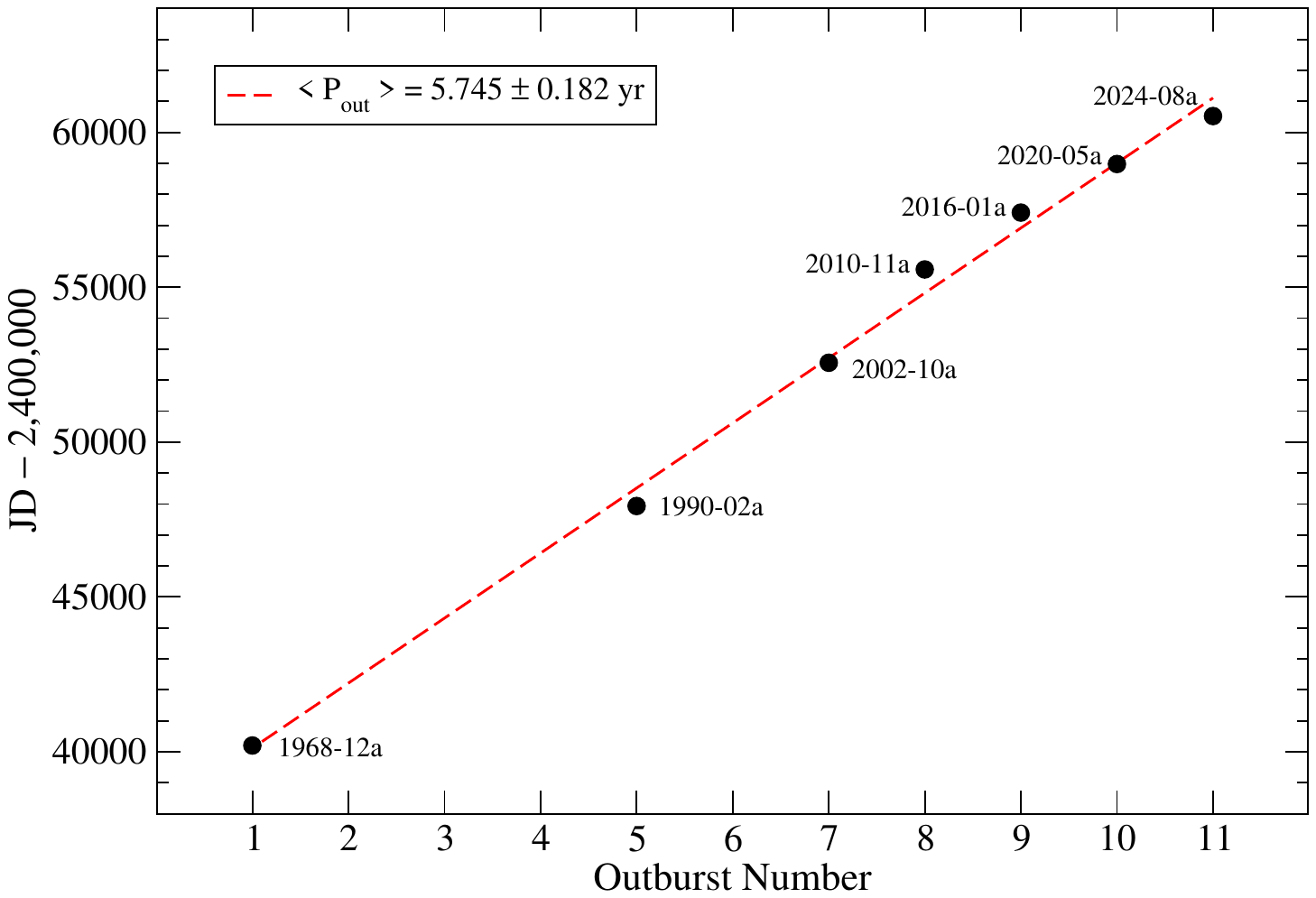}
\caption{The Julian Dates of the observed eruptions of LMCRN 1968-12a plotted
as a function of the expected outburst number. The best-fitting slope
gives the mean recurrence period:
$\langle P_\mathrm{rec}\rangle = 2098.42\pm66.45$~d,
or $5.75\pm0.18$~yr. Expected outbursts Nos. 2, 3, 4, and 6 that most likely
occurred near 1974, 1980, 1985, and 1996 were apparently missed.}
\label{fig6}
\end{figure}

Given that LMCN 2024-08a is the most recent eruption of LMCRN 1968-12a, it is worth reviewing the timings
of all previous observed eruptions in order to estimate the mean recurrence period. Figure~\ref{fig6} shows the
times of all 7 known eruptions of LMCRN 1968-12a plotted as a function of the most likely outburst number
(it appears that eruptions in or around 1974, 1980, 1985, and 1996 were likely missed).
The slope of the best-fitting line yields a mean outburst period of
$\langle P_{rec}\rangle = 5.75\pm0.18$~yr.

In reviewing the peak magnitude and decline rate estimates of all previous eruptions,
for the analysis to follow
I adopt $V_\mathrm{max}=11.5\pm0.3$ and $t_2\simeq3.5$~d as the overall best estimates of the light-curve
parameters for LMCRN 1968-12a.

\subsubsection{LMCN 2025-03a}
LMCN 2025-03a is a peculiar transient source that was seen to brighten
from a relatively
bright quiescent level of $m\sim16.8$ to a peak brightness at
$m\sim12.2$, before
slowly returning to its quiescent level over the next several
months \citep{Mikolajczyk2025}.
Fortunately, the eruption was observed by the ATLAS survey. The resulting
light-curve is shown in Figure~\ref{fig5}. From these data we were able
to estimate a peak magnitude of $o=12.2$ and
an initial decline rate of 0.077~mag~d$^{-1}$,
corresponding to $t_2=26\pm3$.
The bright quiescent magnitude and relatively small amplitude of the
eruption are reminiscent of the Be+WD nova system M31N 2017-01e
\cite{Chamoli2025}.

\subsubsection{LMCN 2025-09a}
LMCN 2024-09a was discovered
by the Gravitational-wave Optical Transient Observer (GOTO) facility
at $L=12.8$ on
2025 September 17.669~UT \citep{O'Neill2025}. The object was subsequently seen
to brighten to L=12.32 on September 18.6743~UT.
Observations by the ATLAS survey covered the eruption,
which had a slow rise over approximately 2 weeks,
reaching a maximum of $o=11.6$ on 2025 September 29.59~UT.
The light-curve, which is shown in Figure~\ref{fig5}, yields
$t_2=56\pm5$~d.
A spectrum obtained on September 18.3878~UT shows
relatively narrow Balmer and \ion{Fe}{2} lines consistent with those
seen in typical \ion{Fe}{2} class novae.

\section{Observed Light-curve Properties}

Of the 66 candidate nova eruptions just discussed,
18 have insufficient photographic coverage to
allow their light-curves to be characterized. Of the remaining 48 eruptions,
9 are recurrences of the 4 known RNe in the LMC
(6 associated with LMCN 1968-12a, and one each for the other 3 RNe).
Thus, we are left with light-curve data for
39 unique nova progenitors that have
sufficient photometric data to enable peak magnitudes and rates
of decline to be determined.
The resulting light-curve properties for these systems
are summarized in Table~\ref{tab2}.
We have also included the spectroscopic classification as defined
in \cite{Williams1992}, when it is available.

\subsection{The Rates of Decline}

The rates of decline are typically parameterized
in terms of the number of days a nova takes
to fade either 2 or 3 magnitudes from maximum
light, $t_2$ or $t_3$.
In some instances both parameters
have been measured and reported for a given nova.
In the analysis to follow I will be
focusing on the $t_2$ time, which is more easily determined and has
become the standard in recent years. As pointed out by \cite{Shara2018},
it is also
approximately equal to the mass loss timescale, $t_\mathrm{ml}$,
that will be used in Section 5.2 as an input to the \citet{Yaron2005} nova models.

In cases where only one of the $t_2$ or $t_3$ times has been published 
(e.g., typically only the $t_3$ time was reported for the older novae),
I have employed the empirical
transformations between $t_3$ and $t_2$ recently determined
by \cite{Shafter2026b} to determine the corresponding parameter. The resulting
transformed value is given in square brackets in Table~\ref{tab2}.
When uncertainties in $t_2$ and $t_3$ have not been reported explicitly, the
uncertainty has been estimated based on the scatter in the light-curves,
and is typically in the range of 10--20\%.

\subsection{The Peak Absolute Magnitudes}

Before we can explore the fundamental properties of the LMC nova
population, the absolute magnitudes at the peak of the eruption
must be determined. This task is simplified because
we can assume the novae lie at the well-determined
distance of the LMC.
I adopt a distance modulus for the LMC of $(m - M)_o = 18.477\pm0.026$
\cite{Pietrzynski2019}, and assume
a line-of-sight foreground reddening of $E(B-V) = 0.06\pm0.02$ mag
\citep{Oestreicher1995,Gochermann2002,Chen2022}.
Assuming a ratio of total to selective extinction of $R=3.1$, we find that
the absolute visual magnitude at the peak of the eruption is given by
$M_{V,\mathrm{o}} = m_{V,\mathrm{obs}} - 18.663\pm0.067$, where $m_{V,\mathrm{obs}}$ is
the visual apparent magnitude observed at maximum light.
We propagate all uncertainties
when determining the value of $M_{V,\mathrm{o}}$, but
in virtually all cases, the error estimates for $m_{V,\mathrm{obs}}$
dominate over the modest uncertainty in the distance modulus.
Given that not all observations of LMC novae have been taken in the $V$ filter,
we must make corrections, as necessary, to convert the available
photometry to the $V$ band.

\subsubsection{Filter Transformations}

Photometric observations of LMC novae have been acquired over time
using a variety of instruments and bandpasses.
Early data were mostly reported using photographic ($pg$) magnitudes,
while later observations (e.g., ASAS-SN) often employed more
modern photometric systems.
Although in most cases the uncertainty in the measurements themselves
is of order any corrections for bandpass,
we have attempted to correct all peak magnitudes
to the Johnson-Cousins $V$ band \citep{Bessell1990}.

\begin{figure*}
\plottwo{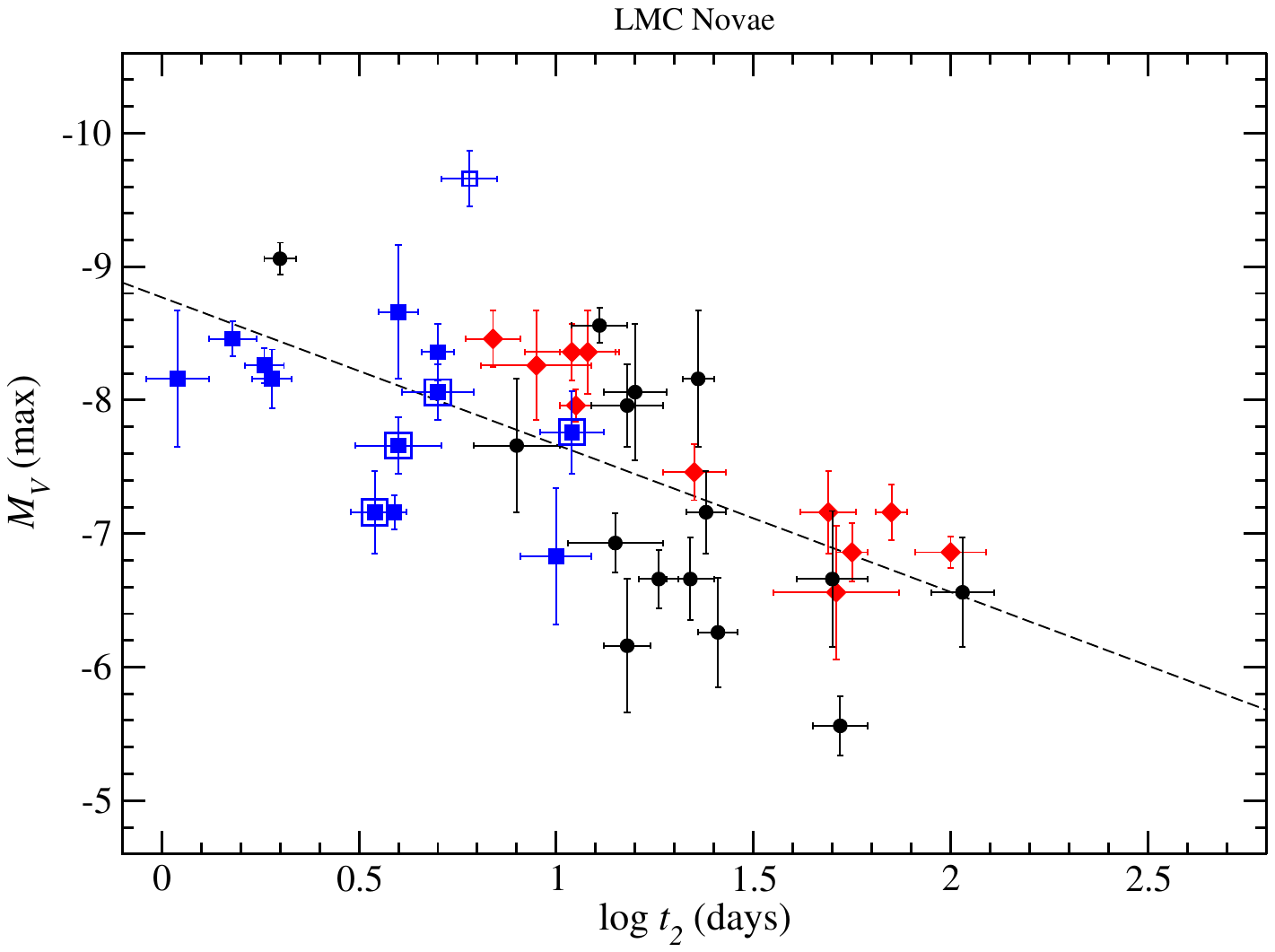}{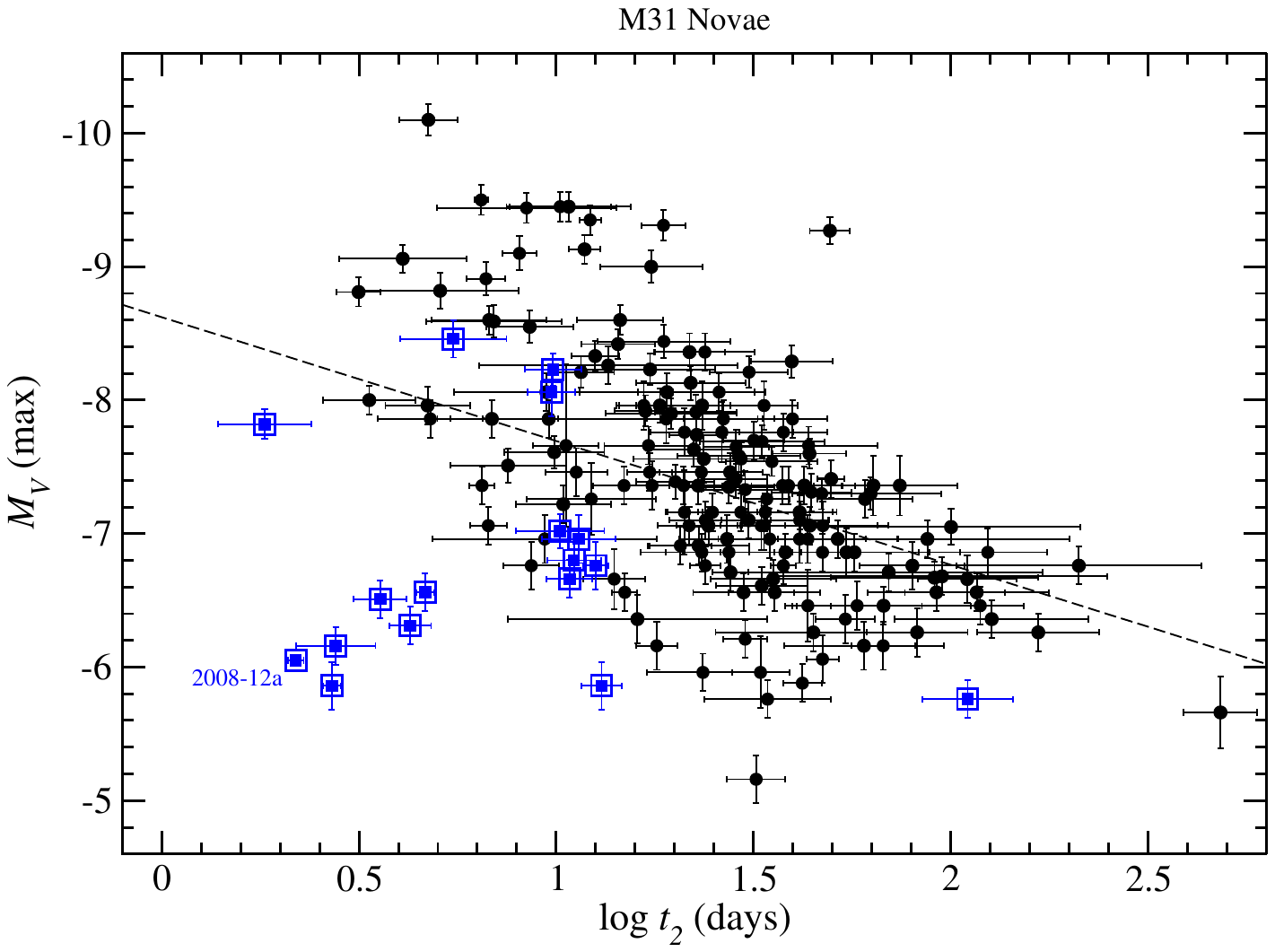}
\caption{Left panel: The MMRD plot for the LMC
(blue points: He/N novae; red points: \ion{Fe}{2} novae).
Right panel: The MMRD plot for M31 (on the same scale)
based on data from \cite{Clark2024}.
The large blue squares in both plots indicate known RNe.
The dashed lines show the best linear fits with
$M_V = -8.77(\pm0.26) + 1.10(\pm0.22)~\mathrm{log}~t_2$
and
$M_V = -8.62(\pm0.21) + 0.93(\pm0.15)~\mathrm{log}~t_2$
for the LMC and M31, respectively.}
\label{fig7}
\end{figure*}

We begin by adopting the following colors for typical novae near maximum light
as determined by \cite{Craig2025}:
$B-V=0.20\pm0.06$, $V-R=0.20\pm0.05$ and $R-I=0.13\pm0.04$
(from their Silver sample). Thus, for $B$ and $R$-band measurements
we have $V=B - (0.2\pm0.06)$, $V=R + (0.20\pm0.05)$,
and $V=I + (0.33\pm0.06)$, respectively.
Assuming the effective wavelenghth of the ATLAS $o$ bandpass is
similar to $R$, we have $V\simeq o + (0.20\pm0.05)$.
In the case of the early photographic ($pg$) magnitudes,
we make use of the relation derived by \cite{Arp1961}:
$B-pg \simeq 0.29 - 0.18(B-V)$, which,
for novae near maximum light, gives $V \simeq pg + 0.1$.
For the $g$ magnitudes reported by ASAS-SN,
we use the following relation from \cite{Jester2005}:
$g-V\simeq0.60(B-V) -0.12$,
which, for novae near peak, gives $V\simeq g$, and no correction is needed.
Similarly, for the GAIA measurements we assume $V\simeq G$.
Finally, for unfiltered or visual estimates,
we also assume they are equivalent within the uncertainties of the measurements
to the $V$ magnitude. Transformation uncertainties of 0.05~mag are assumed when
converting from $g$, $G$, and unfiltered visual magnitudes to the $V$ band.

\subsection{The Calibrated Nova Light-curve Properties}

The final calibrated light-curve properties for the 39 LMC nova
progenitor systems are presented in Table~\ref{tab3}.
In the case of the 4 recognized RN systems, we have combined the observed
light-curve properties for the individual eruptions from Table~\ref{tab2}
to form a representative light-curve for that particular RN.
The calibrated light-curve properties given in Table~\ref{tab3} form
the primary input data for the analysis to follow. We begin by
determining an MMRD relation for our LMC sample of novae, followed by a comparison of
the LMC $M_V$ and $t_2$ distributions with their counterparts in M31 and the Galaxy.

\section{The LMC MMRD relation}

The MMRD relation is an empirical
correlation between the peak luminosity of a nova and its subsequent
rate of decline. It was first described as a ``Life-Luminosity Relation''
by \citep{McLaughlin1945}
in his classic paper showing that the absolute magnitudes
of a large sample of Galactic, M31 and LMC novae were correlated with
their $t_3$ times. Specifically, the brightest novae were seen to
generally fade the most rapidly.
Over the years, the relation has been recalibrated many times, and has
been the subject of mounting criticism as modern surveys have
uncovered an increasing number of
faint yet fast novae that violate
the relation \citep[e.g., see,][]{Kasliwal2011,Shara2016,Shafter2023}.

The most effective approach for evaluating the MMRD relation is to
apply it to equidistant samples of novae in nearby galaxies where
the absolute magnitudes can be most reliably determined. With this
goal in mind, \cite{Clark2024} explored the MMRD relation in M31
using a large and homogeneous
sample of $R$-band light-curves, mostly obtained by K. Hornoch
with the 0.65-m reflector at Ond\v{r}ejov in the Czech Republic.
The MMRD relation in the LMC has been explored previously by
\cite{Capaccioli1990a,Capaccioli1990b} and in Paper~I; however,
given the wealth of photometric data for LMC novae that has been amassed 
in recent years, it is worth revisiting the MMRD relation for the
LMC and comparing it with that recently determined for M31.

The updated MMRD relation for the LMC shown in the left panel of Figure~\ref{fig7}
is based upon the 39 novae with reliable light-curve parameters
given in Table~\ref{tab3}. A clear trend between the peak luminosity
and the rate of decline is evident, as expected, with the most luminous novae
generally fading the fastest. For comparison, we have reproduced
in the right panel of Figure~\ref{fig7}
the MMRD relation for M31 found by \cite{Clark2024},
where the scale and range
of the axes of the two figures have been adjusted to match. It is clear that the
average $t_2$ time of the LMC novae is shorter (faster) than
the corresponding average of the M31 sample. Although less obvious,
the average luminosity of the LMC novae is also brighter.

\begin{figure}
\plotone{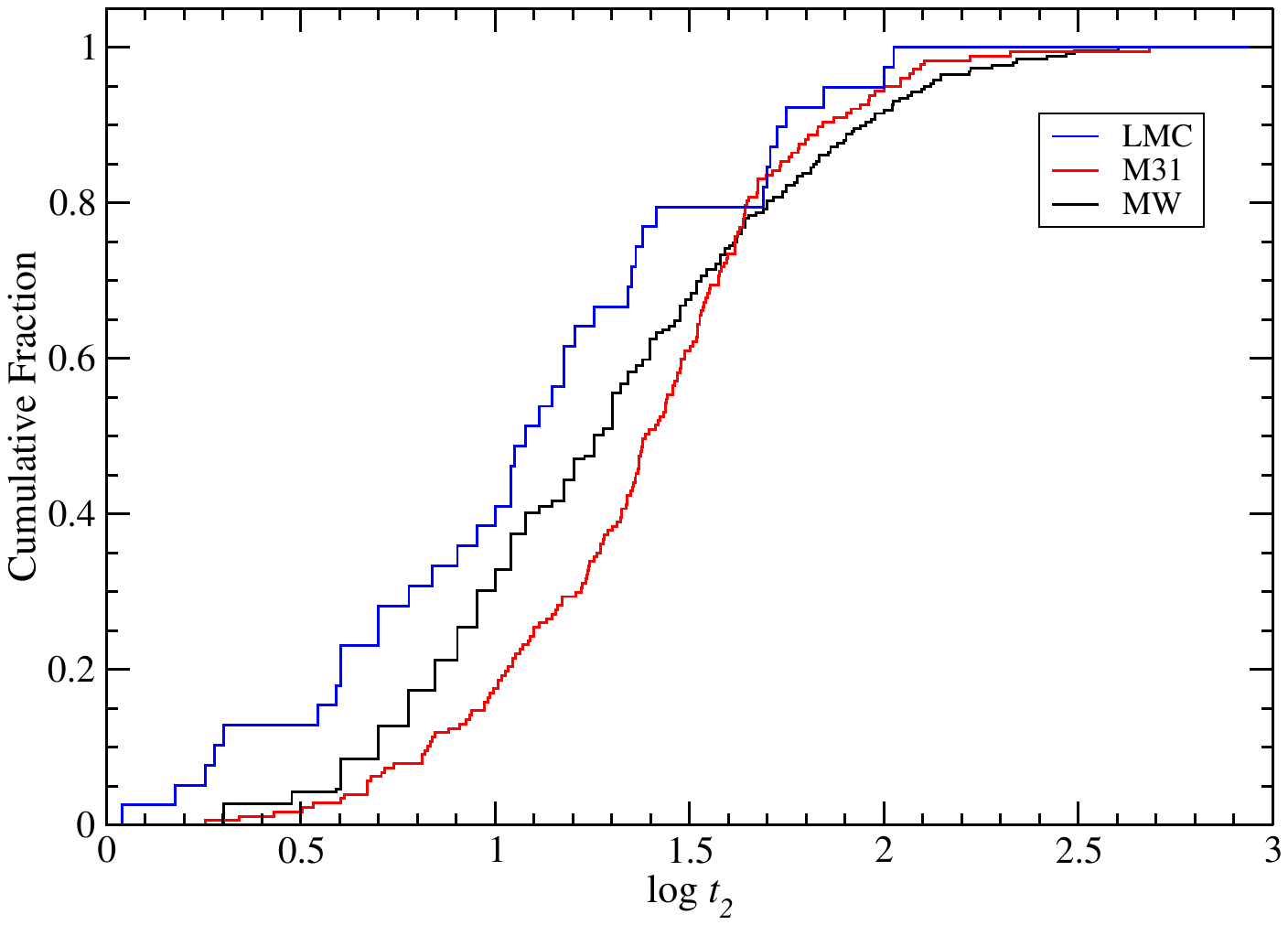}
\caption{The cumulative fractions of the $t_2$ times for the LMC nova sample compared with those
of M31 and the Galaxy (MW). The LMC and Galactic samples yield a K-S probability
$p=0.159$ and thus differ at only $\sim1.4\sigma$. On the other hand, both
the LMC and Galactic samples differ from the M31 sample by $3.3\sigma$
(K-S $p=9.8\times10^{-4}$) and
$2.9\sigma$ (K-S $p=3.9\times10^{-3}$), respectively.}
\label{fig8}
\end{figure}

\begin{figure}
\plotone{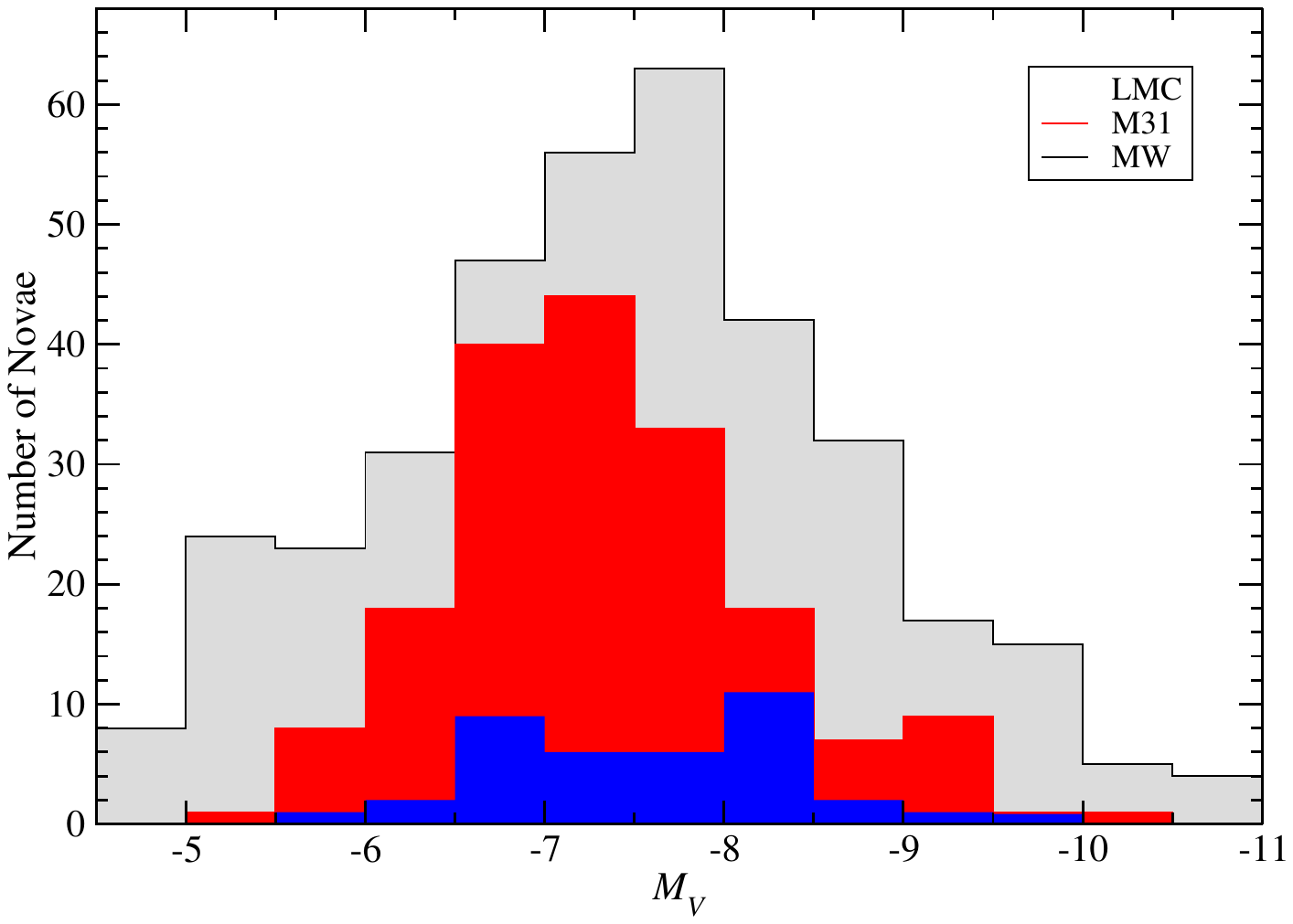}
\caption{The absolute visual magnitude distribution for the LMC nova
sample compared with the corresponding distributions for M31 and the Galaxy (MW).
The relatively small LMC sample does not differ significantly from either the M31 or Galactic
samples ($\sim1.2\sigma$ and $\sim1.3\sigma$, respectively), but the
larger Galactic and M31 samples differ at $>3\sigma$.}
\label{fig9}
\end{figure}

To better illustrate the differences in the decline times and the
peak luminosities of the LMC novae compared with M31 and the Galaxy,
we have computed cumulative distributions of the $t_2$ times
and overlapping histograms of the absolute magnitude
distributions for each galaxy.
The cumulative $t_2$ times are shown in Figure~\ref{fig8}, while
the absolute magnitude distributions are compared in Figure~\ref{fig9}.
Kolmogorov--Smirnov (K-S) tests reveal that the
$t_2$ distributions for the LMC and Galactic nova samples differ from
the M31 $t_2$ distribution with $p\sim10^{-3}$ and $p\sim4\times10^{-3}$,
respectively. However,
the difference between the LMC and Galactic samples
was not significant ($p\simeq0.16$).

A $\chi^2$ test has been used to explore the absolute magnitude
distributions shown in Figure~\ref{fig9}.
The $M_V$ distribution for the LMC nova sample (which is relatively small,
limiting the statistical power) does not differ significantly from
the M31 and Galactic nova samples
($\sim1.2\sigma$ and $\sim1.3\sigma$, respectively).
On the other hand, the larger
M31 and Galactic samples differ significantly ($p=1.93\times10^{-4}$), or
at $\sim3.7\sigma$. However, much of this difference occurs at the
faint end of the distribution. If we exclude novae with $M_V\geq-5.5$
(which are likely largely overlooked in M31), the difference
becomes barely significant ($\sim2.5\sigma$).

The principal takeaway from the MMRD analysis is that the nova population
in the LMC consists of novae that are both brighter and faster on average
compared with the nova population of M31. This result is consistent
with the difference in stellar population between the two galaxies. The
specific star formation rate in the LMC is significantly higher
than that seen in M31, with the LMC currently undergoing vigorous
episodic star formation \citep{Harris2009}. Conversely,
M31 is in a more quiescent phase typical of massive spirals
that have already formed most of their stars. Population synthesis models,
such as those of \cite{Yungelson1997} and \cite{Chen2016}, consistently
show that a younger average stellar population resulting from a
recent history of star formation produces novae that are brighter,
faster, with shorter recurrence times, compared with older populations.

\section{Fundamental Properties of LMC novae}

We now turn to a discussion of the fundamental properties of the
LMC nova population. The analysis closely follows the procedure
employed by \cite{Shafter2026a} in their study of the nova population in M31,
where the nova models presented by \cite{Yaron2005} were inverted
to use the observed photometric properties (peak luminosity
and rate of decline) of novae as input parameters
to estimate the WD masses, accretion rates, ejecta velocities,
and recurrence times for a large sample of M31 novae.
Here, I employ the Yaron et al. models to estimate the
fundamental properties of the 39 LMC novae in Table~\ref{tab3}, which
have photometric coverage sufficient
to characterize their peak luminosities and rates of decline ($t_2$ times).

The analysis
requires that both the bolometric luminosity
and the mass-loss timescale,
$t_\mathrm{ml}$, be specified as input parameters.
The bolometric luminosity, $L_4$ (in units of $10^4$~$L_\odot$),
has been determined from the absolute visual magnitude at maximum
light, $M_V$, as follows:
\begin{equation}
L_4 = 10^{(-4-0.4[M_V + \mathrm{BC} - M_\mathrm{bol,\odot}])},
\end{equation}
where
$M_\mathrm{bol,\odot}=4.74$ \citep{Bessell1998,Torres2010}, and I have adopted
$\mathrm{BC}=-0.02\pm0.05$, which is
characteristic of main-sequence stars with $B-V=0.17$ \citep{Pecaut2013} --
the approximate color of novae near maximum light.
The uncertainty in the luminosity of a given nova, $\sigma_{L_4}$,
has been determined by propagating the uncertainties in $M_V$ and BC.
Specifically, $\sigma_{L_4} = 0.4~\mathrm{ln}(10)~L_4~(\sigma_{M_V}^2 +
\sigma_\mathrm{BC}^2)^{0.5}$ (in practice, the uncertainty in $M_V$ dominates).
Finally, following \cite{Shara2018} and \cite{Shafter2026a},
we adopt $t_\mathrm{ml}\equiv t_2$.

The grid of models given in \cite{Yaron2005} yields observed
nova properties as functions of $M_\mathrm{WD}$ and $\log \dot M$
for 3 representative values of the WD temperature $T_\mathrm{WD}$:
10, 30 and 50 $\times 10^6$~K.
Following the reasoning outlined in \cite{Shafter2026a}, I have adopted
\( T_{\text{WD}} = 10 \times 10^6 \, \text{K} \)
for the analysis, motivated by the work of \citet{Townsley2004}.
Specifically, their Figure~8 (upper panel) shows that for accretion rates
\(\log \dot{M} (M_\odot~\mathrm{yr}^{-1}) \lessim -8\) typical of nova systems, WD
temperatures are expected to be \(\lessim 10 \times 10^6 \, \text{K}\).
Since the analysis covers \(\log \dot{M}\) from \(-12.0\) to
\(-7.0\), the \( T_{\text{WD}} = 10^7 \, \text{K} \) models best
match the physical conditions of nova outbursts in the present study.

\begin{figure}
\plotone{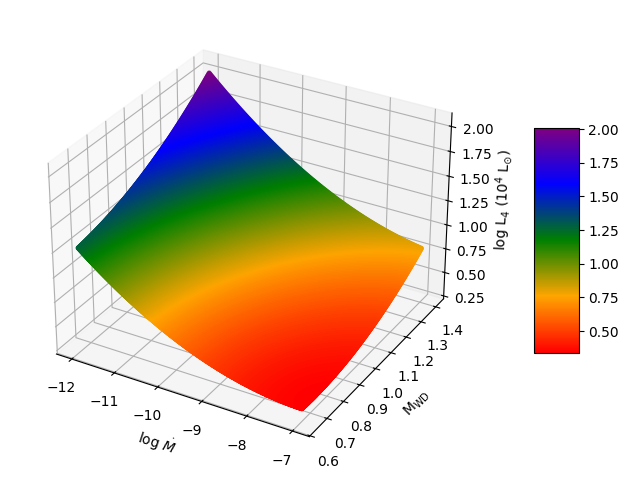}
\plotone{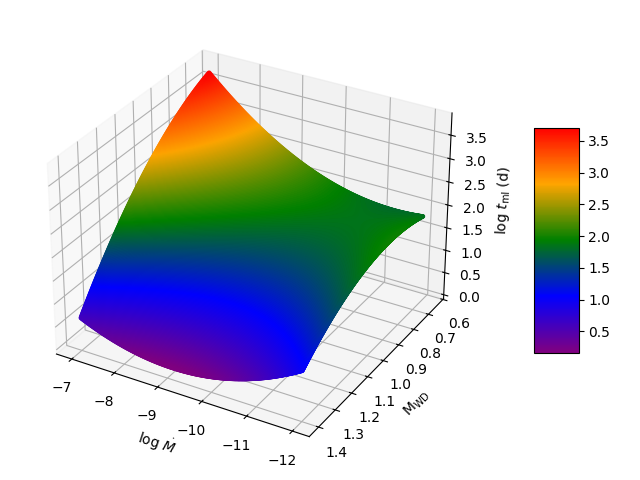}
\caption{The interpolated model grids from \cite{Shafter2026a} for our key input
parameters, the bolometric luminosity 
at the peak of the eruption, $L_4$, (top panel), and 
the mass-loss timescale,
$t_\mathrm{ml}$ (bottom panel) as functions of
$\log \dot M$ ($M_{\odot}$~yr$^{-1}$) and
$M_\mathrm{WD}$ ($M_{\odot}$).
The interpolations were performed by fitting second-order polynomials
to the nova models of \citet{Yaron2005} as described in Section 5.1.
}    
\label{fig10}
\end{figure}

\begin{figure}
\plotone{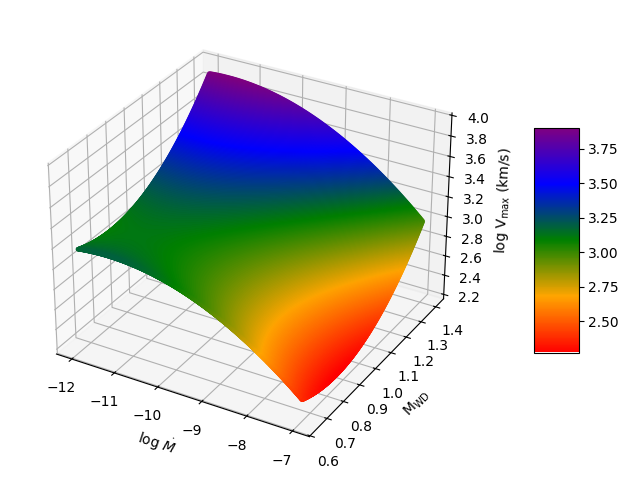}
\plotone{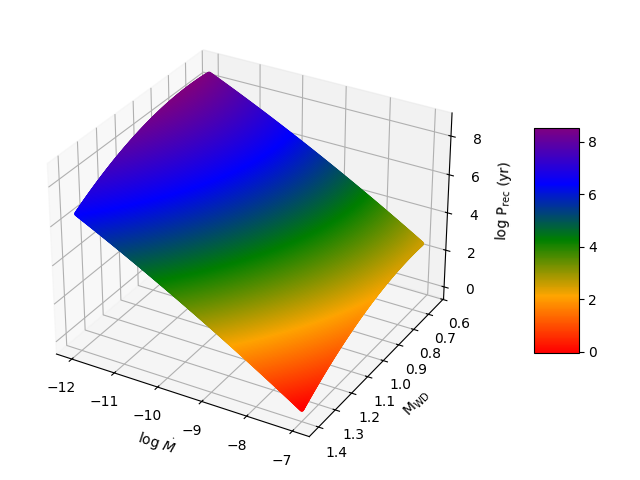}
\caption{Same as Figure~\ref{fig10}, but for
the auxiliary parameters, the maximum expansion velocity 
of the nova ejecta, $V_\mathrm{max}$ and the predicted
recurrence time, $\log P_\mathrm{rec}$ are shown in the
top and bottom panels, respectively.
} 
\label{fig11}
\end{figure}

\subsection{Interpolation of the Model Grid}

The model grid given in Table~3 of \citet{Yaron2005}
is rather coarse, spanning six values of $\log \dot{M}$ (from
$-12$ to $-7$) for just four representative values of $M_\mathrm{WD}$
(0.65, 1.00, 1.25, 1.40 $M_\odot$).
Following \cite{Shafter2026a} the coarse grid for
the $T_\mathrm{WD}=10\times10^6$~K models has been interpolated
using second-order polynomial surface fits
for each model nova parameter
(i.e., $L_4$, $t_\mathrm{ml}$, $V_\mathrm{max}$, $P_\mathrm{rec}$)
in log space.
The uncertainty in the model fit
has been estimated from the residual error
between observed and
predicted values at the original grid points.
This error is then propagated through the fit coefficients to
estimate the standard error at a given interpolated point.
The interpolated grids, which are identical to those computed for the
\cite{Shafter2026a} study, are reproduced in Figures~\ref{fig10} and \ref{fig11}.

\subsection{Determination of Nova Properties}

The WD masses and log accretion rates have been estimated
for each of the LMC novae by
inverting the interpolated $L_4$ and $t_\mathrm{ml}$ grids for
the observed values of $L_4$ (through Eqn.~1)
and the observed $t_2$ time.

As described in \cite{Shafter2026a}, the effective
uncertainty in $L_{4,i}$ and $t_{\mathrm{ml},i}$ at a given grid point is
given by:
\begin{equation}
\Delta_{L_{4,i}} = \sqrt{(\sigma_{L_4})^2 + \delta_{L_{4,i}}^2},
\quad
\Delta_{t_{{\mathrm ml},i}} = \sqrt{(\sigma_{t_2})^2 + \delta_{t_{{\mathrm ml},i}}^2},
\end{equation}
\noindent
where $\sigma_{L_4}$ and $\sigma_{t_2}$ are the observed uncertainties
and $\delta_{L_4,i}$ and
$\delta_{t_\mathrm{ml},i}$
are the uncertainties in the model interpolations.

\begin{figure*}
\plottwo{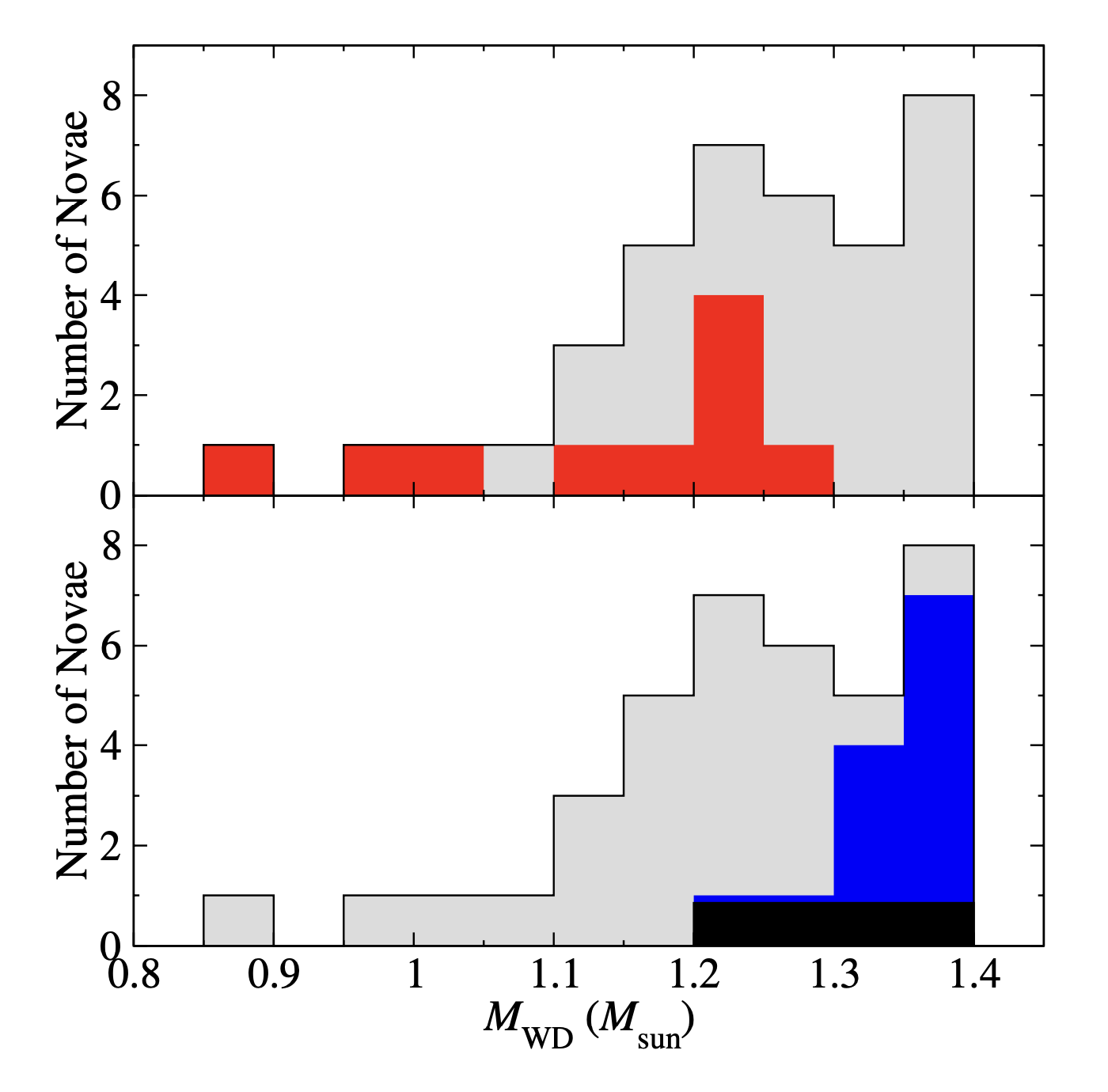}{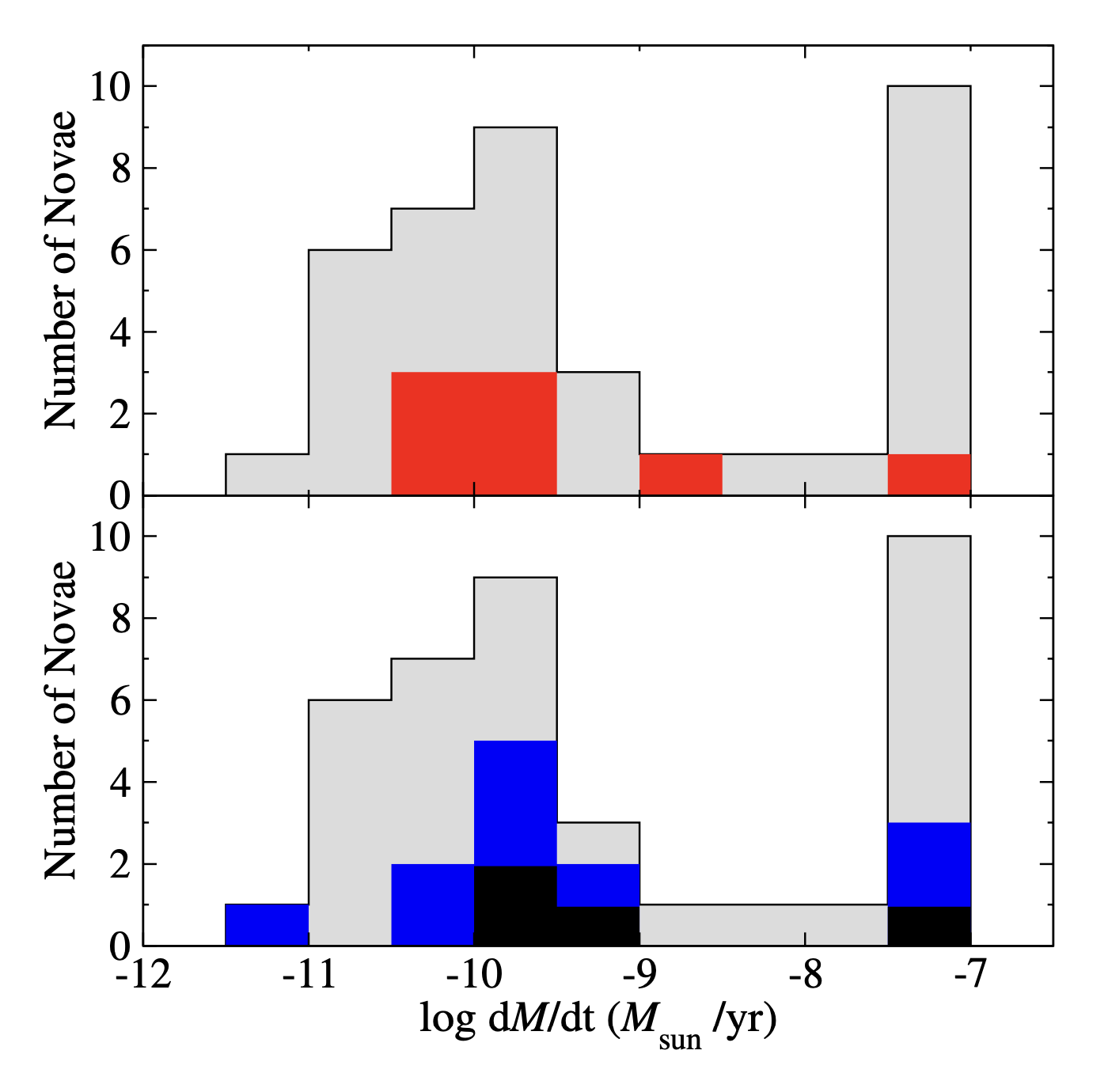}
\plottwo{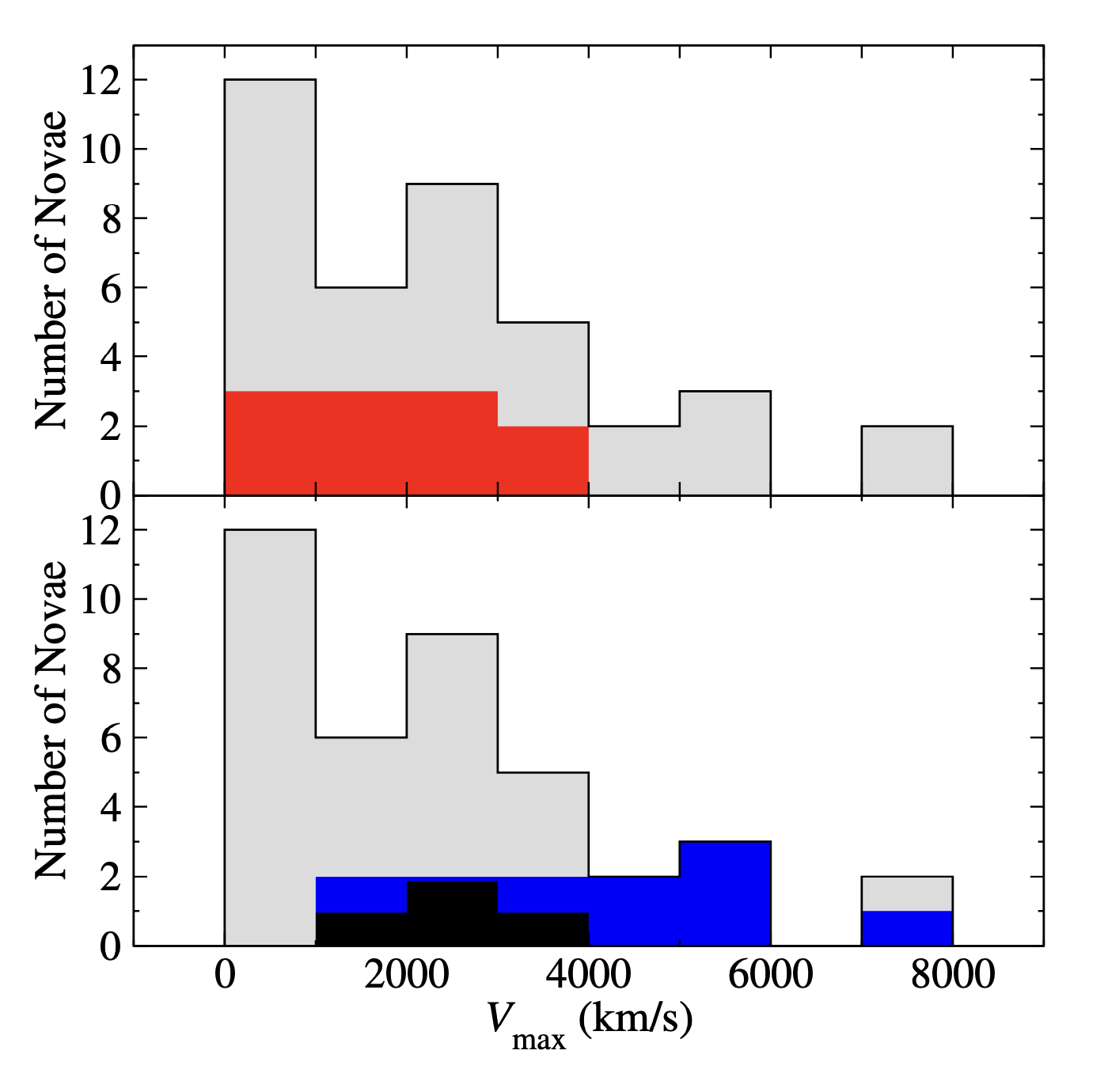}{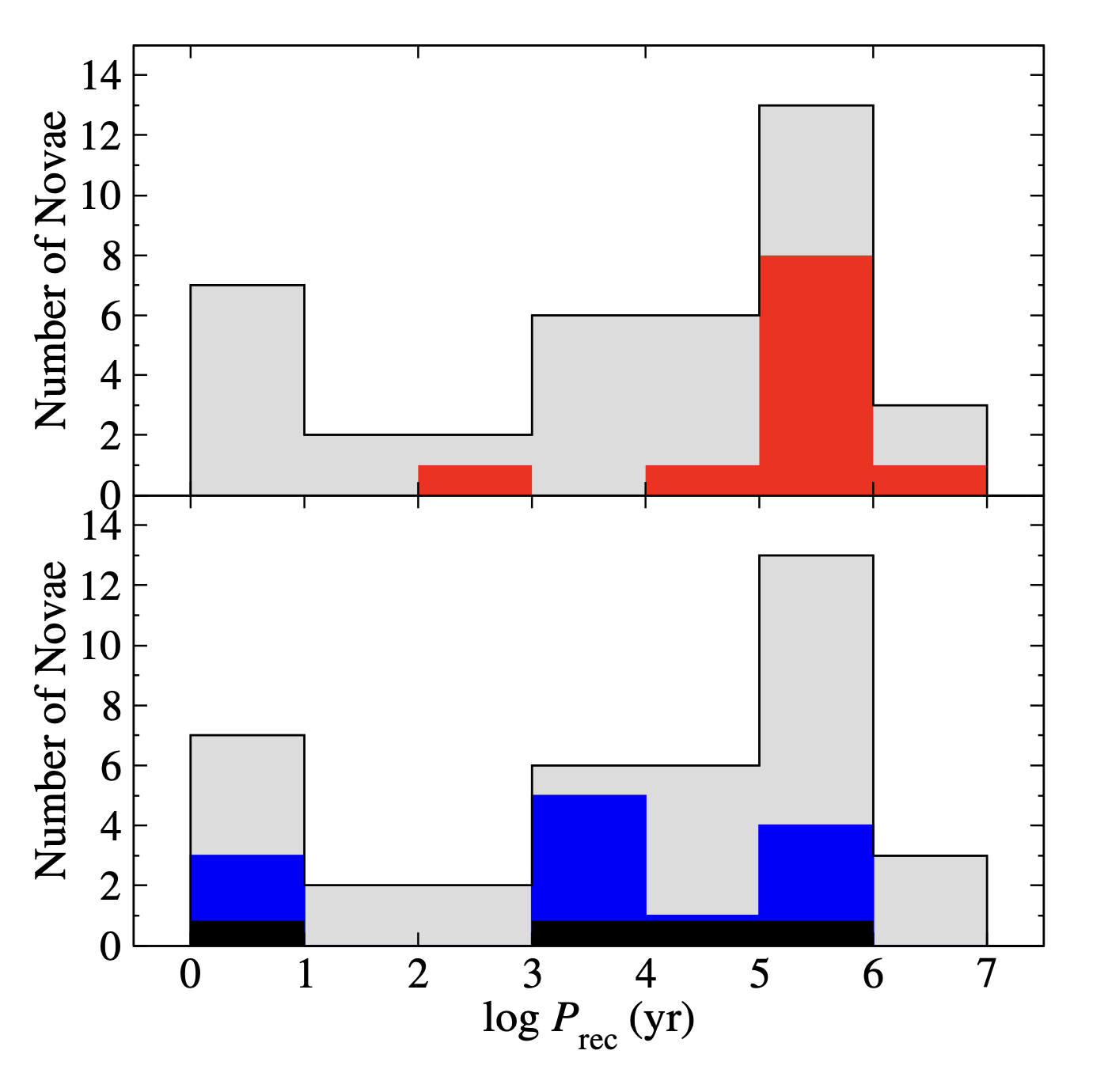}
\caption{The distributions of the WD mass (upper left panel), the mass accretion rate (upper right panel)
the maximum expansion velocity, $V_\mathrm{max}$, (lower left panel), and
predicted recurrence time, $P_\mathrm{rec}$, (lower right panel),
for the 39 LMC novae analyzed with the \cite{Yaron2005} models.
The top panels in each figure show the distributions of \ion{Fe}{2} novae (red),
while the bottom panels show the He/N novae (blue) and known RNe (black).}
\label{fig12}
\end{figure*}

The best estimates of $\log \dot{M}$ and $M_\mathrm{WD}$
are obtained by searching the grid for all points in an effective error box
that simultaneously satisfy the following constraints:
\begin{equation}
|L_{4,i} - L_4| < \Delta_{L_{4,i}} \quad \text{and} \quad
|t_{{\mathrm{ml},i}} - t_2| < \Delta t_{{\mathrm{ml},i}}.
\end{equation}

In the rare cases where no solutions were found
to simultaneously satisfy these criteria,
the observational uncertainties
($\sigma_{L_4}$ and $\sigma_{t_2}$) were incremented by multiples
of their original values until a solution was found.

The best estimates of
$M_{\text{WD}}$ and $\log \dot{M}$ were then determined by
searching the error box for the values that
minimize the weighted Euclidean norm:

\begin{equation}
\chi_i = \sqrt{ \left( \frac{L_{4,i} - L_4}
{\Delta_{L_{4,i}}} \right)^2 + \left(
\frac{t_{\mathrm{ml},i} - t_2}
{\Delta_{t_{{\mathrm ml},i}}} \right)^2 }.
\end{equation}

Uncertainties in the optimum values of $\log \dot{M}$ and $M_\mathrm{WD}$
were then taken to be the RMS deviations
of all points satisfying the uncertainty constraints
from the optimum value.
This nonparametric method measures the true spread of all
acceptable solutions inside the effective error box, while accounting for
how tightly or loosely the points cluster.

\begin{figure*}
\plotone{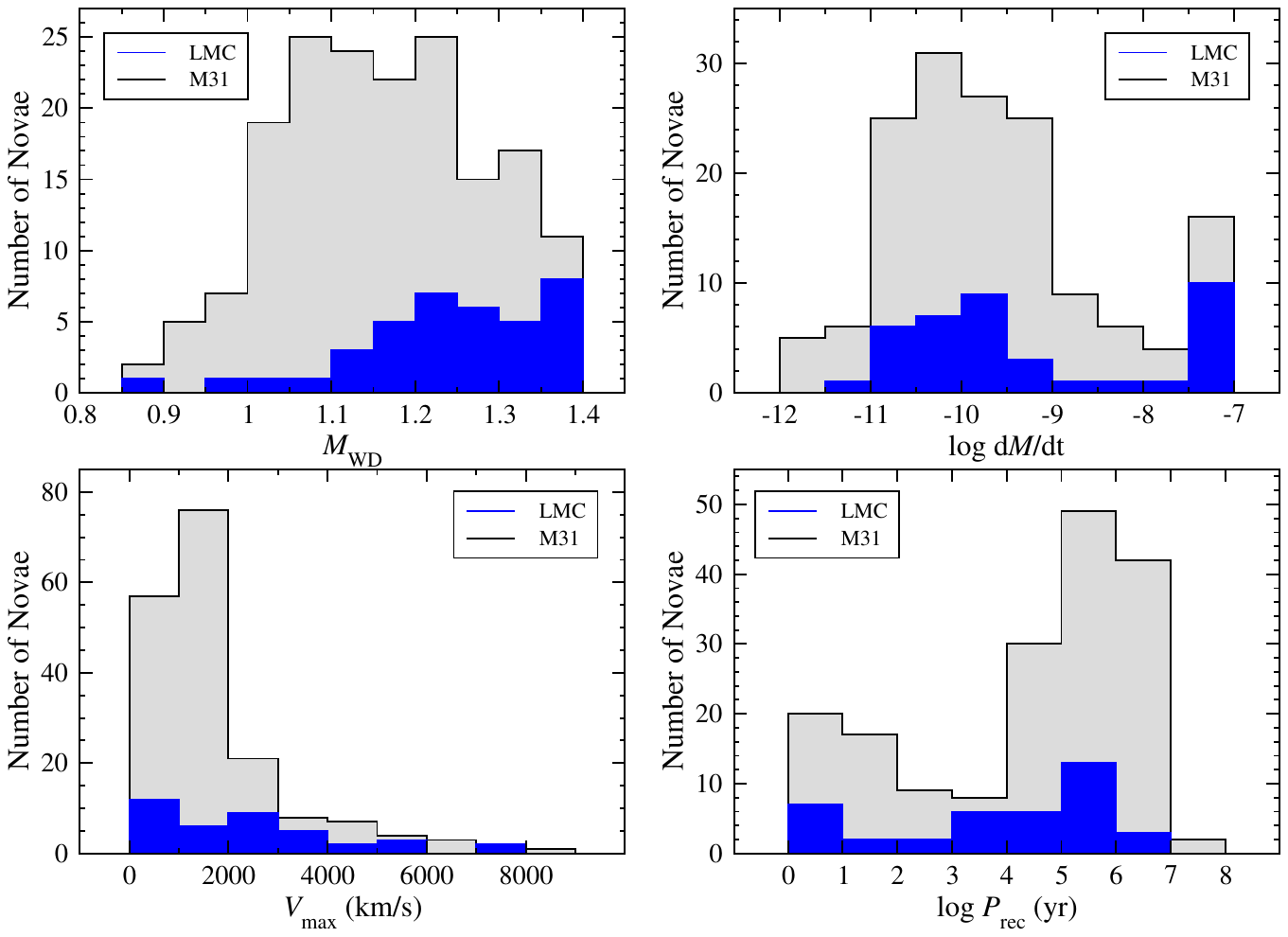}
\caption{The derived distributions for $M_\mathrm{WD}$, $\log \dot M$, $V_\mathrm{max}$ and $P_\mathrm{rec}$
for the LMC novae (blue histograms) compared with the distributions for M31 novae (grey histograms) from \cite{Shafter2026a}.
}
\label{fig13}
\end{figure*}

\begin{figure*}
\plotone{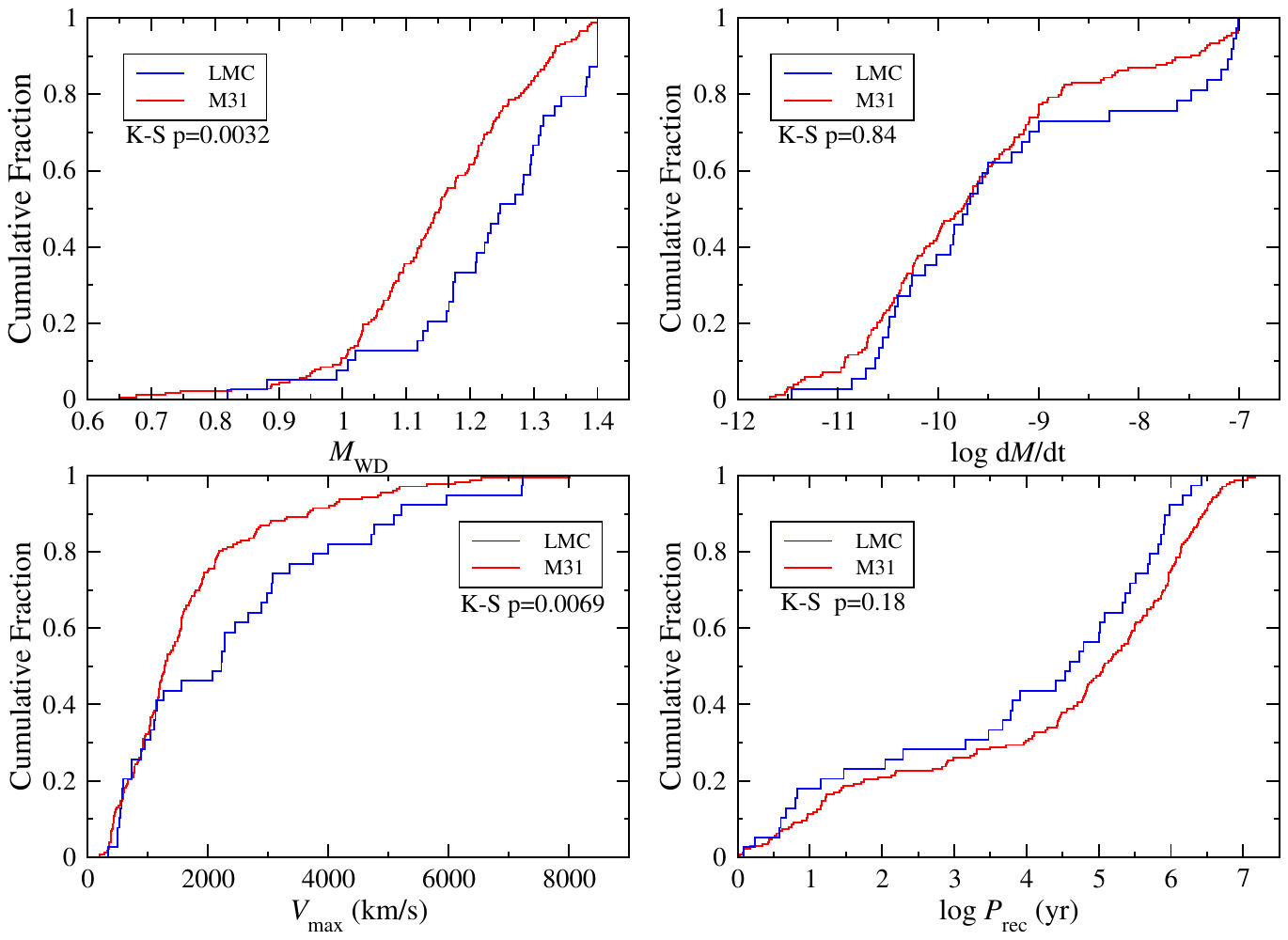}
\caption{The cumulative distributions of the four properties (M31: red; LMC: blue).
No significant differences were found between the $\log \dot M$ and $P_\mathrm{rec}$ distributions, but
K-S tests show that the WD mass and $V_\mathrm{max}$ distributions differ by
$\sim2.95\sigma$ and $\sim2.69\sigma$, respectively.
}
\label{fig14}
\end{figure*}

Once the optimum values of $M_\mathrm{WD}$, $\log \dot M$, and their
associated uncertainties were determined, they were then used as inputs to the
interpolated model grids shown in Figure~\ref{fig11}
to predict the maximum expansion velocity,
$V_\mathrm{max}$, and recurrence time, $P_\mathrm{rec}$.

The best estimates of all four fundamental nova properties:
$M_\mathrm{WD}$, $\log \dot{M}$, $V_\mathrm{max}$ and $P_\mathrm{rec}$,
for our sample of 39 LMC novae from Table~\ref{tab3}
are presented individually
in Tables~\ref{tab4} and \ref{tab5}. Values of $M_\mathrm{WD}$ and $\log \dot{M}$
that lie at the edge of the model grid are shown as lower limits.
Spectroscopic classes, which are available for the majority of the novae,
have also been included.

\subsection{Observed Distributions of Nova Parameters}

Figure~\ref{fig12} shows
the observed distributions for each
of the four nova parameters. To highlight the differences with respect to
spectroscopic class, the figure for each parameter
contains two panels, with the top panel
comparing the distribution of \ion{Fe}{2} novae with the complete sample,
and the bottom panel showing the corresponding comparison for He/N
novae and the four known RNe.
The means and RMS variations of the 4 distributions are summarized
in Table~\ref{tab6}.
As seen with M31 novae \citep{Shafter2026a}, the He/N and RNe
are found more frequently among the systems containing higher mass WDs.
They also tend to be associated with higher expansion velocities and
shorter recurrence times, but there doesn't appear to be any
differentiation with respect to the mass accretion rate.

To determine if the apparent differences between the \ion{Fe}{2} and He/N
novae are statistically significant, the
cumulative distributions of the \ion{Fe}{2} and He/N
subsamples for each nova parameter have been compared with K-S tests.
The tests returned probabilities of
$p=3.2 \times 10^{-4}$ ($3.6\sigma$), $p=0.71$ ($0.4\sigma$), $p=0.11$ ($1.6\sigma$), and
$p=7.2 \times 10^{-3}$ ($2.7\sigma$), for $M_\mathrm{WD}$,
$\log \dot M$, $V_\mathrm{max}$ and $P_\mathrm{rec}$, respectively.
Thus, only the WD mass and recurrence time
distributions for \ion{Fe}{2} and He/N novae differ at $>2\sigma$.

\subsection{Comparison with the M31 Nova Population}

To put the properties of the LMC nova population into
perspective, it is instructive to compare the observed distributions
of nova parameters with the corresponding distributions for the M31
nova population.
Figure~\ref{fig13} shows such a comparison, where the
parameter distributions for the M31 novae are taken from \cite{Shafter2026a}.
At first glance, it appears that the LMC nova population
contains WDs that are on average more massive than their M31
counterparts. Another apparent difference is seen in the
expansion velocity distributions where the M31 population
appears to have a marked excess of novae with predicted
expansion velocities $V_\mathrm{max}\lessim2000$~km~s$^{-1}$
compared with the LMC nova sample. Finally, a comparison of the
recurrence time distributions for the two galaxies suggest 
a possible dearth of LMC novae with predicted recurrence times
shorter than $\sim100$~yr when compared with M31 novae.

A more quantitative approach to comparing the properties of
the LMC and M31 nova populations is to consider the cumulative
distributions of the four nova parameters, as shown in Figure~\ref{fig14}.
The differences between the $M_\mathrm{WD}$ and $V_\mathrm{max}$ distributions
can be clearly seen.
The resulting K-S probabilities are $p=0.005$ and $p=0.0009$,
establishing that the $M_\mathrm{WD}$ and $V_\mathrm{max}$ distributions for
the LMC and M31 nova samples differ at $\sim2.8\sigma$ and $\sim3.3\sigma$,
respectively. On the other hand, despite a suspicion that the recurrence
time distributions for the LMC and M31 novae might differ meaningfully,
no significant difference was found ($p=0.97$), nor as expected, was a difference
found between the $\log \dot M$ distributions ($p=0.84$) of the two galaxies.

\begin{figure*} 
\plottwo{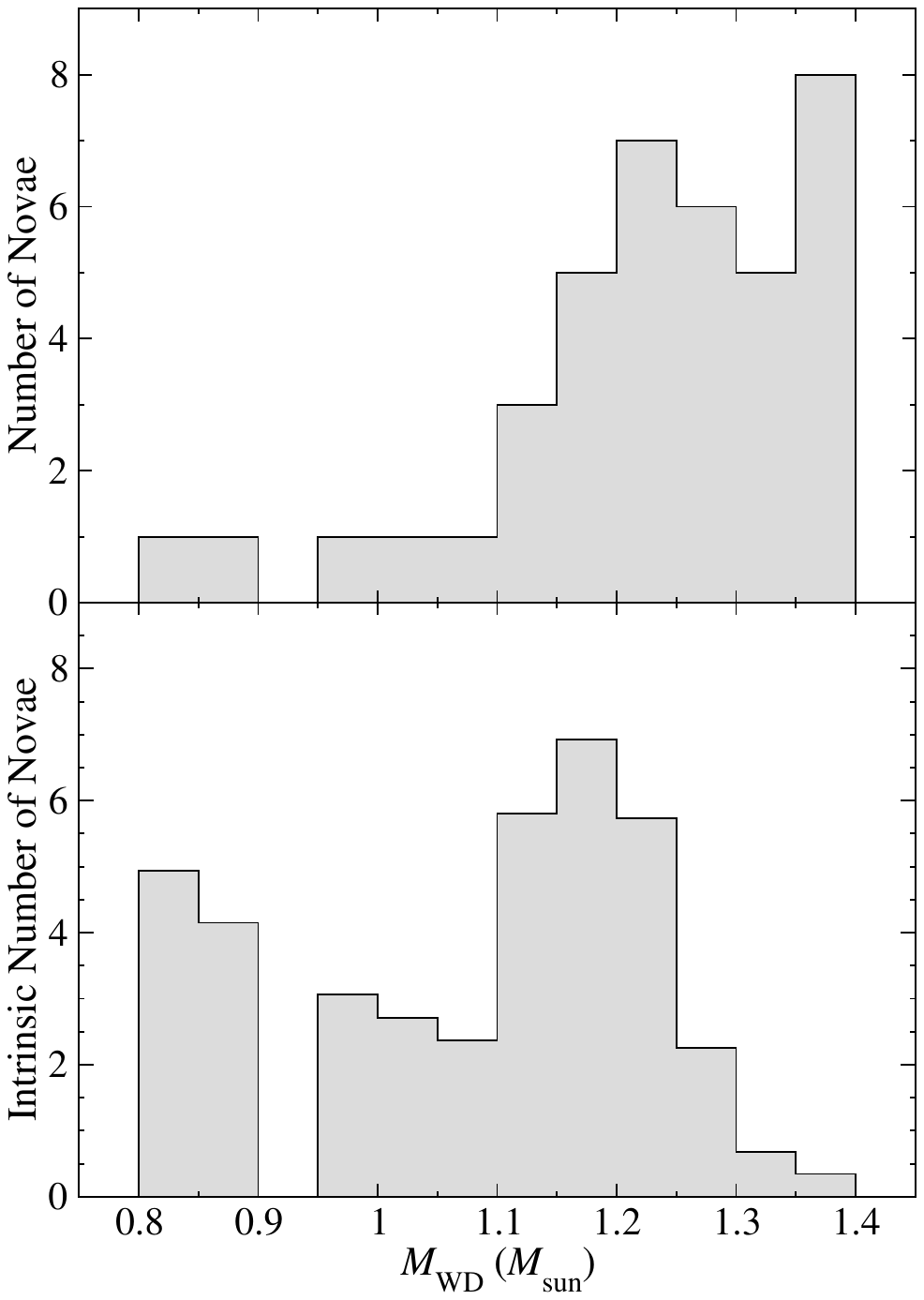}{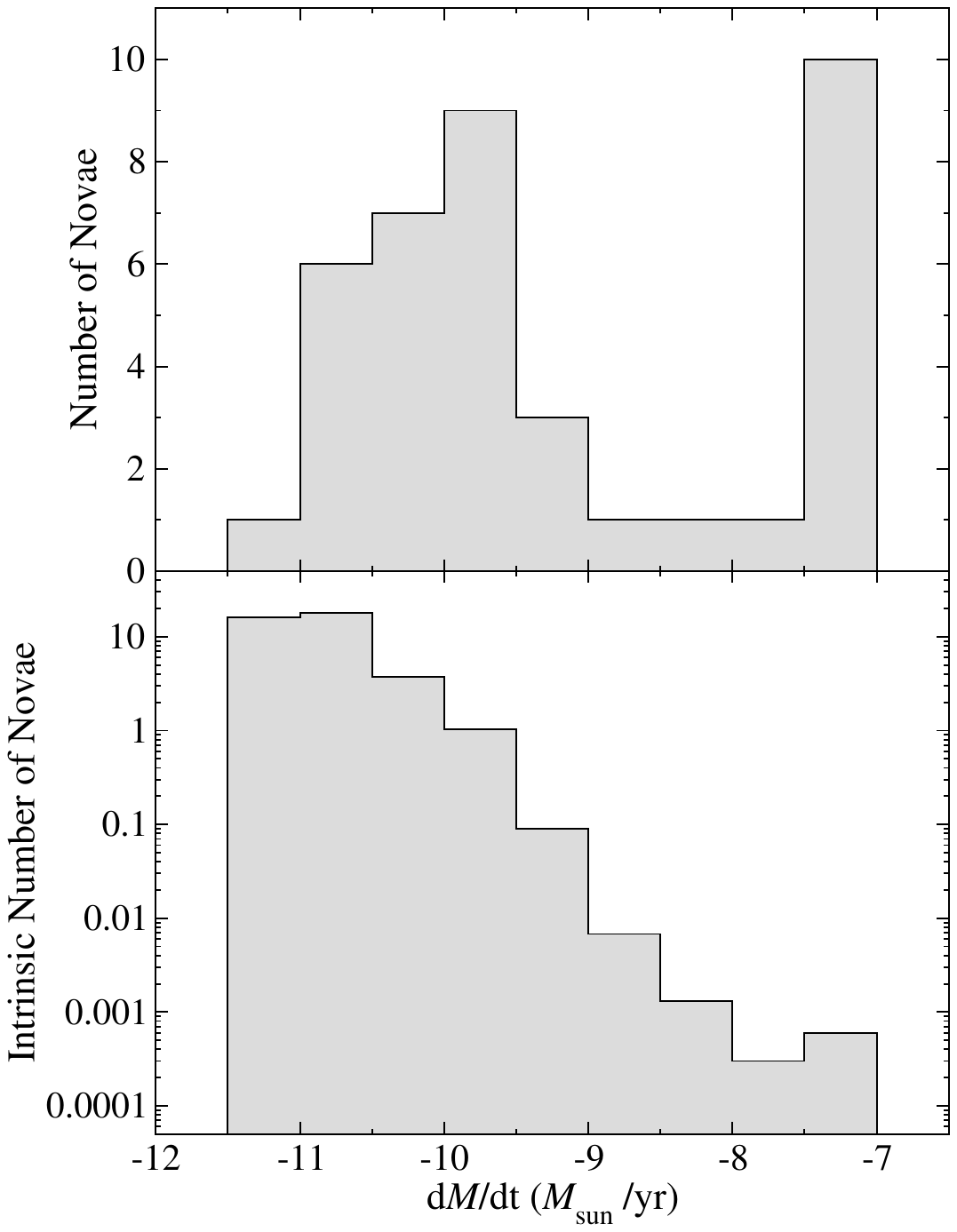} 
\caption{The intrinsic distributions for $M_\mathrm{WD}$ and $\log \dot M$ (bottom left and bottom right, respectively,
compared with the observed distributions which have been reproduced in the top panels of each figure.}
\label{fig15}
\end{figure*}

\subsection{Intrinsic $M_\mathrm{WD}$ and $\log \dot M$ Distributions}

The observed distributions for $M_\mathrm{WD}$ and $\log \dot M$ are biased
toward novae with short recurrence times and thus do not accurately
reflect the intrinsic distributions for the population of nova progenitor
binaries. 
The intrinsic (unbiased) mean
for the $M_\mathrm{WD}$ and $\log {\dot M}$ 
distributions can be estimated by simply by weighting the observed WD
masses, $M_{\mathrm{WD},i}$, and log accretion rates, $[\log \dot M]_i$,
by $P_{\mathrm{rec},i}$. Adopting recurrence time weights
$w_{\mathrm{rec},i} = P_{\mathrm{rec},i}/\sum_{i=1}^{N} P_{\mathrm{rec},i}$, we have:
$\langle M_\mathrm{WD} \rangle _\mathrm{int} = \sum_{i=1}^{N} w_{\mathrm{rec},i}~M_{\mathrm{WD},i}$ and
$\langle \log \dot M \rangle _\mathrm{int} = \sum_{i=1}^{N} w_{\mathrm{rec},i}~[\log \dot M]_i$.
When applied to the full sample of $N=39$ novae in
Tables~\ref{tab4} and \ref{tab5} we find
$\langle M_\mathrm{WD} \rangle _\mathrm{int} = 1.13~M_{\odot}$,
and
$\langle \dot M \rangle _\mathrm{int} = 3.5\times10^{-11}~M_{\odot}$~yr$^{-1}$.
The corresponding values for the sample of M31 novae studied by \cite{Shafter2026a}
is $\langle M_\mathrm{WD} \rangle _\mathrm{int} = 1.08~M_{\odot}$ and
$\langle \dot M \rangle _\mathrm{int} = 2\times10^{-11}~M_{\odot}$~yr$^{-1}$.

In addition to the direct estimation of the intrinsic distribution means,
we can follow the
procedure outlined in \cite{Shara2018} and later adopted by \cite{Shafter2026a}.
Specifically, application of Equation~6 from \cite{Shafter2026a} to
the observed $M_\mathrm{WD}$ and $\log \dot M$ distributions of our
LMC nova sample results
in the intrinsic distributions shown in Figure~\ref{fig15}.
For comparison, the observed distributions have been reproduced in the
top panels of each figure.
The intrinsic $M_\mathrm{WD}$ distribution is relatively flat across
WD masses between 0.65 and 1.25~$M_\odot$, with an overall mean
of $1.10~M_\odot$. For the $\log {\dot M}$ distribution
we find an overall mean of $2.3\times10^{-11}~M_{\odot}$~yr$^{-1}$.
Both of these values are consistent as expected with the recurrence time weighted
means determined above.

As with the observed distributions, the intrinsic distributions for the LMC
are characterized by slightly higher mean values compared with
the corresponding distributions of novae in M31 \citep{Shafter2026a}.
The intrinsic $\log \dot M$ distribution is heavily biased toward
low mass accretion rates, as was found for M31 novae. As pointed
out by Shafter \& Hornoch, the low accretion rates, $\log \dot M\lessim-10$,
are characteristic of the rates expected during the extended hibernation phase
between outbursts.

\section{Discussion}

It has long been argued that novae in the LMC are, on average, both more luminous and
faster evolving compared with novae in M31 \citep{Capaccioli1990a,DellaValle1991,Shafter2013}.
The 18 new LMC nova light-curves presented in the present work have nearly doubled the number of
well-observed LMC novae available for study since the compilation of Paper~I
was published more than a decade ago.
This substantial increase in light-curve data
has significantly strengthened the conclusions of earlier work.

The large collection of nova light-curves now available in the LMC has allowed,
for the first time, an exploration of the fundamental properties of LMC novae
analogous to the recent study of M31 novae by \cite{Shafter2026a}.
Specifically, application of the \cite{Yaron2005} models to the available LMC data
has yielded estimates for the WD mass and accretion rate ($M_\mathrm{WD}$ and $\log \dot M$),
as well as the maximum velocity of the ejecta, $V_\mathrm{max}$, and the 
predicted recurrence time, $P_\mathrm{rce}$. The results of this analysis have
demonstrated that the LMC nova population is characterized by progenitor systems that
contain on average more massive WDs compared with the nova
population in M31.

The higher average WD masses in LMC novae have resulted in
expected ejecta velocities that are higher than those predicted
for M31. Unfortunately,
it is difficult to compare the observed expansion
velocities between novae in the two galaxies given the
limited spectroscopy and the variation in the time after
peak when available spectra have been obtained.
A rough proxy for expansion velocity is the spectroscopic type,
where novae with emission lines having FWHM$\grtsim2500$~km~s$^{-1}$
are almost always associated with He/N systems.

\subsection{Nova Spectroscopic Classification}

The concept of a spectroscopic type for novae
originated with \cite{Williams1992} who argued
that novae could be broadly divided into three
classes: (1) an \ion{Fe}{2} class composed of novae displaying
prominent \ion{Fe}{2} emission features of various multiplets
shortly after eruption. These novae tend to be relatively
slowly evolving with relatively narrow emission lines
(typically FWHM$\lessim2500$~km~s$^{-1}$); (2) a
He/N class consisting of novae displaying strong and broad
He and N emission lines (FWHM$\grtsim2500$~km~s$^{-1}$)
with relatively weak or absent \ion{Fe}{2} emission.
As expected given the broad emission lines,
these novae tend to evolve rapidly;
(3) a ``Hybrid" (or broad-lined \ion{Fe}{2}) class
composed of relatively rare
novae that share features of both classes.

In recent years, the notion that novae belong to any particular
spectroscopic class has been called into question. In particular,
\cite{Aydi2024a} have argued that all novae transition from
an early He/N phase, to an \ion{Fe}{2} phase,
and then back to a late He/N phase,
with the key distinction between novae being the length
of time spent in a given phase. Importantly, they argue
that novae that have been traditionally classified as belonging
to the He/N class simply have passed
through the first two stages rapidly (within hours to days)
of peak. Regardless of whether \ion{Fe}{2} and He/N novae represent
distinct classes, or whether they reflect differences in
the rate of spectral evolution, empirically, it remains the case
that rapidly evolving novae (e.g., $t_2\lessim5$~d) are
almost always observed to have persistent He/N characteristics
(particularly, broad lines) shortly after maximum light.
Thus, from a practical
standpoint, spectra obtained roughly one to two weeks
after eruption can be used to segregate novae into either
an \ion{Fe}{2} class (slow evolving, narrow-lined novae)
and a He/N class (rapidly evolving, broad lined novae).
In this picture, Hybrid systems could be understood as
novae transitioning between the \ion{Fe}{2} and late He/N phases.

Among novae with available spectroscopic classifications,
He/N and Hybrid systems make up
only $\sim20$\% of all novae in M31 \citep{Shafter2011},
with a similar percentage found among Galactic novae
\citep{DellaValle1998,Poggiani2021}.
In contrast, as we have found in the present study (see Table~\ref{tab1}),
more than 50\% of LMC novae with available spectra
are members of the He/N class.
This difference is consistent with what is expected given the
higher proportion of high-mass WDs among the LMC nova progenitors.

\subsection{LMC Recurrent Novae}

In addition to the increasing percentage of He/N class novae,
the relatively high-mass WDs characteristic of LMC nova progenitors have
translated into a higher fraction of RN eruptions, which typically
require $M_\mathrm{WD}\grtsim1.3~M_\odot$ \citep{Wolf2013,Kato2014}. Of the
$\sim$1360 nova eruptions observed to date in M31,
a total of 87 eruptions have been observed from 23 known RNe \citep{Shafter2026c}.
By contrast, in the LMC there have been a total of 63 observed
nova eruptions of which 13 are associated with the 4 known RNe. Thus,
the fraction of RN eruptions in the LMC is $f_{\rm LMC}=13/63=0.2063\pm0.0510$
compared with just
$f_{\rm M31}=87/1360=0.0640\pm0.0066$
in M31, where the uncertainties,
$\sigma_f=\sqrt{f(1-f)/N}$,
follow from the binomial variance.

The observed ratio of RN eruption fractions between
the LMC and M31 is $R=3.22$. To test whether
this difference could simply be due to random
sampling, a $\chi^2$ test was performed under
the null hypothesis that both galaxies share
the same intrinsic RNe fraction
$f_\mathrm{int}=100/1423\approx0.0703$. The
expected numbers of RNe are then 4.43 in the
LMC and 95.57 in M31. This test
yields $\chi^2=16.57$ ($p=4.7\times10^{-5}$).
Thus, we find that the underlying population
fractions of RN eruptions differ significantly between the LMC and
M31 at roughly the 4$\sigma$ level. It is worth noting, however, that some
of the observed difference in the RN eruption fractions
may result from unknown selection biases in the
rate of nova discoveries in the two galaxies.

Three of the LMC novae have shown periodicities that likely represent the
orbital periods of the progenitor binaries: 1.26~d, 2.65~d, and 2.85~d for
LMCRN 1968-12a, LMCRN 1996-11a, and LMCN 2016-04a, respectively. The first two
of these systems are known RNe.
In all three cases, the putative orbital periods exceed a day, implying that the secondary stars
in these systems are evolved and that the accretion rates are expected to be
relatively high \citep{Pala2022}. It would therefore seem likely that
the third system, LMCN 2016-04a, may be an unrecognized RN as well. Unfortunately,
the available light-curve information is very limited (see Figure~\ref{fig3}). We
have estimated an upper limit of $t_2=50$~d, but the decline rate could
be much faster. In this case, WD mass may significantly exceed
the value of $M_\mathrm{WD}=1.16~M_\odot$ derived in our analysis.

It is also interesting to note that the recognized LMC RN are generally more luminous than
those seen in M31. This difference is apparent in the MMRD relation
shown in Figure~\ref{fig7} where the "fast and faint" quadrant normally
populated by RNe is mostly devoid of novae. Why this is the case is unclear.
In contrast, M31 has a significant population of novae in the
lower left quadrant of the MMRD plane, most, if not all, of which are
known RN systems. The 4 known RNe in the LMC have similar $t_2$ times
(generally less than 10~d) compared with M31 RNe, but are significantly more luminous than
their M31 counterparts. The higher luminosities seen in the LMC RNe
require lower accretion rates (resulting in higher degeneracy in the accreted
layer at the onset of the TNR) that translate into
unrealistically long expected recurrence times (see Table~\ref{tab5}).

Before leaving the discussion of the LMC RNe, it is worth commenting briefly
on the observed recurrence times. With only 4 LMC RNe known, it is not
possible to meaningfully compare the
recurrence time distribution with that for M31 and
the Galaxy. That said, it is interesting that only one of the LMC RNe,
LMCRN 1968-12a,
has an observed recurrence time less than 20~yr, whereas of the 23 known RNe in M31,
16 out of 23 ($\sim70$\%) have $P_\mathrm{rec}<20$~yr. The recurrence times
are largely determined by the WD mass and accretion rate, with systems having
the highest mass WDs
accreting at highest rates having the shortest recurrence times.
Given that the WD masses in LMC novae are higher on average compared
with the M31 novae, it would be tempting to speculate that M31 may
have a more significant population of novae with high accretion rates.
However, this possibility seems unlikely given the accretion rate
distributions shown in Figures\ref{fig13} and \ref{fig14},
which show no significant difference between the two galaxies.

\subsection{The LMC Nova Rate}

A determination of the annual nova rate in the LMC requires a knowledge
of the detection efficiency for novae over a specified time interval, and
typically involves the analysis of data from a survey of known limiting magnitude and sampling frequency.
Such an analysis is beyond the scope of the present study.
The best current estimate for the rate of novae in the LMC is based on the
analysis of data from the OGLE survey, and is given by $R=2.4\pm0.8$~yr$^{-1}$ \citep{Mroz2016a}.

For comparison with other galaxies it has become standard to normalize the nova rate
to the $K$-band luminosity of the parent galaxy. For the LMC, we adopt $V=0.40\pm0.06$ \citep{Devaucouleurs1991},
$A_V=0.19\pm0.06$ (see Section 3.2), and a mean $V-K=2.32$ typical of irregular galaxies \citep{Aaronson1978},
yielding an integrated $K$ magnitude of $K=-2.1\pm0.1$. Given the distance modulus from Section 3.2,
we find an absolute magnitude, $K=-20.6\pm0.1$.
Adopting $K_\odot=3.27$ \citep{Willmer2018} for the absolute $K$ magnitude of the Sun
yields a $K$-band luminosity of $L_K = (3.5\pm0.3) \times10^9~L_{\odot,K}$ for the LMC,
and a corresponding $\mathrm{LSNR}_K = 6.9\pm2.4$ novae per year
per $10^{10}$ solar luminosities in the $K$ band.
This LSNR is somewhat higher than current estimates for M31 and the Galaxy \citep[e.g., see][]{DellaValle2023}
suggesting that nova eruptions are more frequent in the LMC, as expected given the WD masses derived in this study.
As a cautionary note, determining accurate integrated $K$ magnitudes of nearby galaxies is challenging given
their typically large angular sizes, making background subtraction difficult. As a result, the corresponding
LSNRs can be quite uncertain.

Given that a total of 32 nova eruptions have been observed in the LMC since
2000, the {\it observed\/} nova rate in the LMC is $\sim1.3$~yr$^{-1}$.
Thus, it appears that  approximately half of the
novae that likely have erupted in the LMC this century have been missed. Over the most
recent decade, the completeness has improved, with
the number of novae discovered passing 70\% of the number
expected based on the current nova rate estimate. Now that
the Rubin telescope is coming online, it is likely that very few, if any, novae
will go missing.

\section{Conclusions}

A century of monitoring the LMC has
provided a sufficiently large sample of novae for a meaningful comparison with their counterparts
in M31 and the Galaxy.
A review of the 66 nova candidates observed between 1926 and 2025
has yielded a total of 48 eruptions having sufficient data to allow reasonable
estimates of their peak magnitudes and rates of decline. Nine of these eruptions
are recurrences of 4 known RNe, leaving a total of 39 unique nova
progenitors in the LMC suitable for further analysis.

The principal conclusions that follow from these data are summarized below.

(1) Novae in the LMC generally evolve more rapidly than do novae in M31 and the Galaxy.
Cumulative distributions of the $t_2$ times for novae in all three galaxies
show that the LMC novae appear to decline more rapidly on average than do
novae in M31 or the Milky Way.
For the LMC novae, the difference with the
M31 novae is significant at the $3.3\sigma$ level, while that with
Galactic novae is not statistically significant ($1.4\sigma$).
As shown by \cite{Shafter2026a}, Galactic novae
are significantly faster than novae in M31. Cumulative distributions of the $t_2$ times shown
in Figure~\ref{fig8} of the present paper show that novae in the two galaxies
differ with a significance of $2.9\sigma$.

(2) The distributions of peak luminosities for LMC, M31 and Galactic novae are difficult to compare.
The LMC sample is relatively small compared to M31 and the Galaxy, and the Galactic nova sample
presented in \cite{Schaefer2025} has a long low-luminosity tail that, if present, would be difficult
to detect in the nova samples from the LMC, and especially from M31. Although the average absolute magnitude of
the LMC nova sample is slightly more luminous compared with that for M31 or the Galaxy,
formal $\chi^2$ tests fail to show any statistically significant differences.

(3) The light-curve parameters (peak luminosities and rates of decline) have been
used in conjunction with the \cite{Yaron2005} models to estimate the WD masses and accretion rates
for the 39 LMC novae in our sample. In addition to estimates for $M_\mathrm{WD}$ and $\log \dot M$,
the models have also been used to predict the maximum velocity of the nova ejecta, $V_\mathrm{max}$,
and the recurrence time of the nova, $P_\mathrm{rec}$.

The WD mass distribution is characterized by a mean, $\langle M_\mathrm{WD} \rangle = 1.23~M_\odot$,
and RMS deviation about the mean of $0.17~M_\odot$. The mean WD mass for the LMC novae
is significantly higher than that found by \cite{Shafter2026a} for the M31 novae
($\langle M_\mathrm{WD} \rangle = 1.15~M_\odot$). A K-S test on the
cumulative $M_\mathrm{WD}$ distributions for the LMC and M31 novae yields $p=3.21\times10^{-3}$,
establishing that the distributions differ at the $2.9\sigma$ level.
On the other hand, the mean masses for RNe in the LMC ($\langle M_\mathrm{WD} \rangle = 1.31\pm0.08~M_\odot$)
and M31 ($\langle M_\mathrm{WD} \rangle = 1.33\pm0.08~M_\odot$) are very similar.

(4) Another significant difference between the LMC and M31 nova populations
was found for the predicted maximum ejecta velocity, $V_\mathrm{max}$, which
differs at the $2.7\sigma$ level with the LMC novae having somewhat higher average
ejecta velocities. The higher percentage of He/N novae in the LMC is consistent
with this finding.

(5) When corrected for recurrence time bias, the mean of the WD mass distribution
falls to $\langle M_\mathrm{WD} \rangle_\mathrm{int} = 1.13~M_\odot$. The $\sim0.1~M_\odot$
lower mean intrinsic WD mass is similar to the difference between the observed and intrinsic
mean WD mass found for the M31 nova population \citep{Shafter2026a}.

(6) The generally higher WD masses found among the LMC novae suggest a
higher fraction of RN eruptions compared with that seen in M31 ($\sim$21\% vs $\sim$6.4\% of the nova population).
Although this difference is significant at the $4\sigma$ level, selection biases in the
discovery rates between the two galaxies could plausibly account for much of the observed difference.

(7) The currently accepted LMC nova rate is $R=2.4\pm0.8$~yr$^{-1}$ \citep{Mroz2016a}.
Although attempting a new nova rate estimate is beyond the scope of this work,
the number of novae observed over the past decade provides
a lower limit on the rate of novae in the LMC of $R\grtsim1.7$~yr$^{-1}$.

Despite the relatively low absolute nova rate in the LMC compared with
M31 and the Milky Way, its luminosity-specific nova rate is likely higher than either galaxy
\citep{DellaValle2020,Shafter2023}.

Looking ahead, the unprecedented depth and cadence of the Vera C. Rubin Observatory
\citep{Ivezic2019}, which is currently coming online,
promises to revolutionize the study of novae in the LMC and other extragalactic systems.

\begin{acknowledgements}
This paper is dedicated to the pioneering generation of 20th-century observers
whose tireless photographic and visual monitoring paved the way for the study of
novae in the LMC.

I thank K. Hornoch, D. C. Leonard, and an anonymous referee for valuable comments.
\end{acknowledgements}

\newpage

\startlongtable
\begin{deluxetable*}{lccccccl}
\tabletypesize{\footnotesize}
\thispagestyle{empty}
\pagenumbering{gobble}
\tablenum{1}
\tablecolumns{8}
\tablecaption{LMC Novae\tablenotemark{a}\label{tab1}}
\tablehead{\colhead{Nova} & \colhead{R.A.} & \colhead{Decl.} & \colhead{JD\tablenotemark{b}} & \colhead{$m_\mathrm{dis}$\tablenotemark{b}} & \colhead{Filter} & \colhead{References\tablenotemark{c}} & \colhead{Notes} \\ \colhead{(LMCN)} & \colhead{(deg)} & \colhead{(deg)} & \colhead{($-2,400,000$)} & \colhead{} & \colhead{} & \colhead{} &\colhead{}
}
\startdata
1926-09a & 78.727917 &$ -66.802861$& 24786.50 & 12.4 &$ pg  $& 1,2,3  & RY Dor \cr
1935-09a & 59.816250 &$ -67.776583$& 28049.00 & 11.0 &$ pg  $& 2,3  & SN in NGC 1511 \cr
1936-02a & 76.895250 &$ -66.656944$& 28223.50 & 10.8 &$ pg  $& 2,3  & N Dor 1936; Scanty data \cr
1937-11a & 89.375000 &$ -68.918333$& 28860.50 & 10.6 &$ pg  $& 2,3  & RN; YY Dor \cr
1948-12a & 84.575000 &$ -70.348611$& 32891.00 & 13.3 &$ pg  $& 2,3  & Scanty data \cr
1951-01a & 78.204167 &$ -69.970278$& 33651.00 & 11.9 &$ H\alpha$ & 2,3  & N Mensae 1951 \cr
1952:    & 81.975000 &$ -66.083333$& 34000.00 & $<11.4$&$ V $& 2  & Very limited data \cr
1966:    & 82.625000 &$ -71.766667$& 39000.00 & $<11.1$&$ V $& 2  & Very limited data \cr
1968-12a & 77.508292 &$ -71.664583$& 40205.01 & 10.9 &$ pg  $& 4  & RN, Well-studied system \cr
1970-03a & 83.302708 &$ -70.583833$& 40653.71 & 12.0: &$ V  $& 5 & Scanty data; Hybrid \cr
1970-11a & 83.889958 &$ -70.772611$& 40972.50 & 12.5: &$ V  $& 6  & Scanty data; \ion{Fe}{2} \cr
1971-03a & 74.576917 &$ -68.097111$& 41039.50 & 12.3: &$ V  $& 6 & Scanty data \cr
1971-08a & 85.148500 &$ -66.671528$& 41079.92 & 13.0 &$ V   $& 7 & RN; discovered late \cr
1972-08a & 82.114250 &$ -68.814250$& 41551.50 & 11.0 &$ pg  $& 8  &  Scanty data \cr
1973-09a & 78.826583 &$ -69.656361$& 41640.96 & 11.6 &$ pg  $& 9,10  &  Scanty data \cr
1977-02a & 91.439583 &$ -68.636861$& 43200.50 & 12.7 &$ V   $& 11,12  & Scanty data \cr
1977-03a & 76.295292 &$ -70.150417$& 43214.50 & 10.6 &$ V   $& 13  & Well observed; \ion{Fe}{2} \cr
1978-03a & 76.467750 &$ -65.884083$& 43596.50 & $<12.0$&$ V $& 14  & Maximum light missed; Hybrid? \cr
1978-11a & 75.248542 &$ -67.212444$& 43814.50 & 16.0 &$ V   $& 15  & Observed far postmaximum \cr
1981-09a & 83.038667 &$ -70.369889$& 44877.88 & 11.8 &$ pg  $& 16  & Scanty data; Hybrid \cr
1987-09a & 80.959667 &$ -70.012417$& 47056.02 & 9.6  &$ V   $& 17 & Good coverage\cr
1988-03a & 83.872083 &$ -70.358139$& 47242.00 & 11.2 &$ V   $& 18,19 & Extensive data; \ion{Fe}{2} \cr
1988-10a & 77.005333 &$ -68.627111$& 47448.00 & 11.3 &$ V   $& 20 &  Good coverage; He/N \cr
1990-01a & 80.840958 &$ -69.496778$& 47908.00 & 11.5 &$ V   $& 21 &  Extensive data; He/N \cr
1990-02a & 77.492833 &$ -71.664278$& 47936.60 & 11.2 &$ V   $& 22 & RN; 2nd eruption of LMCN 1968-12a; He/N \cr
1991-04a & 75.937458 &$ -70.303750$& 48362.50 & 12.3 &$ V   $& 23  & Very luminous; Hybrid \cr
1992-11a & 79.832583 &$ -68.909722$& 48940.10 & 10.7 &$ V   $& 24 &  Good coverage; \ion{Fe}{2} \cr
1995-02a & 81.709708 &$ -70.023278$& 49773.65 & 10.7 &$ V   $& 25  & Supersoft X-ray source (SSS); \ion{Fe}{2} \cr
1996-11a & 78.375000 &$ -68.633333$& 50399.17 & $<12.4$ &$ V$& 26   & MACHO detection \cr
1997-06a & 76.111250 &$ -67.643889$& 50615.89 & 13.5 &$ V   $& 27 & Limited data \cr
1998-12a & 83.884456 &$ -69.497803$& 51176.19 & 16.9 &$ R   $& 28 & Likely not a nova \cr
1999-09a & 79.982457 &$ -70.464836$& 51431.86 & 12.6 &$ I   $& 29  & Maximum possibly missed \cr
2000-07a & 81.254583 &$ -70.238056$& 51737.90 & 11.2 &$ V   $& 31  & Good coverage; \ion{Fe}{2} \cr
2001-06a & 81.000001 &$ -71.166664$& 52067.46 &  9.7 &$ R   $& 32,33  & Probably not a nova \cr
2001-08a & 72.239585 &$ -69.926582$& 52165.89 & 13.3 &$ I   $& 29  & Rare Cusp-type nova? \cr
2002-02a & 84.193254 &$ -71.592888$& 52332.56 & 10.5 &$ V   $& 34  & Good decline coverage; \ion{Fe}{2}  \cr
2002-10a & 77.508292 &$ -71.664583$& 52558.75 & 11.2 &$ V   $& 35 & RN; 3rd eruption of LMCRN 1968-12a \cr
2003-06a & 77.106628 &$ -68.439529$& 52808.47 & 12.2 &$ V   $& 36  & Fast nova, fair coverage; He/N \cr
2004-10a & 89.176748 &$ -68.909668$& 53298.69 & 10.9:&$ V   $& 37  & RN (LMCRN 1937-11a); He/N \cr
2005-09a & 91.651833 &$ -69.826194$& 53641.88 & 11.4 &$ I   $& 29  & SSS; \ion{Fe}{2}? \cr
2005-11a & 77.636168 &$ -69.209915$& 53693.73 & 12.8 &$ V   $& 30  & Slow nova; \ion{Fe}{2}  \cr
2009-02a & 85.184167 &$ -66.669889$& 54867.57 & 10.6 &$ V   $& 38  & RN (LMCRN 1971-08a); He/N \cr
2009-05a & 82.859874 &$ -67.094444$& 54956.49 & 12.1 &$ V   $& 39  & Slow nova; \ion{Fe}{2}  \cr
2010-11a & 77.493333 &$ -71.664639$& 55521.70 & 12.0 &$ I   $& 40  & RN; 4th eruption of LMCN 1968-12a \cr
2011-08a & 85.951917 &$ -69.325279$& 55787.90 &$<13.1$&$ I  $& 29  & Fast nova, maximum missed    \cr
2012-03a & 73.736751 &$ -70.448975$& 56012.53 &$<10.7$&$ w  $& 41  & Very fast, fair coverage; He/N \cr
2012-10a & 80.087872 &$ -73.095360$& 56225.81 & 13.9 &$ I   $& 42  & OGLE discovery; He/N \cr
2013-10a & 89.493250 &$ -74.902611$& 56581.84 & 11.2 &$ I   $& 43 & OGLE; good coverage\cr
2015-03a & 89.171292 &$ -75.160556$& 57096.62 & 14.3 &$ V   $& 44  & ASASSN-15fd; good coverage; \ion{Fe}{2} \cr
2016-01a & 77.493333 &$ -71.664639$& 57408.71 & $<11.5$ &$ I$& 45  & RN; 5th eruption of LMCN 1968-12a: He/N \cr
2016-04a & 77.635750 &$ -69.358444$& 57507.48 & 12.0 &$ w   $& 46 & MASTER discovery; He/N? \cr
2017-11a & 82.569875 &$ -73.269806$& 58074.76 & 11.8 &$ V   $& 47 & ASASSN-17pf; very slow nova; \ion{Fe}{2} \cr
2018-02a & 78.386292 &$ -68.633444$& 58174.66 & 11.0 &$ V   $& 48 & RN (LMCRN 1996-11a); OGLE discovery; He/N \cr
2018-05a & 96.586583 &$ -69.762861$& 58242.46 & 11.9 &$ R   $& 49 & ASASSN-18jj; AT 2018bej; \ion{Fe}{2}? \cr
2018-07a & 85.443583 &$ -71.810444$& 58287.85 & 10.7 &$ G   $& 50 & ASASSN-18pf; AT 2018dya \cr
2019-07a & 82.402667 &$ -70.165833$& 58691.85 & 12.6 &$ w   $& 51 & AT 2019lvm \cr
2019-11a & 78.723542 &$ -70.163500$& 58796.84 & 12.0 &$ g   $& 52 & AT 2019uni; He/N \cr
2020-05a & 77.493333 &$ -71.664639$& 58976.20 & 13.7 &$ g   $& 53 & RN; 6th eruption of LMCN 1968-12a; LMC V1371 \cr
2020-11a & 72.409417 &$ -68.897500$& 59170.82 & 11.2 &$ g   $& 54 & ASASSN-20oh; Hybrid? \cr
2022-05a & 80.616167 &$ -69.609250$& 59721.39 & 10.1 &$ g   $& 55 & BraTS discovery; confirmed nova \cr
2023-10a & 71.718458 &$ -65.897550$& 60230.75 & 10.4 &$ g   $& 56 & ASASSN-23hd; AT 2023uwa; bright; He/N \cr
2024-03a & 73.766375 &$ -65.932483$& 60387.58 & 10.2 &$ g   $& 57 & ASASSN-24by; AT 2024epj; bright; He/N \cr
2024-04a & 77.336375 &$ -67.139539$& 60401.60 & 11.5 &$ g   $& 58 & ASASSN-24ck; AT 2024fjh; He/N \cr
2024-08a & 77.493292 &$ -71.664639$& 60525.00 & 10.0 &$ W1  $& 59 & RN; 7th eruption of LMC 1968-12a; He/N \cr
2025-03a & 75.904417 &$ -69.402450$& 60764.52 & 14.3 &$ g   $& 60 & ASASSN-25ax; AT 2025ggn; questionable nova\cr
2025-09a & 85.526917 &$ -67.018517$& 60936.17 & 12.8 &$ L   $& 61 & AT 2025xvc; GOTO25hfy; \ion{Fe}{2} \cr
\enddata
\tablenotetext{a}{Data through 2020 are summarized in {\tt https://www.mpe.mpg.de/$\sim$m31novae/opt/lmc/}.}
\tablenotetext{b}{Values refer to the nova discovery.}
\tablenotetext{c}{(1) \cite{Luyten1927},
(2) \cite{Henize1954},
(3) \cite{Buscombe1955},
(4) \cite{Sievers1970},
(5) \cite{MacConnell1970},
(6) \cite{Graham1971a},
(7) \cite{Graham1971b},
(8) \cite{Bateson1972},
(9) \cite{Graham1973},
(10) \cite{Bateson1974},
(11) \cite{Graham1977a},
(12) \cite{Lewis1977},
(13) \cite{Graham1977b},
(14) \cite{Graham1978},
(15) \cite{Pesch1978},
(16) \cite{Maza1981},
(17) \cite{McNaught1987},
(18) \cite{Garradd1988},
(19) \cite{McNaught1988a},
(20) \cite{McNaught1988b},
(21) \cite{McNaught1990},
(22) \cite{Williams1990},
(23) \cite{Liller1991},
(24) \cite{Liller1992},
(25) \cite{Liller1995},
(26) \cite{Shida2004},
(27) \cite{Alcock1997},
(28) \cite{Becker1999},
(29) \cite{Mroz2016a},
(30) \cite{Liller2005b},
(31) \cite{Liller2000},
(32) \cite{Liller2002a},
(33) \cite{Liller2003a},
(34) \cite{Liller2002b},
(35) \cite{Pojmanski2002},
(36) \cite{Liller2003b},
(37) \cite{Liller2004a},
(38) \cite{Liller2009a},
(39) \cite{Liller2009b},
(30) \cite{Mroz2014},
(41) \cite{Seach2012},
(42) \cite{Wyrzykowski2012},
(43) \cite{Wyrzykowski2013},
(44) \cite{Danilet2015},
(45) \cite{Mroz2016b},
(46) \cite{Gorbovskoy2016},
(47) \cite{Chomiuk2018a},
(48) \cite{Chomiuk2018b},
(49) \cite{Chomiuk2018c},
(50) \cite{Stanek2018},
(51) \cite{Jacques2019},
(52) \cite{Pimentel2019},
(53) \cite{Schmeer2020},
(54) \cite{Stanek2020},
(55) \cite{Aydi2022},
(56) \cite{Strader2023},
(57) \cite{Perez-Fournon2024},
(58) \cite{Shore2024a},
(58) \cite{Darnley2024},
(60) \cite{Mikolajczyk2025},
(61) \cite{O'Neill2025}
}
\end{deluxetable*}


\bigskip
\bigskip
\bigskip
\startlongtable
\begin{deluxetable*}{lccccccl}
\tabletypesize{\footnotesize}
\thispagestyle{empty}
\pagenumbering{gobble}
\tablenum{2}
\tablecolumns{8}
\tablecaption{LMC Nova Observed Light-curve Properties\label{tab2}}
\tablehead{\colhead{Nova} & \colhead{$m_\mathrm{max}$} & \colhead{Filter} &  \colhead{$t_2$\tablenotemark{a}} & \colhead{$t_3$\tablenotemark{a}} & \colhead{Type} & \colhead{References\tablenotemark{b}} & \colhead{Notes} \\ \colhead{(LMCN)} & \colhead{(mag)} & \colhead{} & \colhead{(day)} & \colhead{(day)} & \colhead{} & \colhead{} & \colhead{}
}
\startdata
1926-09a  &$12.0\pm0.4$  &$pg$ & $[106\pm19]$  & 200:       & \dots   & 1,2,3          & RY Dor \cr
1936-02a  &$10.5\pm0.5$  &$pg$ & $[16\pm3]$    & 31.6:      & \dots   & 1,2,3          & N Dor 1936 \cr
1937-11a  &$10.7\pm0.3$  &$pg$ & $10\pm2$      & 20:        & \dots   & 1,2,3          & RN (LMCN 2004-10a); YY Dor \cr
1948-12a  &$13.0\pm0.2$  &$pg$ & $[53\pm9]$    & 100:       & \dots   & 1,2,3          & Very slow rise \cr
1968-12a  & 10.9:        &$pg$ & $4.0\pm0.4$ & [9.4]        & \dots   & 3,4,5          & RN (1968,1990,2002,2010,2016,2020,2024) \cr
1971-03a  &$12.0\pm0.3$  &$V $ & $20\pm2$      &$30\pm3$    & \dots   & 3,5,6          & See Fig.~3a \cr
1971-08a  &$<13$         &$V $ & \dots         & \dots      & \dots   & 7              & RN (LMCN 2009-02a); minimal data \cr
1977-03a  &$10.7\pm0.1$  &$V $ & $11.2\pm1$    &$20.7\pm2$  & \ion{Fe}{2}  & 3,5,8,9   & Excellent coverage \cr
1987-09a  &$9.6\pm0.1$   &$V $ & $2.0\pm0.2$   &$5.0\pm0.2$ & \dots   & 3,5,10         & See Fig.~3b \cr
1988-03a  &$11.2\pm0.2$  &$V $ & $22.5\pm4.0$  & 38.4       & \ion{Fe}{2}  & 3,11,12   & Multiwavelength observations \cr
1988-10a  &$10.3\pm0.2$  &$V $ &  5            & 10         & He/N    & 3,13,14,15     & Good coverage \cr
1990-01a  &$10.0\pm0.5$  &$V $ & $4.0\pm0.5$   & $7\pm2$    & He/N    & 16,17,18,19    & Poor coverage of maximum light \cr
1990-02a  & 11.2:        &$V $ &  3.4:         & 5.1:       & He/N    & 3,18,19,20,21  & LMCRN 1968-12a \cr
1991-04a  &$9.0\pm0.2$   &$V $ & $6.0\pm1.0$   &$8.0\pm1.0$ & Hybrid  & 19,22,23       & Brightest nova in LMC \cr
1992-11a  &$10.2\pm0.2$  &$V $ & $6.9\pm1.1$   &$13.7\pm1.6$& \ion{Fe}{2} & 11,24,25,26& Well observed \cr
1995-02a  &$10.3\pm0.2$  &$V $ & $11\pm3$      &$19.6\pm3.2$& \ion{Fe}{2} & 11,27      & Good coverage \cr
1996-11a  &$<12$         &$V $ & 9:            & 14:        & \dots   & 18             & RN (LMCN 2018-02a) \cr
1999-09a  &$12.5\pm0.5$  &$V $ & $15\pm2$      & 19.0       & \dots   & 28,29          & OGLE discovery \cr
2000-07a  &$10.4\pm0.4$  &$V $ & $9.0\pm3$     &$21\pm5$    & \ion{Fe}{2} & 11,18,30,31& Poor coverage of maximum light \cr
2002-02a  &$10.3\pm0.3$  &$V $ & $12\pm2$      & 23:        & \ion{Fe}{2} & 3,18,19,32 & Good coverage \cr
2002-10a  &$11.2\pm0.1$  &$V $ & 2.6:          & [6.4:]     & \dots   & 33             & LMCRN 1968-12a; good coverage of peak \cr
2003-06a  &$11.0\pm0.5$  &$V $ &  8:           & 12:        & He/N:   & 18,19,34       & Poor coverage of maximum light \cr
2004-10a  &$10.9\pm0.3$  &$V $ & $11\pm2$      &$21\pm3$    & He/N    & 19,28,35,36    & LMCRN 1937-11a \cr
2005-09a  &$11.4\pm0.2$  &$I $ &$14\pm4$       &[$28\pm7$]  & \ion{Fe}{2}: & 28,37     & Sporadic coverage \cr
2005-11a  &$11.5\pm0.2$  &$V $ & $70\pm7$      & 100:       & \ion{Fe}{2}  & 28,38,39  & Good coverage \cr
2009-02a  &$10.6\pm0.2$  &$V $ & $5.0\pm1.0$   &$10.4\pm2.1$& He/N    & 39,40,41       & LMCRN 1971-08a \cr
2009-05a  &$12.1\pm0.5$  &$V $ &  $51\pm19$    &[$87\pm16$] & \ion{Fe}{2}  & 40,42     & Dust former; see Fig.~3c \cr
2010-11a  &$11.7\pm0.3$  &$V $ &  3.5          & 5.0        & \dots   & 43             & LMCRN 1968-12a; $P_\mathrm{orb}=1.26432(8)$~d \cr
2012-03a  &$10.5\pm0.5$  &$w $ & $1.1\pm0.2$   & 2.1        & He/N    & 40,44,45       & Extremely fast \cr
2012-10a  &$11.5\pm0.5$  &$I $ & $10\pm2$      & 20:        & He/N    & 28,40,46       & OGLE discovery \cr
2013-10a  &$11.5\pm0.3$  &$V $ &  $24\pm4$     &[$45\pm8$]  &\dots    & 28,40,47,48    & See Fig.~3d \cr
2015-03a  &$11.5\pm0.3$  &$V $ &  $49\pm8$     &[$85\pm13$] & \ion{Fe}{2} & 40,49      & ASASSN-15fd; see Fig.~3e \cr
2016-01a  & 11.5:        &$I $ &  4.6          & 7.0        & He/N    & 50,51,52       & LMCRN 1968-12a \cr
2016-04a  &$12.0\pm0.5$  &$w $ &  $\lessim50.0$&[$\sim86$]  & He/N:   & 53,54,55       & $P_{\rm orb}=2.6508(2)$~d; see Fig.~2f \cr
2017-11a  &$11.8\pm0.1$  &$V $ &  $\grtsim100$ &[$\grtsim160$]& \ion{Fe}{2} & 56       & ASASSN-17pf \cr
2018-02a  &$11.0\pm0.2$  &$V $ & $4.0\pm1.0$   &$8.0\pm1.0$ & He/N    & 57,58,59       & LMCRN 1996-11a; $P_{\rm orb}=2.85$~d \cr
2018-05a  &$11.9\pm0.2$  &$R $ & $18\pm2$      &[$35\pm5$]  & \ion{Fe}{2}:  & 40,60    & ASASSN-18jj; AT 2018bej; see Fig.~4a \cr
2018-07a  &$10.5\pm0.5$  &$G $ & $23\pm2$      &[$44\pm5$]  & \dots   & 61,62          & ASASSN-18pf; AT 2018dya; see Fig.~4b \cr
2019-07a  &$10.7\pm0.3$  &$W$  & $15\pm3$      &[$30\pm6$]  & \dots   & 63,64          & AT 2019lvm; see Fig.~4c \cr
2019-11a  &$10.5\pm0.2$  &$g $ & $1.9\pm0.2$   &[$4.9\pm0.5$]& He/N   & 65,66          & AT 2019uni; see Fig.~4d \cr
2020-05a  &$<13.7$       &$g $ & $\sim6.5$     &[$\sim14$]  & He/N    & 67             & LMCRN 1968-12a; maximum light missed \cr
2022-05a  &$10.1\pm0.1$  &$g $ &  $13\pm2$     &[$26\pm4$]  & \dots   & 68             & See Fig.~4e \cr
2023-10a  &$10.4\pm0.1$  &$g $ &  $1.8\pm0.2$  &[$4.7\pm0.5$]& He/N   & 69             & ASASSN-23hd; AT 2023uwa; see Fig.~4f \cr
2024-03a  &$10.2\pm0.1$  &$g $ & $1.5\pm0.2$   &[$4.0\pm0.5$]& He/N   & 70,71          & ASASSN-24by; AT 2024epj; see Fig.~5a \cr
2024-04a  &$11.5\pm0.1$  &$g $ & $3.9\pm0.3$   &[$9.2\pm0.9$]& He/N   & 72,73          & ASASSN-24ck; AT 2024fjh; see Fig.~5b \cr
2024-08a  & 10.0:        &$UV$ & 2.5:          & 3.8:       & He/N    & 74,75,76       & LMCRN 1968-12a, maximum light missed \cr
2025-03a  &$12.2\pm0.4$ &$o$   & $26\pm3$      &[$48\pm7$]  & \dots   & 77             & See Fig.~5c \cr
2025-09a  &$11.6\pm0.2$ &$o$   & $56\pm5$      &[$95\pm12$] & \ion{Fe}{2}    & 78      & See Fig.~5d \cr
\enddata
\tablenotetext{a}{Values in square brackets have been estimated using the transformations in \cite{Shafter2026b}.}
\tablenotetext{b}{(1) \cite{Buscombe1955},
(2) \cite{Henize1954},
(3) \cite{Subramaniam2002},
(4) \cite{Sievers1970},
(5) \cite{Capaccioli1990a},
(6) \cite{Graham1971a},
(7) \cite{Graham1971b},
(8) \cite{Canterna1977},
(9) \cite{Canterna1981},
(10) \cite{McNaught1987},
(11) \cite{Hearnshaw2004},
(12) \cite{Schwarz1998},
(13) \cite{McNaught1988b},
(14) \cite{Martin1988},
(15) \cite{Sekiguchi1989},
(16) \cite{Dopita1990},
(17) \cite{Vanlandingham1999},
(18) \cite{Liller2004b},
(19) \cite{Liller2005},
(20) \cite{Sekiguchi1990},
(21) \cite{Shore1991},
(22) \cite{Schwarz2001},
(23) \cite{DellaValle1991},
(24) \cite{DellaValle1992},
(25) \cite{Duerbeck1992},
(26) \cite{Liller1992},
(27) \cite{DellaValle1995}.
(28) \cite{Mroz2016a},
(29) \cite{Shida2004},
(30) \cite{Greiner2003},
(31) \cite{Duerbeck2000},
(32) \cite{Mason2005},
(33) \cite{Pojmanski2002},
(34) \cite{Liller2003b},
(35) \cite{Bond2004},
(36) \cite{Mason2004},
(37) \cite{Read2009},
(38) \cite{Liller2007},
(39) \cite{Walter2012},
(40) \cite{Liller2009a},
(41) \cite{Bode2016},
(42) \cite{Liller2009b},
(43) \cite{Mroz2014},
(44) \cite{Seach2012},
(45) \cite{Schwarz2015},
(46) \cite{Wyrzykowski2012},
(47) \cite{Wyrzykowski2013},
(48) \cite{Hachisu2018},
(49) \cite{Stanek2015},
(50) \cite{Munari2016},
(51) \cite{Kuin2020},
(52) \cite{DiMille2016},
(53) \cite{Gorbovskoy2016},
(54) \cite{Bohlsen2016},
(55) {\tt https://ogle.astrouw.edu.pl/ogle4/cvom/lmcn-2016-04a.html},
(56) \cite{Aydi2019a},
(57) \cite{Chomiuk2018b},
(58) \cite{Walter2018},
(59) \cite{Mroz2018},
(60) \cite{Chomiuk2018c},
(61) {\tt https://gsaweb.ast.cam.ac.uk/alerts/alert/Gaia18bvg/},
(62) \cite{Stanek2018},
(63) \cite{Jacques2019},
(64) \cite{Aydi2019b},
(65) \cite{Pimentel2019},
(66) \cite{Aydi2019c}
(67) {\tt https://asas-sn.osu.edu/sky-patrol/coordinate/c8d6c909-61ba-4816-b417-73ca109955b1},
(68) \cite{Aydi2022},
(69) \cite{Strader2023},
(70) \cite{Perez-Fournon2024},
(71) \cite{Merc2024},
(72) \cite{Shore2024a},
(73) \cite{Aydi2024b},
(74) \cite{Darnley2024},
(75) \cite{Shore2024b},
(76) \cite{Basu2025},
(77) \cite{Mikolajczyk2025},
(78) \cite{O'Neill2025}.
}
\end{deluxetable*}   

\startlongtable
\begin{deluxetable*}{lcrl}
\tabletypesize{\footnotesize}
\tablenum{3}
\tablecolumns{4}
\tablecaption{LMC Nova Input Parameters\label{tab3}}
\tablehead{\colhead{Nova} & \colhead{$M_V$} & \colhead{$t_2$} & \colhead{} \\ \colhead{(LMCN)} & \colhead{(max)} & \colhead{(day)} & \colhead{Type}
}
\startdata
1926-09a&$-6.56\pm0.41$&$106.0\pm19.0$&\dots \cr
1936-02a&$-8.06\pm0.51$&$ 16.0\pm 3.0$&\dots \cr
1937-11a&$-7.76\pm0.31$&$ 11.0\pm 2.0$&He/N  \cr
1948-12a&$-5.56\pm0.22$&$ 53.0\pm 9.0$&\dots \cr
1968-12a&$-7.16\pm0.31$&$  3.5\pm 0.5$&He/N  \cr
1971-03a&$-6.66\pm0.31$&$ 22.0\pm 3.0$&\dots \cr
1971-08a&$-8.06\pm0.21$&$  5.0\pm 1.0$&He/N  \cr
1977-03a&$-7.96\pm0.12$&$ 11.2\pm 1.0$&\ion{Fe}{2}  \cr
1987-09a&$-9.06\pm0.12$&$  2.0\pm 0.2$&\dots \cr
1988-03a&$-7.46\pm0.21$&$ 22.5\pm 4.0$&\ion{Fe}{2}  \cr
1988-10a&$-8.36\pm0.21$&$  5.0\pm 0.5$&He/N  \cr
1990-01a&$-8.66\pm0.50$&$  4.0\pm 0.5$&He/N  \cr
1991-04a&$-9.66\pm0.21$&$  6.0\pm 1.0$&\ion{Fe}{2}b \cr
1992-11a&$-8.46\pm0.21$&$  6.9\pm 1.1$&\ion{Fe}{2}  \cr
1995-02a&$-8.36\pm0.21$&$ 11.0\pm 3.0$&\ion{Fe}{2}  \cr
1996-11a&$-7.66\pm0.21$&$  4.0\pm 1.0$&He/N  \cr
1999-09a&$-6.16\pm0.50$&$ 15.0\pm 2.0$&\dots \cr
2000-07a&$-8.26\pm0.41$&$  9.0\pm 3.0$&\ion{Fe}{2}  \cr
2002-02a&$-8.36\pm0.31$&$ 12.0\pm 2.0$&\ion{Fe}{2}  \cr
2003-06a&$-7.66\pm0.50$&$  8.0\pm 2.0$&He/N: \cr
2005-09a&$-6.93\pm0.22$&$ 14.0\pm 4.0$&\ion{Fe}{2}: \cr
2005-11a&$-7.16\pm0.21$&$ 70.0\pm 7.0$&\ion{Fe}{2}  \cr
2009-05a&$-6.56\pm0.50$&$ 51.0\pm19.0$&\ion{Fe}{2}  \cr
2012-03a&$-8.16\pm0.51$&$  1.1\pm 0.2$&He/N  \cr
2012-10a&$-6.83\pm0.51$&$ 10.0\pm 2.0$&He/N  \cr
2013-10a&$-7.16\pm0.31$&$ 24.0\pm 3.0$&\dots \cr
2015-03a&$-7.16\pm0.31$&$ 49.0\pm 8.0$&\ion{Fe}{2}  \cr
2016-04a&$-6.66\pm0.51$&$ 50.0\pm10.0$&He/N: \cr
2017-11a&$-6.86\pm0.12$&$100.0\pm20.0$&\ion{Fe}{2}  \cr
2018-05a&$-6.66\pm0.22$&$ 18.0\pm 2.0$&\ion{Fe}{2}: \cr
2018-07a&$-8.16\pm0.51$&$ 23.0\pm 2.0$&\dots \cr
2019-07a&$-7.96\pm0.31$&$ 15.0\pm 3.0$&\dots \cr
2019-11a&$-8.16\pm0.22$&$  1.9\pm 0.2$&He/N  \cr
2022-05a&$-8.56\pm0.13$&$ 13.0\pm 2.0$&\dots \cr
2023-10a&$-8.26\pm0.13$&$  1.8\pm 0.2$&He/N  \cr
2024-03a&$-8.46\pm0.13$&$  1.5\pm 0.2$&He/N  \cr
2024-04a&$-7.16\pm0.13$&$  3.9\pm 0.3$&He/N  \cr
2025-03a&$-6.26\pm0.41$&$ 26.0\pm 3.0$&\dots \cr
2025-09a&$-6.86\pm0.22$&$ 56.0\pm 5.0$&\ion{Fe}{2}  \cr
\enddata
\end{deluxetable*}

\startlongtable
\begin{deluxetable*}{lrrccccl}
\tablenum{4}
\tablecolumns{8}
\tabletypesize{\footnotesize}
\tablecaption{LMC Nova Parameters\label{tab4}}
\tablehead{\colhead{Nova} & \colhead{$L_4$} & \colhead{$t_2$} & \colhead{log $\dot M$} & \colhead{$M_\mathrm{WD}$} & \colhead{$V_\mathrm{max}$} & \colhead{log $P_\mathrm{rec}$} & \colhead{} \\ \colhead{(LMCN)} & \colhead{($10^4~L_{\odot}$)} & \colhead{(days)} & \colhead{($M_{\odot}$~yr$^{-1}$)}  & \colhead{($M_{\odot}$)} & \colhead{(km~s$^{-1}$)} & \colhead{(yr)} & \colhead{Type}}
\startdata
1926-09a&$  3.37\pm 1.31$&$  106.0\pm  19.0$&$ -8.29\pm 1.07$&$ 1.01\pm 0.19$&$   578\pm 475$&$  3.81\pm1.54$&\dots  \cr
1936-02a&$ 13.43\pm 6.42$&$   16.0\pm   3.0$&$-10.43\pm 0.57$&$ 1.17\pm 0.03$&$  2280\pm 674$&$  5.90\pm0.70$&\dots  \cr
1948-12a&$  1.34\pm 0.30$&$   53.0\pm   9.0$&$ -7.06\pm 0.66$&$ 1.17\pm 0.09$&$   341\pm 209$&$  1.47\pm0.68$&\dots  \cr
1971-03a&$  3.70\pm 1.11$&$   22.0\pm   3.0$&$ -7.03\pm 1.18$&$ 1.28\pm 0.09$&$   531\pm 523$&$  0.80\pm1.04$&\dots  \cr
1977-03a&$ 12.25\pm 1.76$&$   11.2\pm   1.0$&$-10.13\pm 0.24$&$ 1.21\pm 0.02$&$  2449\pm 656$&$  5.37\pm0.32$&\ion{Fe}{2}  \cr
1987-09a&$ 33.73\pm 4.85$&$    2.0\pm   0.2$&$-10.55\pm 0.14$&$\grtsim1.40  $&$  7228\pm2494$&$  4.79\pm0.21$&\dots  \cr
1988-03a&$  7.73\pm 1.65$&$   22.5\pm   4.0$&$ -9.76\pm 0.42$&$ 1.12\pm 0.03$&$  1556\pm 520$&$  5.32\pm0.54$&\ion{Fe}{2}  \cr
1988-10a&$ 17.70\pm 3.79$&$    5.0\pm   0.5$&$-10.26\pm 0.25$&$ 1.31\pm 0.02$&$  3999\pm1069$&$  5.01\pm0.33$&He/N   \cr
1990-01a&$ 23.33\pm10.95$&$    4.0\pm   0.5$&$-10.49\pm 0.45$&$ 1.34\pm 0.03$&$  5093\pm1631$&$  5.08\pm0.58$&He/N   \cr
1991-04a&$ 58.61\pm12.54$&$    6.0\pm   1.0$&$-11.46\pm 0.24$&$ 1.38\pm 0.02$&$  7211\pm3349$&$  5.98\pm0.34$&\ion{Fe}{2}b \cr
1992-11a&$ 19.41\pm 4.15$&$    6.9\pm   1.1$&$-10.50\pm 0.23$&$ 1.28\pm 0.02$&$  3741\pm 985$&$  5.44\pm0.31$&\ion{Fe}{2}  \cr
1995-02a&$ 17.70\pm 3.79$&$   11.0\pm   3.0$&$-10.59\pm 0.24$&$ 1.23\pm 0.03$&$  2985\pm 786$&$  5.82\pm0.32$&\ion{Fe}{2}  \cr
1999-09a&$  2.33\pm 1.09$&$   15.0\pm   2.0$&$\grtsim-7.00  $&$ 1.31\pm 0.05$&$   596\pm 320$&$  0.57\pm0.52$&\dots  \cr
2000-07a&$ 16.14\pm 6.27$&$    9.0\pm   3.0$&$-10.40\pm 0.41$&$ 1.25\pm 0.03$&$  3069\pm 868$&$  5.52\pm0.52$&\ion{Fe}{2}  \cr
2002-02a&$ 17.70\pm 5.30$&$   12.0\pm   2.0$&$-10.63\pm 0.28$&$ 1.22\pm 0.02$&$  2891\pm 761$&$  5.91\pm0.36$&\ion{Fe}{2}  \cr
2003-06a&$  9.29\pm 4.36$&$    8.0\pm   2.0$&$ -9.50\pm 1.22$&$ 1.24\pm 0.05$&$  2275\pm1270$&$  4.41\pm1.58$&He/N:  \cr
2005-09a&$  4.74\pm 1.05$&$   14.0\pm   4.0$&$ -7.32\pm 1.05$&$ 1.29\pm 0.07$&$   726\pm 625$&$  1.15\pm1.07$&\ion{Fe}{2}: \cr
2005-11a&$  5.86\pm 1.25$&$   70.0\pm   7.0$&$-10.02\pm 0.50$&$ 0.88\pm 0.11$&$  1102\pm 364$&$  6.28\pm0.62$&\ion{Fe}{2}  \cr
2009-05a&$  3.37\pm 1.58$&$   51.0\pm  19.0$&$ -7.48\pm 1.34$&$ 1.17\pm 0.14$&$   499\pm 552$&$  2.04\pm1.43$&\ion{Fe}{2}  \cr
2012-03a&$ 14.72\pm 7.04$&$    1.1\pm   0.2$&$ -9.27\pm 0.29$&$\grtsim1.40  $&$  4710\pm1564$&$  3.16\pm0.40$&He/N   \cr
2012-10a&$  4.33\pm 2.07$&$   10.0\pm   2.0$&$ -7.11\pm 1.11$&$ 1.33\pm 0.07$&$   736\pm 666$&$  0.60\pm0.99$&He/N   \cr
2013-10a&$  5.86\pm 1.76$&$   24.0\pm   3.0$&$ -9.17\pm 0.99$&$ 1.13\pm 0.07$&$  1259\pm 704$&$  4.54\pm1.30$&\dots  \cr
2015-03a&$  5.86\pm 1.76$&$   49.0\pm   8.0$&$ -9.71\pm 0.96$&$ 0.99\pm 0.12$&$  1130\pm 555$&$  5.68\pm1.22$&\ion{Fe}{2}  \cr
2016-04a&$  3.70\pm 1.77$&$   50.0\pm  10.0$&$ -7.62\pm 1.38$&$ 1.16\pm 0.14$&$   540\pm 593$&$  2.29\pm1.52$&He/N:  \cr
2017-11a&$  4.45\pm 0.64$&$  100.0\pm  20.0$&$ -9.61\pm 0.87$&$ 0.82\pm 0.14$&$   957\pm 401$&$  5.86\pm1.06$&\ion{Fe}{2}  \cr
2018-05a&$  3.70\pm 0.82$&$   18.0\pm   2.0$&$ -7.01\pm 0.91$&$ 1.30\pm 0.07$&$   566\pm 435$&$  0.67\pm0.82$&\ion{Fe}{2}: \cr
2018-07a&$ 14.72\pm 7.04$&$   23.0\pm   2.0$&$-10.72\pm 0.50$&$ 1.13\pm 0.03$&$  2080\pm 597$&$  6.42\pm0.61$&\dots  \cr
2019-07a&$ 12.25\pm 3.67$&$   15.0\pm   3.0$&$-10.28\pm 0.35$&$ 1.18\pm 0.02$&$  2228\pm 637$&$  5.71\pm0.45$&\dots  \cr
2019-11a&$ 14.72\pm 3.27$&$    1.9\pm   0.2$&$ -9.56\pm 0.37$&$ 1.38\pm 0.01$&$  4764\pm1584$&$  3.67\pm0.50$&He/N   \cr
2022-05a&$ 21.28\pm 3.21$&$   13.0\pm   2.0$&$-10.86\pm 0.19$&$ 1.23\pm 0.02$&$  3062\pm 794$&$  6.16\pm0.26$&\dots  \cr
2023-10a&$ 16.14\pm 2.44$&$    1.8\pm   0.2$&$ -9.68\pm 0.31$&$ 1.39\pm 0.01$&$  5212\pm1695$&$  3.78\pm0.42$&He/N   \cr
2024-03a&$ 19.41\pm 2.93$&$    1.5\pm   0.2$&$ -9.83\pm 0.24$&$\grtsim1.40  $&$  5929\pm1919$&$  3.89\pm0.33$&He/N   \cr
2024-04a&$  5.86\pm 0.88$&$    3.9\pm   0.3$&$ -7.08\pm 0.91$&$\grtsim1.40  $&$  1047\pm 765$&$  0.09\pm0.75$&He/N   \cr
2025-03a&$  2.56\pm 0.99$&$   26.0\pm   3.0$&$\grtsim-7.00  $&$ 1.27\pm 0.03$&$   491\pm 190$&$  0.82\pm0.24$&\dots  \cr
2025-09a&$  4.45\pm 0.99$&$   56.0\pm   5.0$&$ -9.00\pm 0.93$&$ 1.02\pm 0.11$&$   893\pm 517$&$  4.73\pm1.25$&\ion{Fe}{2}  \cr
\enddata
\end{deluxetable*}

\startlongtable
\begin{deluxetable*}{lrrcccccl}
\tablenum{5}
\tablecolumns{9}
\tabletypesize{\footnotesize}
\tablecaption{LMC Recurrent Nova Parameters\label{tab5}}
\tablehead{\colhead{Nova} & \colhead{$L_4$} & \colhead{$t_2$} & \colhead{log $\dot M$} & \colhead{$M_\mathrm{WD}$} & \colhead{$V_\mathrm{max}$} & \colhead{$P_\mathrm{rec}$} & $P_\mathrm{rec}$ (obs) & \colhead{} \\ \colhead{(LMCRN)} & \colhead{($10^4~L_{\odot}$)} & \colhead{(days)} & \colhead{($M_{\odot}$~yr$^{-1}$)}  & \colhead{($M_{\odot}$)} & \colhead{(km~s$^{-1}$)} & \colhead{(yr)} & \colhead{(yr)} & \colhead{Type}}
\startdata
1937-11a&$ 10.19\pm 3.05$&$   11.0\pm   2.0$&$ -9.84\pm 0.43$&$ 1.21\pm 0.02$&$  2223\pm 678$&   1.04E$+05$&$66.9$&He/N   \cr
1968-12a&$  5.86\pm 1.76$&$    3.5\pm   0.5$&$ -7.18\pm 0.94$&$\grtsim1.40  $&$  1146\pm 866$&   1.73     &$5.75$&He/N   \cr
1971-08a&$ 13.43\pm 2.87$&$    5.0\pm   1.0$&$ -9.88\pm 0.30$&$ 1.30\pm 0.02$&$  3357\pm 929$&  4.03E$+04$&$37.7$&He/N   \cr
1996-11a&$  9.29\pm 1.99$&$    4.0\pm   1.0$&$ -9.09\pm 0.76$&$ 1.32\pm 0.03$&$  2667\pm1268$&  2.98E$+03$&$21.3$&He/N   \cr
\enddata
\end{deluxetable*}

\startlongtable
\begin{deluxetable*}{lrrrrr}
\tablenum{6}
\tablecolumns{6}
\tabletypesize{\scriptsize}
\tablecaption{Nova Distribution Parameters\label{tab6}}
\tablehead{\colhead{Parameter} & \colhead{\ion{Fe}{2}} & \colhead{He/N\tablenotemark{a}} & \colhead{RNe} & \colhead{All LMC} & \colhead{All M31}
}
\startdata
$\langle M_\mathrm{WD} \rangle$ ($M_{\odot}$) & $1.11\pm0.16$ & $1.35\pm0.06$ & $1.31\pm0.08$ & $1.23\pm0.14$ & $1.15\pm0.14$  \cr
$\langle \mathrm{log}~{\dot M} \rangle$ ($M_{\odot}~\mathrm{yr}^{-1}$) & $-9.80\pm0.91$ & $-9.29\pm1.37$ & $-9.00\pm1.26$ & $-9.19\pm1.41$ & $-9.27\pm1.41$ \cr   
$\langle V_\mathrm{max} \rangle$ (km~s$^{-1}$) & $1934\pm1114$ & $3700\pm2031$ & $2348\pm928$ & $2414\pm1913$ & $1690\pm1361$ \cr
$\langle \mathrm{log}~P_\mathrm{rec} \rangle$ (yr) & $5.27\pm1.14$ & $3.43\pm1.95$ & $3.33\pm2.16$ & $3.90\pm2.05$ & $4.39\pm2.06$ \cr
\enddata
\tablenotetext{a}{Including Hybrid (\ion{Fe}{2}b) novae}
\end{deluxetable*}

\newpage

\bibliography{lmc_novae2}{}

@ARTICLE{Basu2025,
       author = {{Basu}, Judhajeet and {Anupama}, G.~C. and {Ness}, Jan-Uwe and {Singh}, Kulinder Pal and {Barway}, Sudhanshu and {Chamoli}, Shatakshi},
        title = "{Survival of the Accretion Disk in LMC Recurrent Nova 1968-12a: UV─X-Ray Case Study of the 2024 Eruption}",
      journal = {\apj},
         year = 2025,
        month = dec,
       volume = {994},
       number = {2},
          eid = {229},
        pages = {229},
          doi = {10.3847/1538-4357/ae13e1},
archivePrefix = {arXiv},
       eprint = {2510.11231},
 primaryClass = {astro-ph.HE},
       adsurl = {https://ui.adsabs.harvard.edu/abs/2025ApJ...994..229B}
}

@ARTICLE{Kato2014,
       author = {{Kato}, Mariko and {Saio}, Hideyuki and {Hachisu}, Izumi and {Nomoto}, Ken'ichi},
        title = "{Shortest Recurrence Periods of Novae}",
      journal = {\apj},
         year = 2014,
        month = oct,
       volume = {793},
       number = {2},
          eid = {136},
        pages = {136},
          doi = {10.1088/0004-637X/793/2/136},
archivePrefix = {arXiv},
       eprint = {1404.0582},
 primaryClass = {astro-ph.SR},
       adsurl = {https://ui.adsabs.harvard.edu/abs/2014ApJ...793..136K}
}

@ARTICLE{Wolf2013,
       author = {{Wolf}, William M. and {Bildsten}, Lars and {Brooks}, Jared and {Paxton}, Bill},
        title = "{Hydrogen Burning on Accreting White Dwarfs: Stability, Recurrent Novae, and the Post-nova Supersoft Phase}",
      journal = {\apj},
         year = 2013,
        month = nov,
       volume = {777},
       number = {2},
          eid = {136},
        pages = {136},
          doi = {10.1088/0004-637X/777/2/136},
archivePrefix = {arXiv},
       eprint = {1309.3375},
 primaryClass = {astro-ph.SR},
       adsurl = {https://ui.adsabs.harvard.edu/abs/2013ApJ...777..136W}
}

@ARTICLE{Bessell1998,
       author = {{Bessell}, M.~S. and {Castelli}, F. and {Plez}, B.},
        title = "{Model atmospheres broad-band colors, bolometric corrections and temperature calibrations for O - M stars}",
      journal = {\aap},
         year = 1998,
        month = may,
       volume = {333},
        pages = {231-250},
       adsurl = {https://ui.adsabs.harvard.edu/abs/1998A&A...333..231B}
}

@ARTICLE{Aaronson1978,
       author = {{Aaronson}, M.},
        title = "{The morphological distribution of bright galaxies in the UVK color plane.}",
      journal = {\apjl},
         year = 1978,
        month = may,
       volume = {221},
        pages = {L103-L107},
          doi = {10.1086/182674},
       adsurl = {https://ui.adsabs.harvard.edu/abs/1978ApJ...221L.103A}
}

@BOOK{Devaucouleurs1991,
       author = {{de Vaucouleurs}, Gerard and {de Vaucouleurs}, Antoinette and {Corwin}, Herold G., Jr. and {Buta}, Ronald J. and {Paturel}, Georges and {Fouque}, Pascal},
        title = "{Third Reference Catalogue of Bright Galaxies}",
         year = 1991,
       adsurl = {https://ui.adsabs.harvard.edu/abs/1991rc3..book.....D}
}

@ARTICLE{DellaValle2023,
       author = {{Della Valle}, Massimo and {Shafter}, Allen W. and {Starrfield}, Sumner},
        title = "{Recent Extragalactic Nova Rate Determinations and their Implications}",
      journal = {Research Notes of the American Astronomical Society},
         year = 2023,
        month = apr,
       volume = {7},
       number = {4},
          eid = {62},
        pages = {62},
          doi = {10.3847/2515-5172/acc937},
       adsurl = {https://ui.adsabs.harvard.edu/abs/2023RNAAS...7...62D}
}

@ARTICLE{Willmer2018,
       author = {{Willmer}, Christopher N.~A.},
        title = "{The Absolute Magnitude of the Sun in Several Filters}",
      journal = {\apjs},
         year = 2018,
        month = jun,
       volume = {236},
       number = {2},
          eid = {47},
        pages = {47},
          doi = {10.3847/1538-4365/aabfdf}, 
archivePrefix = {arXiv},
       eprint = {1804.07788},
 primaryClass = {astro-ph.SR},
       adsurl = {https://ui.adsabs.harvard.edu/abs/2018ApJS..236...47W}
}

@ARTICLE{DellaValle1998,
       author = {{Della Valle}, Massimo and {Livio}, Mario},
        title = "{The Spectroscopic Differences between Disk and Thick-Disk/Bulge Novae}",
      journal = {\apj},
         year = 1998,
        month = oct,
       volume = {506},
       number = {2},
        pages = {818-823},
          doi = {10.1086/306275},
       adsurl = {https://ui.adsabs.harvard.edu/abs/1998ApJ...506..818D}
}

@INPROCEEDINGS{Poggiani2021,
       author = {{Poggiani}, R.},
        title = "{Galactic and Extragalactic Novae - A Multiwavelenght Review}",
    booktitle = {The Golden Age of Cataclysmic Variables and Related Objects V},
         year = 2021,
       volume = {2-7},
        month = feb,
          eid = {26},
        pages = {26},
          doi = {10.22323/1.368.0026},
       adsurl = {https://ui.adsabs.harvard.edu/abs/2021gacv.workE..26P}
}

@ARTICLE{DellaValle2020,
       author = {{Della Valle}, Massimo and {Izzo}, Luca},
        title = "{Observations of galactic and extragalactic novae}",
      journal = {\aapr},
         year = 2020,
        month = jul,
       volume = {28},
       number = {1},
          eid = {3},
        pages = {3},
          doi = {10.1007/s00159-020-0124-6},
archivePrefix = {arXiv},
       eprint = {2004.06540},
 primaryClass = {astro-ph.SR},
       adsurl = {https://ui.adsabs.harvard.edu/abs/2020A&ARv..28....3D}
}

@ARTICLE{Ivezic2019,
       author = {{Ivezi{\'c}}, {\v{Z}}eljko and {Kahn}, Steven M. and {Tyson}, J. Anthony and {Abel}, Bob and {Acosta}, Emily and {Allsman}, Robyn and {Alonso}, David and {AlSayyad}, Yusra and {Anderson}, Scott F. and {Andrew}, John and {Angel}, James Roger P. and {Angeli}, George Z. and {Ansari}, Reza and {Antilogus}, Pierre and {Araujo}, Constanza and {Armstrong}, Robert and {Arndt}, Kirk T. and {Astier}, Pierre and {Aubourg}, {\'E}ric and {Auza}, Nicole and {Axelrod}, Tim S. and {Bard}, Deborah J. and {Barr}, Jeff D. and {Barrau}, Aurelian and {Bartlett}, James G. and {Bauer}, Amanda E. and {Bauman}, Brian J. and {Baumont}, Sylvain and {Bechtol}, Ellen and {Bechtol}, Keith and {Becker}, Andrew C. and {Becla}, Jacek and {Beldica}, Cristina and {Bellavia}, Steve and {Bianco}, Federica B. and {Biswas}, Rahul and {Blanc}, Guillaume and {Blazek}, Jonathan and {Blandford}, Roger D. and {Bloom}, Josh S. and {Bogart}, Joanne and {Bond}, Tim W. and {Booth}, Michael T. and {Borgland}, Anders W. and {Borne}, Kirk and {Bosch}, James F. and {Boutigny}, Dominique and {Brackett}, Craig A. and {Bradshaw}, Andrew and {Brandt}, William Nielsen and {Brown}, Michael E. and {Bullock}, James S. and {Burchat}, Patricia and {Burke}, David L. and {Cagnoli}, Gianpietro and {Calabrese}, Daniel and {Callahan}, Shawn and {Callen}, Alice L. and {Carlin}, Jeffrey L. and {Carlson}, Erin L. and {Chandrasekharan}, Srinivasan and {Charles-Emerson}, Glenaver and {Chesley}, Steve and {Cheu}, Elliott C. and {Chiang}, Hsin-Fang and {Chiang}, James and {Chirino}, Carol and {Chow}, Derek and {Ciardi}, David R. and {Claver}, Charles F. and {Cohen-Tanugi}, Johann and {Cockrum}, Joseph J. and {Coles}, Rebecca and {Connolly}, Andrew J. and {Cook}, Kem H. and {Cooray}, Asantha and {Covey}, Kevin R. and {Cribbs}, Chris and {Cui}, Wei and {Cutri}, Roc and {Daly}, Philip N. and {Daniel}, Scott F. and {Daruich}, Felipe and {Daubard}, Guillaume and {Daues}, Greg and {Dawson}, William and {Delgado}, Francisco and {Dellapenna}, Alfred and {de Peyster}, Robert and {de Val-Borro}, Miguel and {Digel}, Seth W. and {Doherty}, Peter and {Dubois}, Richard and {Dubois-Felsmann}, Gregory P. and {Durech}, Josef and {Economou}, Frossie and {Eifler}, Tim and {Eracleous}, Michael and {Emmons}, Benjamin L. and {Fausti Neto}, Angelo and {Ferguson}, Henry and {Figueroa}, Enrique and {Fisher-Levine}, Merlin and {Focke}, Warren and {Foss}, Michael D. and {Frank}, James and {Freemon}, Michael D. and {Gangler}, Emmanuel and {Gawiser}, Eric and {Geary}, John C. and {Gee}, Perry and {Geha}, Marla and {Gessner}, Charles J.~B. and {Gibson}, Robert R. and {Gilmore}, D. Kirk and {Glanzman}, Thomas and {Glick}, William and {Goldina}, Tatiana and {Goldstein}, Daniel A. and {Goodenow}, Iain and {Graham}, Melissa L. and {Gressler}, William J. and {Gris}, Philippe and {Guy}, Leanne P. and {Guyonnet}, Augustin and {Haller}, Gunther and {Harris}, Ron and {Hascall}, Patrick A. and {Haupt}, Justine and {Hernandez}, Fabio and {Herrmann}, Sven and {Hileman}, Edward and {Hoblitt}, Joshua and {Hodgson}, John A. and {Hogan}, Craig and {Howard}, James D. and {Huang}, Dajun and {Huffer}, Michael E. and {Ingraham}, Patrick and {Innes}, Walter R. and {Jacoby}, Suzanne H. and {Jain}, Bhuvnesh and {Jammes}, Fabrice and {Jee}, M. James and {Jenness}, Tim and {Jernigan}, Garrett and {Jevremovi{\'c}}, Darko and {Johns}, Kenneth and {Johnson}, Anthony S. and {Johnson}, Margaret W.~G. and {Jones}, R. Lynne and {Juramy-Gilles}, Claire and {Juri{\'c}}, Mario and {Kalirai}, Jason S. and {Kallivayalil}, Nitya J. and {Kalmbach}, Bryce and {Kantor}, Jeffrey P. and {Karst}, Pierre and {Kasliwal}, Mansi M. and {Kelly}, Heather and {Kessler}, Richard and {Kinnison}, Veronica and {Kirkby}, David and {Knox}, Lloyd and {Kotov}, Ivan V. and {Krabbendam}, Victor L. and {Krughoff}, K. Simon and {Kub{\'a}nek}, Petr and {Kuczewski}, John and {Kulkarni}, Shri and {Ku}, John and {Kurita}, Nadine R. and {Lage}, Craig S. and {Lambert}, Ron and {Lange}, Travis and {Langton}, J. Brian and {Le Guillou}, Laurent and {Levine}, Deborah and {Liang}, Ming and {Lim}, Kian-Tat and {Lintott}, Chris J. and {Long}, Kevin E. and {Lopez}, Margaux and {Lotz}, Paul J. and {Lupton}, Robert H. and {Lust}, Nate B. and {MacArthur}, Lauren A. and {Mahabal}, Ashish and {Mandelbaum}, Rachel and {Markiewicz}, Thomas W. and {Marsh}, Darren S. and {Marshall}, Philip J. and {Marshall}, Stuart and {May}, Morgan and {McKercher}, Robert and {McQueen}, Michelle and {Meyers}, Joshua and {Migliore}, Myriam and {Miller}, Michelle and {Mills}, David J.},
        title = "{LSST: From Science Drivers to Reference Design and Anticipated Data Products}",
      journal = {\apj},
         year = 2019,
        month = mar,
       volume = {873},
       number = {2},
          eid = {111},
        pages = {111},
          doi = {10.3847/1538-4357/ab042c},
archivePrefix = {arXiv},
       eprint = {0805.2366},
 primaryClass = {astro-ph},
       adsurl = {https://ui.adsabs.harvard.edu/abs/2019ApJ...873..111I}
}

@ARTICLE{Schaefer2025,
       author = {{Schaefer}, Bradley E.},
        title = "{Comprehensive Listing of 208 Nova White Dwarf Masses as the Primary Determinant of Spectral Class and Light-curve Class}",
      journal = {\apj},
         year = 2025,
        month = nov,
       volume = {993},
       number = {2},
          eid = {232},
        pages = {232},
          doi = {10.3847/1538-4357/ae0616}, 
       adsurl = {https://ui.adsabs.harvard.edu/abs/2025ApJ...993..232S}
}

@ARTICLE{Pala2022,
       author = {{Pala}, A.~F. and {G{\"a}nsicke}, B.~T. and {Belloni}, D. and {Parsons}, S.~G. and {Marsh}, T.~R. and {Schreiber}, M.~R. and {Breedt}, E. and {Knigge}, C. and {Sion}, E.~M. and {Szkody}, P. and {Townsley}, D. and {Bildsten}, L. and {Boyd}, D. and {Cook}, M.~J. and {De Martino}, D. and {Godon}, P. and {Kafka}, S. and {Kouprianov}, V. and {Long}, K.~S. and {Monard}, B. and {Myers}, G. and {Nelson}, P. and {Nogami}, D. and {Oksanen}, A. and {Pickard}, R. and {Poyner}, G. and {Reichart}, D.~E. and {Rodriguez Perez}, D. and {Shears}, J. and {Stubbings}, R. and {Toloza}, O.},
        title = "{Constraining the evolution of cataclysmic variables via the masses and accretion rates of their underlying white dwarfs}",
      journal = {\mnras},
         year = 2022,
        month = mar,
       volume = {510},
       number = {4},
        pages = {6110-6132},
          doi = {10.1093/mnras/stab3449},
archivePrefix = {arXiv},
       eprint = {2111.13706},
 primaryClass = {astro-ph.SR},
       adsurl = {https://ui.adsabs.harvard.edu/abs/2022MNRAS.510.6110P}
}

@ARTICLE{Chen2022,
   author = {{Chen}, B.-Q. and {Guo}, H.-L. and {Gao}, J. and {Liu}, X.-W. and {Zhang}, H.-W. and {Jiang}, B.-W. and {Li}, Y.},
    title = "{Dust distributions in the Magellanic Clouds}",
  journal = {\mnras},
   volume = {511},
   number = {1},
    pages = {1317-1329},
     year = {2022},
    month = mar,
   doi = {10.1093/mnras/stac072},
   adsurl = {https://ui.adsabs.harvard.edu/abs/2022MNRAS.511.1317C}
}

@ARTICLE{Oestreicher1995,
   author = {{Oestreicher}, M.~O. and {Gochermann}, J. and {Schmidt-Kaler}, Th.},
    title = "{The spatial distribution of the LMC foreground stars}",
  journal = {\aap},
   volume = {294},
    pages = {57-},
     year = {1995},
   adsurl = {https://ui.adsabs.harvard.edu/abs/1995A%26A...294...57O}
}

@ARTICLE{Gochermann2002,
   author = {{Gochermann}, J. and {Schmidt-Kaler}, Th.},
    title = "{Massive luminous early type stars in the LMC. I. The reddening of individual stars and the LMC reddening law}",
  journal = {\aap},
   volume = {391},
    pages = {187-193},
     year = {2002},
    month = aug,
   doi = {10.1051/0004-6361:20020771},
   adsurl = {https://ui.adsabs.harvard.edu/abs/2002A%26A...391..187G}
}

@ARTICLE{Shafter2026c,
       author = {{Shafter}, Allen W. and {Hornoch}, Kamil},
        title = "{The Recurrent Nova Population in M31}",
      journal = {arXiv e-prints},
         year = {2026c},
        month = apr,
          eid = {arXiv:2604.17637},
        pages = {arXiv:2604.17637},
archivePrefix = {arXiv},
       eprint = {2604.17637},
 primaryClass = {astro-ph.SR},
       adsurl = {https://ui.adsabs.harvard.edu/abs/2026arXiv260417637S}
}

@ARTICLE{Townsley2004,
  author = {{Townsley}, D. M. and {Bildsten}, L.},
  title = {Thermal Evolution of Accreting White Dwarfs: Effects of
           Convective Overshoot and Mixing}, 
  journal = {\apj},
  year = {2004},
  volume = {600},
  pages = {390--403},
  doi = {10.1086/379701},
  adsurl = {https://ui.adsabs.harvard.edu/abs/2004ApJ...600..390T}
}

@ARTICLE{Pecaut2013,
       author = {{Pecaut}, Mark J. and {Mamajek}, Eric E.},
        title = "{Intrinsic Colors, Temperatures, and Bolometric Corrections of Pre-main-sequence Stars}",
      journal = {\apjs},
         year = 2013,
        month = sep,
       volume = {208},
       number = {1},
          eid = {9},
        pages = {9},
          doi = {10.1088/0067-0049/208/1/9},
archivePrefix = {arXiv},
       eprint = {1307.2657},
 primaryClass = {astro-ph.SR},
       adsurl = {https://ui.adsabs.harvard.edu/abs/2013ApJS..208....9P}
}

@ARTICLE{Torres2010,
       author = {{Torres}, Guillermo},
        title = "{On the Use of Empirical Bolometric Corrections for Stars}",
      journal = {\aj},
         year = 2010,
        month = nov,
       volume = {140},
       number = {5},
        pages = {1158-1162}, 
          doi = {10.1088/0004-6256/140/5/1158},
archivePrefix = {arXiv},
       eprint = {1008.3913},
 primaryClass = {astro-ph.SR},
       adsurl = {https://ui.adsabs.harvard.edu/abs/2010AJ....140.1158T}
}

@ARTICLE{Aydi2019c,
       author = {{Aydi}, E. and {Buckley}, D.~A.~H. and {Chomiuk}, L. and {Kawash}, A. and {Orio}, M. and {Sokolovsky}, K.~V. and {Strader}, J. and {Woudt}, P.},
        title = "{SALT spectroscopic observations of the LMC nova PNV J05145365-7009486}",
      journal = {The Astronomer's Telegram},
         year = {2019c},
        month = nov,
       volume = {13288},
        pages = {1},
       adsurl = {https://ui.adsabs.harvard.edu/abs/2019ATel13288....1A}
}

@ARTICLE{Craig2025,
       author = {{Craig}, Peter and {Aydi}, Elias and {Chomiuk}, Laura and {Strader}, Jay and {Stone}, Ashley and {Sokolovsky}, Kirill V. and {Mukai}, Koji and {Kawash}, Adam and {Fl{\'o}}, Joan Guarro and {Boussin}, Christophe and {Charbonnel}, St{\'e}phane and {Garde}, Olivier},
        title = "{Revisiting the classics: on the optical colours of novae as standard crayons}",
      journal = {\mnras},
         year = 2025,
        month = apr,
       volume = {538},
       number = {4},
        pages = {2339-2357},
          doi = {10.1093/mnras/staf385},
archivePrefix = {arXiv},
       eprint = {2412.15108},
 primaryClass = {astro-ph.HE},
       adsurl = {https://ui.adsabs.harvard.edu/abs/2025MNRAS.538.2339C}
}

@ARTICLE{Andrillat1983,
       author = {{Andrillat}, Y. and {Dennefeld}, M.},
        title = "{A far red spectrum of Nova LMC 1981.}",
      journal = {\aap},
         year = 1983,
        month = aug,
       volume = {124},
        pages = {143-146},
       adsurl = {https://ui.adsabs.harvard.edu/abs/1983A&A...124..143A}
}

@ARTICLE{Yungelson1997,
       author = {{Yungelson}, L. and {Livio}, M. and {Tutukov}, A.},
        title = "{On The Rate of Novae in Galaxies of Different Types}",
      journal = {\apj},
         year = 1997,
        month = may,
       volume = {481},
       number = {1},
        pages = {127-131},
          doi = {10.1086/304020},
       adsurl = {https://ui.adsabs.harvard.edu/abs/1997ApJ...481..127Y}
}

@ARTICLE{Arp1961,
       author = {{Arp}, Halton},
        title = "{U - B and B - V Colors of Black Bodies.}",
      journal = {\apj},
         year = 1961,
        month = may,
       volume = {133},
        pages = {874},
          doi = {10.1086/147090},
       adsurl = {https://ui.adsabs.harvard.edu/abs/1961ApJ...133..874A}
}

@ARTICLE{Shara2018,
       author = {{Shara}, Michael M. and {Prialnik}, Dina and {Hillman}, Yael and {Kovetz}, Attay},
        title = "{The Masses and Accretion Rates of White Dwarfs in Classical and Recurrent Novae}",
      journal = {\apj},
         year = 2018,
        month = jun,
       volume = {860},
       number = {2},
          eid = {110},
        pages = {110},
          doi = {10.3847/1538-4357/aabfbd},
archivePrefix = {arXiv},
       eprint = {1804.06880},
 primaryClass = {astro-ph.SR},
       adsurl = {https://ui.adsabs.harvard.edu/abs/2018ApJ...860..110S}
}

@ARTICLE{Chamoli2025,
       author = {{Chamoli}, Shatakshi and {Basu}, Judhajeet and {Barway}, Sudhanshu and {Anupama}, G.~C. and {Swain}, Vishwajeet and {Bhalerao}, Varun},
        title = "{Challenging Classical Paradigms: Recurrent Nova M31N 2017-01e, a BeWD System in M31?}",
      journal = {\apj},
         year = 2025,
        month = oct,
       volume = {991},
       number = {2},
          eid = {174},
        pages = {174},
          doi = {10.3847/1538-4357/adf843},
archivePrefix = {arXiv},
       eprint = {2508.02227},
 primaryClass = {astro-ph.SR},
       adsurl = {https://ui.adsabs.harvard.edu/abs/2025ApJ...991..174C}
}

@ARTICLE{Orio1999,
       author = {{Orio}, M. and {Greiner}, J.},
        title = "{LMC 1995, the third supersoft X-ray nova}",
      journal = {\aap},
         year = 1999,
        month = apr,
       volume = {344},
        pages = {L13-L16},
       adsurl = {https://ui.adsabs.harvard.edu/abs/1999A&A...344L..13O}
}

@ARTICLE{Chen2016,
       author = {{Chen}, Hai-Liang and {Woods}, T.~E. and {Yungelson}, L.~R. and {Gilfanov}, M. and {Han}, Zhanwen},
        title = "{Modelling nova populations in galaxies}",
      journal = {\mnras},
         year = 2016,
        month = may,
       volume = {458},
       number = {3},
        pages = {2916-2927},
          doi = {10.1093/mnras/stw458},
archivePrefix = {arXiv},
       eprint = {1602.07849},
 primaryClass = {astro-ph.SR},
       adsurl = {https://ui.adsabs.harvard.edu/abs/2016MNRAS.458.2916C}
}

@ARTICLE{Harris2009,
       author = {{Harris}, Jason and {Zaritsky}, Dennis},
        title = "{The Star Formation History of the Large Magellanic Cloud}",
      journal = {\aj},
         year = 2009,
        month = nov,
       volume = {138},
       number = {5},
        pages = {1243-1260},
          doi = {10.1088/0004-6256/138/5/1243},
archivePrefix = {arXiv},
       eprint = {0908.1422},
 primaryClass = {astro-ph.CO},
       adsurl = {https://ui.adsabs.harvard.edu/abs/2009AJ....138.1243H}
}

@ARTICLE{Kochanek2017,
       author = {{Kochanek}, C.~S. and {Shappee}, B.~J. and {Stanek}, K.~Z. and {Holoien}, T.~W.-S. and {Thompson}, Todd A. and {Prieto}, J.~L. and {Dong}, Subo and {Shields}, J.~V. and {Will}, D. and {Britt}, C. and {Perzanowski}, D. and {Pojma{\'n}ski}, G.},
        title = "{The All-Sky Automated Survey for Supernovae (ASAS-SN) Light Curve Server v1.0}",
      journal = {\pasp},
         year = 2017,
        month = oct,
       volume = {129},
       number = {980},
        pages = {104502},
          doi = {10.1088/1538-3873/aa80d9},
archivePrefix = {arXiv},
       eprint = {1706.07060},
 primaryClass = {astro-ph.SR},
       adsurl = {https://ui.adsabs.harvard.edu/abs/2017PASP..129j4502K}
}

@ARTICLE{Williams1992,
       author = {{Williams}, Robert E.},
        title = "{The Formation of Novae Spectra}",
      journal = {\aj},
         year = 1992,
        month = aug,
       volume = {104},
        pages = {725},
          doi = {10.1086/116268},
       adsurl = {https://ui.adsabs.harvard.edu/abs/1992AJ....104..725W}
}

@ARTICLE{Clark2024,
       author = {{Clark}, J. Grace and {Hornoch}, Kamil and {Shafter}, Allen W. and {Ku{\v{c}}{\'a}kov{\'a}}, Hana and {Vra{\v{s}}til}, Jan and {Ku{\v{s}}nir{\'a}k}, Peter and {Wolf}, Marek},
        title = "{Exploring the Maximum Magnitude versus Rate of Decline Relation for Novae in M31}", 
      journal = {\apjs},
         year = 2024, 
        month = jun,
       volume = {272},
       number = {2},
          eid = {28},
        pages = {28},
          doi = {10.3847/1538-4365/ad3c39},
       adsurl = {https://ui.adsabs.harvard.edu/abs/2024ApJS..272...28C}
}

@ARTICLE{Rector2022,
       author = {{Rector}, Travis A. and {Shafter}, Allen W. and {Burris}, William A. and {Walentosky}, Matthew J. and {Viafore}, Kendall D. and {Strom}, Allison L. and {Cool}, Richard J. and {Sola}, Nicole A. and {Crayton}, Hannah and {Pilachowski}, Catherine A. and {Jacoby}, George H. and {Corbett}, Danielle L. and {Rene}, Michelle and {Hernandez}, Denise},
        title = "{The Rate and Spatial Distribution of Novae in M31 as Determined by a 20 Year Survey}",
      journal = {\apj},
         year = 2022,
        month = sep,
       volume = {936},
       number = {2},
          eid = {117},
        pages = {117},
          doi = {10.3847/1538-4357/ac87ad},
archivePrefix = {arXiv},
       eprint = {2207.05689},
 primaryClass = {astro-ph.GA},
       adsurl = {https://ui.adsabs.harvard.edu/abs/2022ApJ...936..117R}
}

@ARTICLE{Shafter2011,
       author = {{Shafter}, A.~W. and {Darnley}, M.~J. and {Hornoch}, K. and {Filippenko}, A.~V. and {Bode}, M.~F. and {Ciardullo}, R. and {Misselt}, K.~A. and {Hounsell}, R.~A. and {Chornock}, R. and {Matheson}, T.},
        title = "{A Spectroscopic and Photometric Survey of Novae in M31}",
      journal = {\apj},
         year = 2011,
        month = jun,
       volume = {734},
       number = {1},
          eid = {12},
        pages = {12},
          doi = {10.1088/0004-637X/734/1/12},
archivePrefix = {arXiv},
       eprint = {1104.0222},
 primaryClass = {astro-ph.GA},
       adsurl = {https://ui.adsabs.harvard.edu/abs/2011ApJ...734...12S}
}

@ARTICLE{Darnley2006,
       author = {{Darnley}, M.~J. and {Bode}, M.~F. and {Kerins}, E. and {Newsam}, A.~M. and {An}, J. and {Baillon}, P. and {Belokurov}, V. and {Calchi Novati}, S. and {Carr}, B.~J. and {Cr{\'e}z{\'e}}, M. and {Evans}, N.~W. and {Giraud-H{\'e}raud}, Y. and {Gould}, A. and {Hewett}, P. and {Jetzer}, Ph. and {Kaplan}, J. and {Paulin-Henriksson}, S. and {Smartt}, S.~J. and {Tsapras}, Y. and {Weston}, M.},
        title = "{Classical novae from the POINT-AGAPE microlensing survey of M31 - II. Rate and statistical characteristics of the nova population}",
      journal = {\mnras},
         year = 2006,
        month = jun,
       volume = {369},
       number = {1},
        pages = {257-271},
          doi = {10.1111/j.1365-2966.2006.10297.x},
archivePrefix = {arXiv},
       eprint = {astro-ph/0509493},
 primaryClass = {astro-ph},
       adsurl = {https://ui.adsabs.harvard.edu/abs/2006MNRAS.369..257D}
}

@ARTICLE{Shafter2001,
       author = {{Shafter}, Allen W. and {Irby}, Bryan K.},
        title = "{On the Spatial Distribution, Stellar Population, and Rate of Novae in M31}",
      journal = {\apj},
         year = 2001,
        month = dec,
       volume = {563},
       number = {2},
        pages = {749-767},
          doi = {10.1086/324044},
       adsurl = {https://ui.adsabs.harvard.edu/abs/2001ApJ...563..749S}
}

@ARTICLE{Capaccioli1989,
       author = {{Rosino}, L. and {Capaccioli}, M. and {D'Onofrio}, M. and {Della Valle}, M.},
        title = "{Fifty-Two Novae in M31 Discovered and Observed at Asiago from 1971 to 1986}", 
      journal = {\aj},
         year = 1989,
        month = jan,
       volume = {97},
        pages = {83},
          doi = {10.1086/114959},
       adsurl = {https://ui.adsabs.harvard.edu/abs/1989AJ.....97...83R}
}

@ARTICLE{Ciardullo1987,
       author = {{Ciardullo}, Robin and {Ford}, Holland C. and {Neill}, James D. and {Jacoby}, George H. and {Shafter}, Allen W.},
        title = "{The Spatial Distribution and Population of Novae in M31}",
      journal = {\apj},
         year = 1987,
        month = jul,
       volume = {318},
        pages = {520},
          doi = {10.1086/165388},
       adsurl = {https://ui.adsabs.harvard.edu/abs/1987ApJ...318..520C}
}

@ARTICLE{Rosino1973,
       author = {{Rosino}, L.},
        title = "{Novae in M 31 discovered and observed at Asiago from 1963 to 1970}",
      journal = {\aaps},
         year = 1973,
        month = mar,
       volume = {9},
        pages = {347},
       adsurl = {https://ui.adsabs.harvard.edu/abs/1973A&AS....9..347R}
}

@ARTICLE{Arp1956,
       author = {{Arp}, H.~C.},
        title = "{Novae in the Andromeda nebula.}",
      journal = {\aj},
         year = 1956,
        month = feb,
       volume = {61},
        pages = {15-34},
          doi = {10.1086/107284},
       adsurl = {https://ui.adsabs.harvard.edu/abs/1956AJ.....61...15A}
}

@ARTICLE{Hubble1929,
       author = {{Hubble}, E.~P.},
        title = "{A spiral nebula as a stellar system, Messier 31.}",
      journal = {\apj},
         year = 1929,
        month = mar,
       volume = {69},
        pages = {103-158},
          doi = {10.1086/143167},
       adsurl = {https://ui.adsabs.harvard.edu/abs/1929ApJ....69..103H}
}

@ARTICLE{Nail1951,
       author = {{McKibben}, Virginia Nail},
        title = "{Three Notes on Variable Stars in the Small Magellanic Cloud}",
      journal = {Harvard College Observatory Bulletin},
         year = 1951,
        month = sep,
       volume = {920},
        pages = {13-15},
       adsurl = {https://ui.adsabs.harvard.edu/abs/1951BHarO.920...13M}
}

@ARTICLE{Leavitt1908,
       author = {{Leavitt}, Henrietta S.},
        title = "{1777 variables in the Magellanic Clouds}",
      journal = {Annals of Harvard College Observatory},
         year = 1908,
        month = jan,
       volume = {60},
        pages = {87-108.3},
       adsurl = {https://ui.adsabs.harvard.edu/abs/1908AnHar..60...87L}
}

@ARTICLE{Hubble1926,
       author = {{Hubble}, E.~P.},
        title = "{A spiral nebula as a stellar system: Messier 33.}",
      journal = {\apj},
         year = 1926,
        month = may,
       volume = {63},
        pages = {236-274},
          doi = {10.1086/142976},
       adsurl = {https://ui.adsabs.harvard.edu/abs/1926ApJ....63..236H}
}

@ARTICLE{Pietrzynski2019,
       author = {{Pietrzyński}, G. and {Graczyk}, D. and {Gallenne}, A. and {Gieren}, W. and {Thompson}, I.~B. and {Pilecki}, B. and {Karczmarek}, P. and {Górski}, M. and {Suchomska}, K. and {Taormina}, M. and {Zgirski}, B. and {Wielgórski}, P. and {Kołaczkowski}, Z. and {Konorski}, P. and {Villanova}, S. and {Nardetto}, N. and {Kervella}, P. and {Bresolin}, F. and {Kudritzki}, R.-P. and {Storm}, J. and {Smolec}, R. and {Narloch}, W.},
        title = "{A distance to the Large Magellanic Cloud that is precise to one per cent}",
      journal = {Nature},
         year = 2019,
        month = mar,
       volume = {567},
       number = {7747},
        pages = {200-203},
          doi = {10.1038/s41586-019-0999-4},
archivePrefix = {arXiv},
       eprint = {1903.08096},
 primaryClass = {astro-ph.SR},
       adsurl = {https://ui.adsabs.harvard.edu/abs/2019Natur.567..200P}
}

@ARTICLE{Shara2016,
       author = {{Shara}, Michael M. and {Doyle}, Trisha F. and {Lauer}, Tod R. and {Zurek}, David and {Neill}, J.~D. and {Madrid}, Juan P. and {Miko{\l}ajewska}, Joanna and {Welch}, D.~L. and {Baltz}, Edward A.},
        title = "{A Hubble Space Telescope Survey for Novae in M87. I. Light and Color Curves, Spatial Distributions, and the Nova Rate}",
      journal = {\apjs},
         year = 2016,
        month = nov,
       volume = {227},
       number = {1},
          eid = {1},
        pages = {1},
          doi = {10.3847/0067-0049/227/1/1},
archivePrefix = {arXiv},
       eprint = {1602.00758},
 primaryClass = {astro-ph.SR},
       adsurl = {https://ui.adsabs.harvard.edu/abs/2016ApJS..227....1S}
}

@ARTICLE{Kasliwal2011,
       author = {{Kasliwal}, M.~M. and {Cenko}, S.~B. and {Kulkarni}, S.~R. and {Ofek}, E.~O. and {Quimby}, R. and {Rau}, A.},
        title = "{Discovery of a New Photometric Sub-class of Faint and Fast Classical Novae}", 
      journal = {\apj},
         year = 2011,
        month = jul,
       volume = {735},
       number = {2},
          eid = {94},
        pages = {94},
          doi = {10.1088/0004-637X/735/2/94},
archivePrefix = {arXiv},
       eprint = {1003.1720},
 primaryClass = {astro-ph.SR},
       adsurl = {https://ui.adsabs.harvard.edu/abs/2011ApJ...735...94K}
}

@ARTICLE{Shafter2023,
       author = {{Shafter}, Allen W. and {Clark}, J. Grace and {Hornoch}, Kamil},
        title = "{Concerning the Verity of the MMRD Relation for Novae}",
      journal = {Research Notes of the American Astronomical Society},
         year = 2023,
        month = sep,
       volume = {7},
       number = {9},
          eid = {191},
        pages = {191},
          doi = {10.3847/2515-5172/acf5e8},
archivePrefix = {arXiv},
       eprint = {2309.00739},
 primaryClass = {astro-ph.SR},
       adsurl = {https://ui.adsabs.harvard.edu/abs/2023RNAAS...7..191S}
}

@ARTICLE{McLaughlin1945,
       author = {{Mclaughlin}, Dean B.},
        title = "{The Relation between Light-Curves and Luminosities of Novae}",
      journal = {\pasp},
         year = 1945,
        month = apr,
       volume = {57},
       number = {335},
        pages = {69},
          doi = {10.1086/125689},
       adsurl = {https://ui.adsabs.harvard.edu/abs/1945PASP...57...69M}
}

@ARTICLE{Aydi2019b,
       author = {{Aydi}, E. and {Strader}, J. and {Chomiuk}, L. and {Kawash}, A. and {Sokolovsky}, K.~V. and {Shishkovsky}, L.},
        title = "{Spectroscopic classification of TCP J05293666-7009566 as a likely classical nova in the LMC}",
      journal = {The Astronomer's Telegram},
         year = {2019b},
        month = jul,
       volume = {12963},
        pages = {1},
       adsurl = {https://ui.adsabs.harvard.edu/abs/2019ATel12963....1A}
}

@ARTICLE{Mikolajczyk2025,
       author = {{Mikolajczyk}, P.~J. and {Wyrzykowski}, L. and {Kotysz}, K. and {Hambsch}, F.~J. and {Bronikowski}, M.},
        title = "{Photometric follow-up of extragalactic nova candidate AT2025ggn with BHTOM.space global telescope network}",
      journal = {Transient Name Server AstroNote},
         year = 2025,
        month = aug,
       volume = {255},
        pages = {1},
       adsurl = {https://ui.adsabs.harvard.edu/abs/2025TNSAN.255....1M}
}

@ARTICLE{Aydi2020,
       author = {{Aydi}, E. and {Buckley}, D.~A.~H. and {Chomiuk}, L. and {Harvey}, E.~J. and {Izzo}, L. and {Kawash}, A. and {Orio}, M. and {Sokolovsky}, K.~V. and {Strader}, J.},
        title = "{SALT spectroscopic classification of ASASSN-20oh as a classical nova in the LMC}",
      journal = {The Astronomer's Telegram},
         year = 2020,
        month = nov,
       volume = {14203},
        pages = {1},
       adsurl = {https://ui.adsabs.harvard.edu/abs/2020ATel14203....1A}
}

@ARTICLE{Graham1973,
       author = {{Graham}, J.~A.},
        title = "{Nova in Large Magellanic Cloud.}",
      journal = {\iaucirc},
         year = 1973,
        month = jan,
       volume = {2605},
        pages = {1},
       adsurl = {https://ui.adsabs.harvard.edu/abs/1973IAUC.2605....1G}
}

@ARTICLE{Havlen1972,
       author = {{Havlen}, R.~J. and {West}, R.~M. and {Westerlund}, B.~E.},
        title = "{Nova Mensae 1970b in the Large Magellanic Cloud.}",
      journal = {\aap},
         year = 1972,
        month = feb,
       volume = {16},
        pages = {404},
       adsurl = {https://ui.adsabs.harvard.edu/abs/1972A&A....16..404H}
}

@ARTICLE{Becker1999,
       author = {{Becker}, A. and {Winkler}, F. and {Paul}, D. and {MACHO Collaboration}},
        title = "{Variable in the Large Magellanic Cloud}",
      journal = {\iaucirc},
         year = 1999,
        month = jan,
       volume = {7087},
        pages = {1},
       adsurl = {https://ui.adsabs.harvard.edu/abs/1999IAUC.7087....1B}
}

@article{Yaron2005,
    author = {Yaron, O. and Prialnik, D. and Shara, M. M. and Kovetz, A.},
    title = {The Mass and Composition Dependence of Nova Outbursts},
    journal = {The Astrophysical Journal},
    volume = {623},
    number = {1},
    pages = {398--410},
    year = {2005},
    doi = {10.1086/428436}
}

@ARTICLE{Gilmore1991,
       author = {{Gilmore}, A.~C. and {Liller}, W. and {Cooper}, T. and {Overbeek}, D.},
        title = "{Nova in the Large Magellanic Cloud 1991}",
      journal = {\iaucirc},
         year = 1991,
        month = apr,
       volume = {5250},
        pages = {1},
       adsurl = {https://ui.adsabs.harvard.edu/abs/1991IAUC.5250....1G}
}

@ARTICLE{Shore2024b,
       author = {{Shore}, Steven N. and {Charbonnel}, Stephane and {Garde}, Olivier and {Le Du}, Pascal and {Mulato}, Lionel and {Petit}, Thomas},
        title = "{Spectroscopic observations of the latest outburst of Nova LMC 1968}",
      journal = {The Astronomer's Telegram},
         year = {2024b},
        month = aug,
       volume = {16761},
        pages = {1},
       adsurl = {https://ui.adsabs.harvard.edu/abs/2024ATel16761....1S}
}

@article{Bessell1990,
  author = {Bessell, M. S.},
  title = {UBVRI Passbands},
  journal = {Publications of the Astronomical Society of the Pacific},
  volume = {102},
  pages = {1181--1199},
  year = {1990},
  month = oct,
  doi = {10.1086/132749},
  adsurl = {https://ui.adsabs.harvard.edu/abs/1990PASP..102.1181B/abstract}
}

@ARTICLE{Jester2005,
       author = {{Jester}, Sebastian and {Schneider}, Donald P. and {Richards}, Gordon T. and {Green}, Richard F. and {Schmidt}, Maarten and {Hall}, Patrick B. and {Strauss}, Michael A. and {Vanden Berk}, Daniel E. and {Stoughton}, Chris and {Gunn}, James E. and {Brinkmann}, Jon and {Kent}, Stephen M. and {Smith}, J. Allyn and {Tucker}, Douglas L. and {Yanny}, Brian},
        title = "{The Sloan Digital Sky Survey View of the Palomar-Green Bright Quasar Survey}",
      journal = {\aj},
         year = 2005,
        month = sep,
       volume = {130},
       number = {3},
        pages = {873-895},
          doi = {10.1086/432466},
archivePrefix = {arXiv},
       eprint = {astro-ph/0506022},
 primaryClass = {astro-ph},
       adsurl = {https://ui.adsabs.harvard.edu/abs/2005AJ....130..873J}
}

@ARTICLE{Strope2010,
       author = {{Strope}, Richard J. and {Schaefer}, Bradley E. and {Henden}, Arne A.},
        title = "{Catalog of 93 Nova Light Curves: Classification and Properties}",
      journal = {\aj},
         year = 2010,
        month = jul,
       volume = {140},
       number = {1},
        pages = {34-62},
          doi = {10.1088/0004-6256/140/1/34},
archivePrefix = {arXiv},
       eprint = {1004.3698},
 primaryClass = {astro-ph.SR},
       adsurl = {https://ui.adsabs.harvard.edu/abs/2010AJ....140...34S}
}

@ARTICLE{Lipunov2010,
       author = {{Lipunov}, Vladimir and {Kornilov}, Victor and {Gorbovskoy}, Evgeny and {Shatskij}, Nikolaj and {Kuvshinov}, Dmitry and {Tyurina}, Nataly and {Belinski}, Alexander and {Krylov}, Alexander and {Balanutsa}, Pavel and {Chazov}, Vadim and {Kuznetsov}, Artem and {Kortunov}, Petr and {Sankovich}, Anatoly and {Tlatov}, Andrey and {Parkhomenko}, A. and {Krushinsky}, Vadim and {Zalozhnyh}, Ivan and {Popov}, A. and {Kopytova}, Taisia and {Ivanov}, Kirill and {Yazev}, Sergey and {Yurkov}, Vladimir},
        title = "{Master Robotic Net}",
      journal = {Advances in Astronomy},
         year = 2010,
        month = jan,
       volume = {2010},
          eid = {349171},
        pages = {349171},
          doi = {10.1155/2010/349171},
archivePrefix = {arXiv},
       eprint = {0907.0827},
 primaryClass = {astro-ph.HE},
       adsurl = {https://ui.adsabs.harvard.edu/abs/2010AdAst2010E..30L}
}

@ARTICLE{Walter2015,
       author = {{Walter}, Frederick},
        title = "{ASASSN-15fd = Nova LMC 2015}",
      journal = {The Astronomer's Telegram},
         year = 2015,
        month = apr,
       volume = {7350},
        pages = {1},
       adsurl = {https://ui.adsabs.harvard.edu/abs/2015ATel.7350....1W}
}

@ARTICLE{Hachisu2018,
       author = {{Hachisu}, Izumi and {Kato}, Mariko},
        title = "{A Light Curve Analysis of Recurrent and Very Fast Novae in Our Galaxy, Magellanic Clouds, and M31}",
      journal = {\apjs},
         year = 2018,
        month = jul,
       volume = {237},
       number = {1},
          eid = {4},
        pages = {4},
          doi = {10.3847/1538-4365/aac833},
archivePrefix = {arXiv},
       eprint = {1805.09932},
 primaryClass = {astro-ph.SR},
       adsurl = {https://ui.adsabs.harvard.edu/abs/2018ApJS..237....4H}
}

@ARTICLE{Prieto2012,
       author = {{Prieto}, J.~L.},
        title = "{Optical Spectroscopy of LMC Nova candidate 2012-03a}",
      journal = {The Astronomer's Telegram},
         year = 2012,
        month = mar,
       volume = {4002},
        pages = {1},
       adsurl = {https://ui.adsabs.harvard.edu/abs/2012ATel.4002....1P}
}

@ARTICLE{Hachisu2015,
       author = {{Hachisu}, Izumi and {Kato}, Mariko},
        title = "{A Light Curve Analysis of Classical Novae: Free-free Emission versus Photospheric Emission}",
      journal = {\apj},
         year = 2015,
        month = jan,
       volume = {798},
       number = {2},
          eid = {76},
        pages = {76},
          doi = {10.1088/0004-637X/798/2/76},
archivePrefix = {arXiv},
       eprint = {1410.7888},
 primaryClass = {astro-ph.SR},
       adsurl = {https://ui.adsabs.harvard.edu/abs/2015ApJ...798...76H}
}

@ARTICLE{Mason2004,
       author = {{Mason}, E. and {Ederoclite}, A. and {Stefanon}, M. and {dall}, T.~H. and {Della Valle}, M.},
        title = "{Nova in the Large Magellanic Cloud 2004}",
      journal = {\iaucirc},
         year = 2004,
        month = oct,
       volume = {8424},
        pages = {2},
       adsurl = {https://ui.adsabs.harvard.edu/abs/2004IAUC.8424....2M}
}

@ARTICLE{Duerbeck2000,
       author = {{Duerbeck}, H.~W. and {Pompei}, E.},
        title = "{Nova in the Large Magellanic Cloud 2000}",
      journal = {\iaucirc},
         year = 2000,
        month = jul,
       volume = {7457},
        pages = {1},
       adsurl = {https://ui.adsabs.harvard.edu/abs/2000IAUC.7457....1D}
}

@ARTICLE{DellaValle1995,
       author = {{Della Valle}, M. and {Masetti}, N. and {Benetti}, S. and {Kilmartin}, P.~M. and {Gilmore}, A.~C.},
        title = "{Nova in the Large Magellanic Cloud 1995}",
      journal = {\iaucirc},
         year = 1995,
        month = mar,
       volume = {6144},
        pages = {1},
       adsurl = {https://ui.adsabs.harvard.edu/abs/1995IAUC.6144....1D}
}

@ARTICLE{DellaValle1992,
       author = {{Della Valle}, M. and {Kaeufl}, U.},
        title = "{Nova in the Large Magellanic Cloud 1992}",
      journal = {\iaucirc},
         year = 1992,
        month = nov,
       volume = {5653},
        pages = {1},
       adsurl = {https://ui.adsabs.harvard.edu/abs/1992IAUC.5653....1D}
}

@ARTICLE{Duerbeck1992,
       author = {{Duerbeck}, H.~W. and {Grebel}, E.~K. and {Beele}, D.},
        title = "{Nova in the Large Magellanic Cloud 1992}",
      journal = {\iaucirc},
         year = 1992,
        month = nov,
       volume = {5654},
        pages = {1},
       adsurl = {https://ui.adsabs.harvard.edu/abs/1992IAUC.5654....1D}
}

@ARTICLE{Liller1992,
       author = {{Liller}, W. and {Camilleri}, P.},
        title = "{Nova in the Large Magellanic Cloud 1992}",
      journal = {\iaucirc},
         year = 1992,
        month = nov,
       volume = {5651},
        pages = {1},
       adsurl = {https://ui.adsabs.harvard.edu/abs/1992IAUC.5651....1L}
}

@ARTICLE{Shore1991,
       author = {{Shore}, Steven N. and {Sonneborn}, George and {Starrfield}, Sumner G. and {Hamuy}, M. and {Williams}, R.~E. and {Cassatella}, A. and {Drechsel}, H.},
        title = "{Multiwavelength Observations of Nova LMC 1990 Number 2: The First Extragalactic Recurrent Nova}",
      journal = {\apj},
         year = 1991,
        month = mar,
       volume = {370},
        pages = {193},
          doi = {10.1086/169804},
       adsurl = {https://ui.adsabs.harvard.edu/abs/1991ApJ...370..193S}
}

@ARTICLE{DellaValle1991,
       author = {{Della Valle}, M.},
        title = "{Nova LMC 1991 : evidence for a super-bright nova population.}",
      journal = {\aap},
         year = 1991,
        month = dec,
       volume = {252},
        pages = {L9},
       adsurl = {https://ui.adsabs.harvard.edu/abs/1991A&A...252L...9D}
}

@ARTICLE{Williams1991,
       author = {{Williams}, R.~E. and {Hamuy}, M. and {Phillips}, M.~M. and {Heathcote}, S.~R. and {Wells}, Lisa and {Navarrete}, M.},
        title = "{The Evolution and Classification of Postoutburst Novae Spectra}",
      journal = {\apj},
         year = 1991,
        month = aug,
       volume = {376},
        pages = {721},
          doi = {10.1086/170319},
       adsurl = {https://ui.adsabs.harvard.edu/abs/1991ApJ...376..721W}
}

@ARTICLE{Dopita1990,
       author = {{Dopita}, M.~A. and {Rawlings}, S.~J.},
        title = "{Nova in the Large Magellanic Cloud 1990 No. 1}",
      journal = {\iaucirc},
         year = 1990,
        month = feb,
       volume = {4964},
        pages = {2},
       adsurl = {https://ui.adsabs.harvard.edu/abs/1990IAUC.4964....2D}
}

@ARTICLE{Garradd1988,
       author = {{Garradd}, G.~J. and {Tregaskis}, B.},
        title = "{Nova in the Large Magellanic Cloud}",
      journal = {\iaucirc},
         year = 1988,
        month = mar,
       volume = {4568},
        pages = {1},
       adsurl = {https://ui.adsabs.harvard.edu/abs/1988IAUC.4568....1G}
}

@ARTICLE{Kuin2020,
       author = {{Kuin}, N.~P.~M. and {Page}, K.~L. and {Mr{\'o}z}, P. and {Darnley}, M.~J. and {Shore}, S.~N. and {Osborne}, J.~P. and {Walter}, F. and {Di Mille}, F. and {Morrell}, N. and {Munari}, U. and {Bohlsen}, T. and {Evans}, A. and {Gehrz}, R.~D. and {Starrfield}, S. and {Henze}, M. and {Williams}, S.~C. and {Schwarz}, G.~J. and {Udalski}, A. and {Szyma{\'n}ski}, M.~K. and {Poleski}, R. and {Soszy{\'n}ski}, I. and {Ribeiro}, V.~A.~R.~M. and {Angeloni}, R. and {Breeveld}, A.~A. and {Beardmore}, A.~P. and {Skowron}, J.},
        title = "{The 2016 January eruption of recurrent Nova LMC 1968}",
      journal = {\mnras},
         year = 2020,
        month = jan,
       volume = {491},
       number = {1},
        pages = {655-679},
          doi = {10.1093/mnras/stz2960},
archivePrefix = {arXiv},
       eprint = {1909.03281},
 primaryClass = {astro-ph.SR},
       adsurl = {https://ui.adsabs.harvard.edu/abs/2020MNRAS.491..655K}
}

@ARTICLE{Pojmanski2002,
       author = {{Pojmanski}, G.},
        title = "{The All Sky Automated Survey. Catalog of Variable Stars.  I. 0 h - 6 hQuarter of the Southern Hemisphere}",
      journal = {\actaa},
         year = 2002,
        month = dec,
       volume = {52},
        pages = {397-427},
          doi = {10.48550/arXiv.astro-ph/0210283},
archivePrefix = {arXiv},
       eprint = {astro-ph/0210283},
 primaryClass = {astro-ph},
       adsurl = {https://ui.adsabs.harvard.edu/abs/2002AcA....52..397P}
}

@ARTICLE{Shafter2026a,
       author = {{Shafter}, Allen W. and {Hornoch}, Kamil},
        title = "{Fundamental Properties of Novae in M31}",
      journal = {\apjs},
         year = {2026a},
        month = mar,
       volume = {283},
       number = {1},
          eid = {24},
        pages = {24},
          doi = {10.3847/1538-4365/ae3a86},
archivePrefix = {arXiv},
       eprint = {2601.11476},
 primaryClass = {astro-ph.SR},
       adsurl = {https://ui.adsabs.harvard.edu/abs/2026ApJS..283...24S}
}

@misc{Shafter2026b,
  author = {{Shafter}, A.~W.},
  title = {Modern Rate-of-Decline Relations for Novae},
  year = {2026b},
  howpublished = {Research Notes of the AAS}
}

@ARTICLE{Boyce1943,
       author = {{Boyce}, Emily Hughes},
        title = "{Eighty-one New Variable Sears; in VSF 524 and 566}",
      journal = {Harvard College Observatory Bulletin},
         year = 1943,
        month = dec,
       volume = {917},
        pages = {1-5},
       adsurl = {https://ui.adsabs.harvard.edu/abs/1943BHarO.917....1B}
}

@ARTICLE{vandenbergh1988b,
       author = {{van den Bergh}, S. and {Hazen}, M.~L. and {Boyce}, E.~H.},
        title = "{Supernova 1935C in NGC 1511}",
      journal = {\iaucirc},
         year = {1988b},
        month = aug,
       volume = {4647},
        pages = {2},
       adsurl = {https://ui.adsabs.harvard.edu/abs/1988IAUC.4647....2V}
}

@ARTICLE{vandenbergh1988a,
       author = {{van den Bergh}, Sidney and {Hazen}, Martha L.},
        title = "{Was Nova Hydri 1935 a Supernova?}",
      journal = {\pasp},
         year = {1988a},
        month = dec,
       volume = {100},
        pages = {1542},
          doi = {10.1086/132362},
       adsurl = {https://ui.adsabs.harvard.edu/abs/1988PASP..100.1542V}
}

@ARTICLE{Shafter2013,
       author = {{Shafter}, A.~W.},
        title = "{Photometric and Spectroscopic Properties of Novae in the Large Magellanic Cloud}",
      journal = {\aj},
         year = 2013,
        month = may,
       volume = {145},
       number = {5},
          eid = {117},
        pages = {117},
          doi = {10.1088/0004-6256/145/5/117},
archivePrefix = {arXiv},
       eprint = {1302.6285},
 primaryClass = {astro-ph.GA},
       adsurl = {https://ui.adsabs.harvard.edu/abs/2013AJ....145..117S}
}

@ARTICLE{Schwarz2015,
       author = {{Schwarz}, Greg J. and {Shore}, Steven N. and {Page}, Kim L. and {Osborne}, Julian P. and {Beardmore}, Andrew P. and {Walter}, Frederick M. and {Bode}, Michael F. and {Drake}, Jeremy J. and {Ness}, Jan-Uwe and {Starrfield}, Sumner and {Van Rossum}, Daniel R. and {Woodward}, Charles E.},
        title = "{Pan-Chromatic Observations of the Remarkable Nova Large Magellanic Cloud 2012}",
      journal = {\aj},
         year = 2015,
        month = mar,
       volume = {149},
       number = {3},
          eid = {95},
        pages = {95},
          doi = {10.1088/0004-6256/149/3/95},
archivePrefix = {arXiv},
       eprint = {1412.6492},
 primaryClass = {astro-ph.SR},
       adsurl = {https://ui.adsabs.harvard.edu/abs/2015AJ....149...95S}
}

@ARTICLE{Bode2016,
       author = {{Bode}, M.~F. and {Darnley}, M.~J. and {Beardmore}, A.~P. and {Osborne}, J.~P. and {Page}, K.~L. and {Walter}, F.~M. and {Krautter}, J. and {Melandri}, A. and {Ness}, J.-U. and {O'Brien}, T.~J. and {Orio}, M. and {Schwarz}, G.~J. and {Shara}, M.~M. and {Starrfield}, S.},
        title = "{Pan-chromatic Observations of the Recurrent Nova LMC 2009a (LMC 1971b)}",
      journal = {\apj},
         year = 2016,
        month = feb,
       volume = {818},
       number = {2},
          eid = {145},
        pages = {145},
          doi = {10.3847/0004-637X/818/2/145},
archivePrefix = {arXiv},
       eprint = {1601.00474},
 primaryClass = {astro-ph.SR},
       adsurl = {https://ui.adsabs.harvard.edu/abs/2016ApJ...818..145B}
}

@ARTICLE{Schwarz1998,
       author = {{Schwarz}, Greg J. and {Hauschildt}, Peter H. and {Starrfield}, S. and {Whitelock}, P.~A. and {Baron}, E. and {Sonneborn}, G.},
        title = "{A multiwavelength study of the early evolution of the classical nova LMC 1988 1}",
      journal = {\mnras},
         year = 1998,
        month = nov,
       volume = {300},
       number = {3},
        pages = {931-944},
          doi = {10.1046/j.1365-8711.1998.01964.x},
       adsurl = {https://ui.adsabs.harvard.edu/abs/1998MNRAS.300..931S}
}

@ARTICLE{Liller2004a,
       author = {{Liller}, W. and {Pearce}, A. and {Monard}, L.~A.~G.},
        title = "{Possible Nova in the Large Magellanic Cloud}",
      journal = {\iaucirc},
         year = {2004a},
        month = oct,
       volume = {8422},
        pages = {1},
       adsurl = {https://ui.adsabs.harvard.edu/abs/2004IAUC.8422....1L}
}

@ARTICLE{Liller2004b,
       author = {{Liller}, W. and {Shida}, R.~Y. and {Jones}, A.~F.},
        title = "{Light Curves for Recent Magellanic Cloud Novae}",
      journal = {Information Bulletin on Variable Stars},
         year = {2004b},
        month = dec,
       volume = {5582},
        pages = {1},
       adsurl = {https://ui.adsabs.harvard.edu/abs/2004IBVS.5582....1L}
}

@ARTICLE{Liller1995,
       author = {{Liller}, W.},
        title = "{Possible Nova in the Large Magellanic Cloud}",
      journal = {\iaucirc},
         year = 1995,
        month = mar,
       volume = {6143},
        pages = {2},
       adsurl = {https://ui.adsabs.harvard.edu/abs/1995IAUC.6143....2L}
}

@ARTICLE{Liller1991,
       author = {{Liller}, W. and {McNaught}, R.~H. and {Hughes}, S.~M. and {Hartley}, M. and {Camilleri}, P. and {Garradd}, G.},
        title = "{Nova in the Large Magellanic Cloud 1991}",
      journal = {\iaucirc},
         year = 1991,
        month = apr,
       volume = {5244},
        pages = {1},
       adsurl = {https://ui.adsabs.harvard.edu/abs/1991IAUC.5244....1L}
}

@ARTICLE{Williams1990,
       author = {{Williams}, R. and {Liller}, W. and {Shara}, M. and {Moffat}, A. and {Wells}, L. and {Heathcote}, S.},
        title = "{Nova in the Large Magellanic Cloud 1990 No. 2}",
      journal = {\iaucirc},
         year = 1990,
        month = feb,
       volume = {4964},
        pages = {1},
       adsurl = {https://ui.adsabs.harvard.edu/abs/1990IAUC.4964....1W}
}

@ARTICLE{McNaught1990,
       author = {{McNaught}, R.~H. and {Garradd}, G.~J. and {Seargent}, D.~A.~J. and {Pearce}, A.},
        title = "{Nova in the Large Magellanic Cloud 1990}",
      journal = {\iaucirc},
         year = 1990,
        month = jan,
       volume = {4946},
        pages = {1},
       adsurl = {https://ui.adsabs.harvard.edu/abs/1990IAUC.4946....1M}
}

@ARTICLE{McNaught1988b,
       author = {{McNaught}, R.~H. and {Garradd}, G.~J. and {Hartley}, M.},
        title = "{Nova in the Large Magellanic Cloud 1988 No. 2}",
      journal = {\iaucirc},
         year = {1988b},
        month = oct,
       volume = {4663},
        pages = {1},
       adsurl = {https://ui.adsabs.harvard.edu/abs/1988IAUC.4663....1M}
}

@ARTICLE{Martin1988,
       author = {{Martin}, G. and {Hamuy}, M. and {Suntzeff}, N. and {McNaught}, R.~H. and {Jones}, A.~F.~A.~L. and {Williams}, P. and {Seargent}, D.~A.~J.},
        title = "{Nova in the Large Magellanic Cloud 1988 No. 2}",
      journal = {\iaucirc},
         year = 1988,
        month = oct,
       volume = {4666},
        pages = {1},
       adsurl = {https://ui.adsabs.harvard.edu/abs/1988IAUC.4666....1M}
}

@ARTICLE{McNaught1988a,
       author = {{McNaught}, R.~H. and {Tregaskis}, B.},
        title = "{Nova in the Large Magellanic Cloud}",
      journal = {\iaucirc},
         year = {1988a},
        month = mar,
       volume = {4569},
        pages = {1},
       adsurl = {https://ui.adsabs.harvard.edu/abs/1988IAUC.4569....1M}
}

@ARTICLE{Maza1981,
       author = {{Maza}, J. and {Wischnjewsky}, M. and {Gonzalez}, L.~E. and {Jekabsons}, P. and {Page}, A.~A. and {Duerbeck}, H.~W. and {Barbier}, R. and {Bouchet}, P. and {Le van Suu}, A. and {Bessell}, M.~S. and {Wood}, P.~R.},
        title = "{Nova in Large Magellanic Cloud}",
      journal = {\iaucirc},
         year = 1981,
        month = oct,
       volume = {3641},
        pages = {1},
       adsurl = {https://ui.adsabs.harvard.edu/abs/1981IAUC.3641....1M}
}

@ARTICLE{Lewis1977,
       author = {{Lewis}, B.~M. and {Walker}, W.~S.~G.},
        title = "{Nova in Large Magellanic Cloud}",
      journal = {\iaucirc},
         year = 1977,
        month = may,
       volume = {3069},
        pages = {2},
       adsurl = {https://ui.adsabs.harvard.edu/abs/1977IAUC.3069....2L}
}

@ARTICLE{Bateson1974,
       author = {{Bateson}, F.~M.},
        title = "{Nova in Large Magellanic Cloud}",
      journal = {Information Bulletin on Variable Stars},
         year = 1974,
        month = jan,
       volume = {860},
        pages = {2},
       adsurl = {https://ui.adsabs.harvard.edu/abs/1974IBVS..860....2B}
}

@ARTICLE{Bateson1972,
       author = {{Bateson}, F.~M.},
        title = "{Nova in Large Magellanic Cloud}",
      journal = {Information Bulletin on Variable Stars},
         year = 1972,
        month = oct,
       volume = {725},
        pages = {1},
       adsurl = {https://ui.adsabs.harvard.edu/abs/1972IBVS..725....1B}
}

@ARTICLE{Luyten1927,
       author = {{Luyten}, W.~J.},
        title = "{A Possible Nova in the Large Magellanic Cloud}",
      journal = {Harvard College Observatory Bulletin},
         year = 1927,
        month = jun,
       volume = {847},
        pages = {8-9},
       adsurl = {https://ui.adsabs.harvard.edu/abs/1927BHarO.847....8L}
}

@ARTICLE{Darnley2024,
       author = {{Darnley}, M.~J. and {Kuin}, N.~P.~M. and {Page}, K.~L.},
        title = "{The 2024 eruption of the Recurrent Nova LMC 1968 as caught by Swift}",
      journal = {The Astronomer's Telegram},
         year = 2024,
        month = aug,
       volume = {16752},
        pages = {1},
       adsurl = {https://ui.adsabs.harvard.edu/abs/2024ATel16752....1D}
}

@ARTICLE{Stanek2020,
       author = {{Stanek}, K.~Z. and {Bersier}, D. and {Brimacombe}, J. and {Kochanek}, C.~S. and {Desai}, D. and {Way}, Z. and {Valley}, P. and {Thompson}, T.~A. and {Shappee}, B.~J. and {Holoien}, T.~W.-S. and {Prieto}, J.~L. and {Dong}, Subo and {Stritzinger}, M.},
        title = "{ASASSN-20oh: Discovery of a Likely Classical Nova in the Large Magellanic Cloud}",
      journal = {The Astronomer's Telegram},
         year = 2020,
        month = nov,
       volume = {14185},
        pages = {1},
       adsurl = {https://ui.adsabs.harvard.edu/abs/2020ATel14185....1S}
}

@ARTICLE{Chomiuk2018a,
       author = {{Chomiuk}, L. and {Strader}, J. and {Stanek}, K.~Z. and {Kochanek}, C.~S. and {Shields}, J.~V. and {Thompson}, T.~A. and {Shappee}, B.~J. and {Holoien}, T.~W.-S. and {Prieto}, J.~L. and {Dong}, Subo and {Morrell}, N. and {Phillips}, M. and {Bohlsen}, T.},
        title = "{ASAS-SN 17pf is a nova in the LMC}",
      journal = {The Astronomer's Telegram},
         year = {2018a},
        month = jan,
       volume = {11132},
        pages = {1},
       adsurl = {https://ui.adsabs.harvard.edu/abs/2018ATel11132....1C}
}

@ARTICLE{Wyrzykowski2013,
       author = {{Wyrzykowski}, L. and {Udalski}, A. and {Kozlowski}, S. and {Skowron}, J.},
        title = "{OGLE-2013-NOVA-02: Bright nova in the LMC}",
      journal = {The Astronomer's Telegram},
         year = 2013,
        month = oct,
       volume = {5491},
        pages = {1},
       adsurl = {https://ui.adsabs.harvard.edu/abs/2013ATel.5491....1W}
}

@ARTICLE{Bond2004,
       author = {{Bond}, H.~E. and {Walter}, F. and {Espinoza}, J. and {Gonzalez}, D. and {Pasten}, A. and {Green}, D.~W.~E.},
        title = "{Nova in the Large Magellanic Cloud 2004}",
      journal = {\iaucirc},
         year = 2004,
        month = oct,
       volume = {8424},
        pages = {1},
       adsurl = {https://ui.adsabs.harvard.edu/abs/2004IAUC.8424....1B}
}

@ARTICLE{Liller2002b,
       author = {{Liller}, W.},
        title = "{Nova in the Large Magellanic Cloud 2002}",
      journal = {\iaucirc},
         year = {2002b},
        month = mar,
       volume = {7841},
        pages = {1},
       adsurl = {https://ui.adsabs.harvard.edu/abs/2002IAUC.7841....1L}
}

@INPROCEEDINGS{Liller2002a,
       author = {{Liller}, William and {Morel}, Mati},
        title = "{The Mysterious Eruption of V2434 - LMC}",
    booktitle = {Classical Nova Explosions},
         year = {2002a},
       editor = {{Hernanz}, Margarita and {Jos{\'e}}, Jordi},
       series = {American Institute of Physics Conference Series},
       volume = {637},
        month = nov,
    publisher = {AIP},
        pages = {472-475},
          doi = {10.1063/1.1518247},
       adsurl = {https://ui.adsabs.harvard.edu/abs/2002AIPC..637..472L}
}

@ARTICLE{pesch1978,
       author = {{Pesch}, P. and {Sanduleak}, N.},
        title = "{Probable Nova in the Large Magellanic Cloud}",
      journal = {\iaucirc},
         year = 1978,
        month = nov,
       volume = {3308},
        pages = {1},
       adsurl = {https://ui.adsabs.harvard.edu/abs/1978IAUC.3308....1P}
}

@ARTICLE{Graham1977b,
       author = {{Graham}, J.~A. and {Rojas}, H. and {Canterna}, R.},
        title = "{Nova in Large Magellanic Cloud}",
      journal = {\iaucirc},
         year = {1977b},
        month = mar,
       volume = {3049},
        pages = {1},
       adsurl = {https://ui.adsabs.harvard.edu/abs/1977IAUC.3049....1G}
}

@ARTICLE{Graham1977a,
       author = {{Graham}, J.~A. and {Rojas}, H.},
        title = "{Nova in Large Magellanic Cloud}",
      journal = {\iaucirc},
         year = {1977a},
        month = mar,
       volume = {3045},
        pages = {1},
       adsurl = {https://ui.adsabs.harvard.edu/abs/1977IAUC.3045....1G}
}

@ARTICLE{Graham1979,
       author = {{Graham}, J.~A.},
        title = "{The premaximum spectrum of a Magellanic Cloud nova.}",
      journal = {\pasp},
         year = 1979,
        month = feb,
       volume = {91},
        pages = {79-82},
          doi = {10.1086/130445},
       adsurl = {https://ui.adsabs.harvard.edu/abs/1979PASP...91...79G}
}

@ARTICLE{Graham1971b,
       author = {{Graham}, J.~A.},
        title = "{Nova in the Large Magellanic Cloud.}",
      journal = {\iaucirc},
         year = {1971b},
        month = jan,
       volume = {2353},
        pages = {1},
       adsurl = {https://ui.adsabs.harvard.edu/abs/1971IAUC.2353....1G}
}

@ARTICLE{MacConnell1970,
       author = {{MacConnell}, D.~J. and {Gomez}, A.},
        title = "{Nova in the Large Magellanic Cloud.}",
      journal = {\iaucirc},
         year = 1970,
        month = jan,
       volume = {2238},
        pages = {1},
       adsurl = {https://ui.adsabs.harvard.edu/abs/1970IAUC.2238....1M}
}

@INPROCEEDINGS{Liller2003a,
       author = {{Liller}, W.},
        title = "{V2434 (LMC) and the ViNa del Mar nova program}",
    booktitle = {Interplay of Periodic, Cyclic and Stochastic Variability in Selected Areas of the H-R Diagram},
         year = {2003a},
       editor = {{Sterken}, C.},
       series = {Astronomical Society of the Pacific Conference Series},
       volume = {292},
        month = mar,
        pages = {101},
       adsurl = {https://ui.adsabs.harvard.edu/abs/2003ASPC..292..101L}
}

@ARTICLE{Liller2000,
       author = {{Liller}, W. and {Stubbings}, R.},
        title = "{Nova in the Large Magellanic Cloud 2000}",
      journal = {\iaucirc},
         year = 2000,
        month = jul,
       volume = {7453},
        pages = {1},
       adsurl = {https://ui.adsabs.harvard.edu/abs/2000IAUC.7453....1L}
}

@ARTICLE{Alcock1997,
       author = {{Alcock et al.}},
        title = "{Nova in the Large Magellanic Cloud 1997}",
      journal = {\iaucirc},
         year = 1997,
        month = oct,
       volume = {6756},
        pages = {1},
       adsurl = {https://ui.adsabs.harvard.edu/abs/1997IAUC.6756....1G}
}

@ARTICLE{Stanek2015,
       author = {{Stanek}, K.~Z. and {Danilet}, A.~B. and {Holoien}, T.~W.-S. and {Kochanek}, C.~S. and {Simonian}, G. and {Basu}, U. and {Goss}, N. and {Beacom}, J.~F. and {Thompson}, T.~A. and {Shappee}, B.~J. and {Prieto}, J.~L. and {Bersier}, D. and {Brimacombe}, J. and {Dong}, Subo and {Falco}, E. and {Wozniak}, P.~R. and {Szczygiel}, D. and {Pojmanski}, G.},
        title = "{ASASSN-15fd: A Possible Nova in the LMC}",
      journal = {The Astronomer's Telegram},
         year = 2015,
        month = mar,
       volume = {7313},
        pages = {1},
       adsurl = {https://ui.adsabs.harvard.edu/abs/2015ATel.7313....1S}
}

@ARTICLE{Wyrzykowski2012,
       author = {{Wyrzykowski}, L. and {Ulaczyk}, K. and {Udalski}, A. and {Kozlowski}, S.},
        title = "{Nova in the LMC}",
      journal = {The Astronomer's Telegram},
         year = 2012,
        month = nov,
       volume = {4540},
        pages = {1},
       adsurl = {https://ui.adsabs.harvard.edu/abs/2012ATel.4540....1W}
}

@ARTICLE{Seach2012,
       author = {{Seach}, J. and {Liller}, W. and {Brimacombe}, J. and {Pearce}, A.},
        title = "{Nova in the large Magellanic cloud 2012 = TCP J04550000-7027150.}",
      journal = {Central Bureau Electronic Telegrams},
         year = 2012,
        month = mar,
       volume = {3071},
        pages = {1},
       adsurl = {https://ui.adsabs.harvard.edu/abs/2012CBET.3071....1S}
}

@ARTICLE{Mroz2014,
       author = {{Mr{\'o}z}, P. and {Poleski}, R. and {Udalski}, A. and {Soszy{\'n}ski}, I. and {Szyma{\'n}ski}, M.~K. and {Kubiak}, M. and {Pietrzy{\'n}ski}, G. and {Wyrzykowski}, {\L}. and {Ulaczyk}, K. and {Koz{\l}owski}, S. and {Pietrukowicz}, P. and {Skowron}, J.},
        title = "{Recurrent and symbiotic novae in data from the Optical Gravitational Lensing Experiment}",
      journal = {\mnras},
         year = 2014,
        month = sep,
       volume = {443},
       number = {1},
        pages = {784-790},
          doi = {10.1093/mnras/stu1181},
archivePrefix = {arXiv},
       eprint = {1405.2007},
 primaryClass = {astro-ph.SR},
       adsurl = {https://ui.adsabs.harvard.edu/abs/2014MNRAS.443..784M}
}

@ARTICLE{Walter2012,
       author = {{Walter}, Frederick M. and {Battisti}, Andrew and {Towers}, Sarah E. and {Bond}, Howard E. and {Stringfellow}, Guy S.},
        title = "{The Stony Brook/SMARTS Atlas of (mostly) Southern Novae}",
      journal = {\pasp},
         year = 2012,
        month = oct,
       volume = {124},
       number = {920},
        pages = {1057},
          doi = {10.1086/668404},
archivePrefix = {arXiv},
       eprint = {1209.1583},
 primaryClass = {astro-ph.SR},
       adsurl = {https://ui.adsabs.harvard.edu/abs/2012PASP..124.1057W}
}

@ARTICLE{Liller2009a,
       author = {{Liller}, W.},
        title = "{Nova in the Large Magellanic Cloud 2009}",
      journal = {\iaucirc},
         year = {2009a},
        month = feb,
       volume = {9019},
        pages = {1},
       adsurl = {https://ui.adsabs.harvard.edu/abs/2009IAUC.9019....1L}
}

@ARTICLE{Liller2009b,
       author = {{Liller}, W. and {Monard}, L.~A.~G.},
        title = "{Nova in the Large Magellanic Cloud 2009 No. 2}",
      journal = {\iaucirc},
         year = {2009b},
        month = may,
       volume = {9042},
        pages = {1},
       adsurl = {https://ui.adsabs.harvard.edu/abs/2009IAUC.9042....1L}
}

@ARTICLE{Liller2007,
       author = {{Liller}, W. and {Heathcote}, B. and {di Scala}, G. and {Allen}, W.},
        title = "{Light and Color Curves of the Unusual Slow Nova LMC 2005}",
      journal = {\jaavso},
         year = 2007,
        month = jan,
       volume = {35},
       number = {2},
        pages = {359},
       adsurl = {https://ui.adsabs.harvard.edu/abs/2007JAVSO..35..359L}
}

@ARTICLE{Read2009,
       author = {{Read}, A.~M. and {Saxton}, R.~D. and {Jonker}, P.~G. and {Kuulkers}, E. and {Esquej}, P. and {Pojmanski}, G. and {Torres}, M.~A.~P. and {Goad}, M.~R. and {Freyberg}, M.~J. and {Modjaz}, M.},
        title = "{XMMSL1 J060636.2-694933: an XMM-Newton slew discovery and Swift/Magellan follow up of a new classical nova in the LMC}",
      journal = {\aap},
         year = 2009,
        month = nov,
       volume = {506},
       number = {3},
        pages = {1309-1317},
          doi = {10.1051/0004-6361/200912082},
archivePrefix = {arXiv},
       eprint = {0908.3989},
 primaryClass = {astro-ph.HE},
       adsurl = {https://ui.adsabs.harvard.edu/abs/2009A&A...506.1309R}
}

@ARTICLE{Liller2003b,
       author = {{Liller}, W. and {Monard}, L.~A.~G. and {Africa}, S. and {Pearce}, A. and {Bond}, H.~E. and {Gonzalez}, S.},
        title = "{Nova in the Large Magellanic Cloud}",
      journal = {\iaucirc},
         year = {2003b},
        month = jun,
       volume = {8160},
        pages = {1},
       adsurl = {https://ui.adsabs.harvard.edu/abs/2003IAUC.8160....1L}
}

@ARTICLE{Greiner2003,
       author = {{Greiner}, J. and {Orio}, M. and {Schartel}, N.},
        title = "{XMM-Newton observations of Nova LMC 2000}",
      journal = {\aap},
         year = 2003,
        month = jul,
       volume = {405},
        pages = {703-710},
          doi = {10.1051/0004-6361:20030602},
archivePrefix = {arXiv},
       eprint = {astro-ph/0304479},
 primaryClass = {astro-ph},
       adsurl = {https://ui.adsabs.harvard.edu/abs/2003A&A...405..703G}
}

@ARTICLE{Mroz2016a,
       author = {{Mr{\'o}z}, P. and {Udalski}, A. and {Poleski}, R. and {Soszy{\'n}ski}, I. and {Szyma{\'n}ski}, M.~K. and {Pietrzy{\'n}ski}, G. and {Wyrzykowski}, {\L}. and {Ulaczyk}, K. and {Koz{\l}owski}, S. and {Pietrukowicz}, P. and {Skowron}, J.},
        title = "{OGLE Atlas of Classical Novae. II. Magellanic Clouds}",
      journal = {\apjs},
         year = {2016a},
        month = jan,
       volume = {222},
       number = {1},
          eid = {9},
        pages = {9},
          doi = {10.3847/0067-0049/222/1/9},
archivePrefix = {arXiv},
       eprint = {1511.06355},
 primaryClass = {astro-ph.SR},
       adsurl = {https://ui.adsabs.harvard.edu/abs/2016ApJS..222....9M}
}

@ARTICLE{Schwarz2001,
       author = {{Schwarz}, Greg J. and {Shore}, S.~N. and {Starrfield}, S. and {Hauschildt}, Peter H. and {Della Valle}, M. and {Baron}, E.},
        title = "{Multiwavelength analyses of the extraordinary nova LMC 1991$^{*}$}",
      journal = {\mnras},
         year = 2001,
        month = jan,
       volume = {320},
       number = {1},
        pages = {103-123},
          doi = {10.1046/j.1365-8711.2001.03960.x},
       adsurl = {https://ui.adsabs.harvard.edu/abs/2001MNRAS.320..103S}
}

@ARTICLE{Sekiguchi1990,
       author = {{Sekiguchi}, K. and {Stobie}, R.~S. and {Buckley}, D.~A.~H. and {Caldwell}, J.~A.~R.},
        title = "{Recurrent nova in the Large Magellanic Cloud - Nova LMC 1990 n0 2.}",
      journal = {\mnras},
         year = 1990,
        month = jul,
       volume = {245},
        pages = {28P},
       adsurl = {https://ui.adsabs.harvard.edu/abs/1990MNRAS.245P..28S}
}

@ARTICLE{Liller2005,
       author = {{Liller}, W. and {Shida}, R.~Y.},
        title = "{Large Magellanic Cloud Novae 13 Days After Peak Brightness}",
      journal = {\jaavso},
         year = 2005,
        month = aug,
       volume = {33},
       number = {2},
        pages = {207-211},
       adsurl = {https://ui.adsabs.harvard.edu/abs/2005JAVSO..33..207L}
}

@ARTICLE{Liller2005b,
       author = {{Liller}, W. and {Allen}, B. and {Pearce}, A.},
        title = "{Nova in the Large Magellanic Cloud 2005}",
      journal = {\iaucirc},
         year = {2005b},
        month = nov,
       volume = {8635},
        pages = {1},
       adsurl = {https://ui.adsabs.harvard.edu/abs/2005IAUC.8635....1L}
}

@ARTICLE{Vanlandingham1999,
       author = {{Vanlandingham}, Karen M. and {Starrfield}, Sumner and {Shore}, Steven N. and {Sonneborn}, George},
        title = "{Elemental abundances for Nova LMC 1990\#1}",
      journal = {\mnras},
         year = 1999,
        month = sep,
       volume = {308},
       number = {2},
        pages = {577-587},
          doi = {10.1046/j.1365-8711.1999.02731.x},
       adsurl = {https://ui.adsabs.harvard.edu/abs/1999MNRAS.308..577V}
}

@ARTICLE{Mason2005,
       author = {{Mason}, E. and {Della Valle}, M. and {Gilmozzi}, R. and {Lo Curto}, G. and {Williams}, R.~E.},
        title = "{Early decline spectra of nova SMC 2001 and nova LMC 2002}",
      journal = {\aap},
         year = 2005,
        month = jun,
       volume = {435},
       number = {3},
        pages = {1031-1042},
          doi = {10.1051/0004-6361:20041351},
archivePrefix = {arXiv},
       eprint = {astro-ph/0504153},
 primaryClass = {astro-ph},
       adsurl = {https://ui.adsabs.harvard.edu/abs/2005A&A...435.1031M}
}

@ARTICLE{Sekiguchi1989,
       author = {{Sekiguchi}, K. and {Kilkenny}, D. and {Winkler}, H. and {Doyle}, J.~G.},
        title = "{Optical spectroscopy of nova LMC 1988 No 2 during its early decline stage.}",
      journal = {\mnras},
         year = 1989,
        month = dec,
       volume = {241},
        pages = {827-837},
          doi = {10.1093/mnras/241.4.827},
       adsurl = {https://ui.adsabs.harvard.edu/abs/1989MNRAS.241..827S}
}

@INPROCEEDINGS{Hearnshaw2004,
       author = {{Hearnshaw}, J.~B. and {Livingston}, C.~M. and {Gilmore}, A.~C. and {Kilmartin}, P.~M.},
        title = "{Light curves and absolute magnitudes of four recent fast LMC novae}",
    booktitle = {IAU Colloquium 193: Variable Stars in the Local Group},
         year = 2004,
       editor = {{Kurtz}, Donald W. and {Pollard}, Karen R.},
       series = {Astronomical Society of the Pacific Conference Series},
       volume = {310},
        month = may,
        pages = {103},
       adsurl = {https://ui.adsabs.harvard.edu/abs/2004ASPC..310..103H}
}

@ARTICLE{McNaught1987,
       author = {{McNaught}, R.~H. and {Garradd}, G.~J.},
        title = "{Nova in the Large Magellanic Cloud}",
      journal = {\iaucirc},
         year = 1987,
        month = sep,
       volume = {4453},
        pages = {1},
       adsurl = {https://ui.adsabs.harvard.edu/abs/1987IAUC.4453....1M}
}

@ARTICLE{Graham1978,
       author = {{Graham}, J.~A. and {Rojas}, H. and {Landolt}, A.},
        title = "{Nova in Large Magellanic Cloud}",
      journal = {\iaucirc},
         year = 1978,
        month = apr,
       volume = {3206},
        pages = {2},
       adsurl = {https://ui.adsabs.harvard.edu/abs/1978IAUC.3206....2G}
}

@ARTICLE{Canterna1981,
       author = {{Canterna}, R. and {Thompson}, L.~F.},
        title = "{Spectral properties of nova LMC 1977 b.}",
      journal = {\pasp},
         year = 1981,
        month = oct,
       volume = {93},
        pages = {581-586},
          doi = {10.1086/130891},
       adsurl = {https://ui.adsabs.harvard.edu/abs/1981PASP...93..581C}
}

@ARTICLE{Canterna1977,
       author = {{Canterna}, R. and {Schwartz}, R.~D.},
        title = "{Photometry of LMC Nova 1977b.}",
      journal = {\apjl},
         year = 1977,
        month = sep,
       volume = {216},
        pages = {L91-L94},
          doi = {10.1086/182518},
       adsurl = {https://ui.adsabs.harvard.edu/abs/1977ApJ...216L..91C}
}

@ARTICLE{Graham1971a,
       author = {{Graham}, J.~A. and {Araya}, G.},
        title = "{Novae in the Magellanic Clouds during the 1970 - 1971 observing season.}",
      journal = {\aj},
         year = {1971a},
        month = nov,
       volume = {76},
        pages = {768-774},
          doi = {10.1086/111194},
       adsurl = {https://ui.adsabs.harvard.edu/abs/1971AJ.....76..768G}
}

@INCOLLECTION{Capaccioli1990b,
       author = {{Capaccioli}, M. and {Della Valle}, M. and {D'Onofrio}, M. and {Rosino}, L.},
        title = "{Maximum Magnitude Versus Rate of Decline for Novae of the Large Magellanic Cloud}",
    booktitle = {IAU Colloquium 122: Physics of Classical Novae},
         year = {1990b},
       editor = {{Cassatella}, Angelo and {Viotti}, Roberto},
       volume = {369},
        pages = {71},
          doi = {10.1007/3-540-53500-4_103},
       adsurl = {https://ui.adsabs.harvard.edu/abs/1990LNP...369...71C}
}

@ARTICLE{Capaccioli1990a,
       author = {{Capaccioli}, Massimo and {Della Valle}, Massimo and {D'Onofrio}, Mauro and {Rosino}, Leonida},
        title = "{Distance of the Large Magellanic Cloud through the Maximum Magnitude versus Rate of Decline Relation for Novae}",
      journal = {\apj},
         year = {1990a},
        month = sep,
       volume = {360},
        pages = {63},
          doi = {10.1086/169096},
       adsurl = {https://ui.adsabs.harvard.edu/abs/1990ApJ...360...63C}
}

@ARTICLE{Subramaniam2002,
       author = {{Subramaniam}, A. and {Anupama}, G.~C.},
        title = "{The local stellar population of nova regions in the Large Magellanic Cloud}",
      journal = {\aap},
         year = 2002,
        month = aug,
       volume = {390},
        pages = {449-471},
          doi = {10.1051/0004-6361:20020742},
archivePrefix = {arXiv},
       eprint = {astro-ph/0203098},
 primaryClass = {astro-ph},
       adsurl = {https://ui.adsabs.harvard.edu/abs/2002A&A...390..449S}
}

@ARTICLE{Sievers1970,
       author = {{Sievers}, J.},
        title = "{A Nova in the Large Magellanic Cloud, BV 1261}",
      journal = {Information Bulletin on Variable Stars},
         year = 1970,
        month = jul,
       volume = {448},
        pages = {1},
       adsurl = {https://ui.adsabs.harvard.edu/abs/1970IBVS..448....1S}
}

@ARTICLE{Henize1954,
       author = {{Henize}, Karl G. and {Hoffleit}, Dorrit and {McKibben Nail}, Virginia},
        title = "{Magellanic Clouds. XI. Survey of the Novae}",
      journal = {Proceedings of the National Academy of Science},
         year = 1954,
        month = jun,
       volume = {40},
       number = {6},
        pages = {365-372},
          doi = {10.1073/pnas.40.6.365},
       adsurl = {https://ui.adsabs.harvard.edu/abs/1954PNAS...40..365H}
}

@ARTICLE{Buscombe1955,
       author = {{Buscombe}, W. and {de Vaucouleurs}, G.},
        title = "{Novae in the Magellanic Clouds and in the Galaxy}",
      journal = {The Observatory},
         year = 1955,
        month = aug,
       volume = {75},
        pages = {170-175},
       adsurl = {https://ui.adsabs.harvard.edu/abs/1955Obs....75..170B}
}

@ARTICLE{Shore2024a,
       author = {{Shore}, Steven N. and {Charbonnel}, Stephane and {Garde}, Olivier and {Le Du}, Pascal and {Mulato}, Lionel and {Petit}, Thomas and {Barker}, Hamish and {Velez}, Peter},
        title = "{Early spectroscopy of the classical LMC nova AT2 024fjh}",
      journal = {The Astronomer's Telegram},
         year = {2024a},
        month = apr,
       volume = {16575},
        pages = {1},
       adsurl = {https://ui.adsabs.harvard.edu/abs/2024ATel16575....1S}
}

@ARTICLE{Merc2024,
       author = {{Merc}, J. and {Love}, T. and {Velez}, P. and {Barker}, H. and {Charbonnel}, S. and {Garde}, O. and {Du}, P. Le and {Mulato}, L. and {Petit}, T.},
        title = "{Spectroscopic classification of ASASSN-24by (AT 2024epj) as a classical nova in the LMC}",
      journal = {The Astronomer's Telegram},
         year = 2024,
        month = mar,
       volume = {16545},
        pages = {1},
       adsurl = {https://ui.adsabs.harvard.edu/abs/2024ATel16545....1M}
}

@ARTICLE{Perez-Fournon2024,
       author = {{Perez-Fournon}, I. and {Poidevin}, F.},
        title = "{LMC nova candidate ASASSN-24by (AT 2024epj): LCOGT follow-up}",
      journal = {The Astronomer's Telegram},
         year = 2024,
        month = mar,
       volume = {16543},
        pages = {1},
       adsurl = {https://ui.adsabs.harvard.edu/abs/2024ATel16543....1P}
}

@ARTICLE{Strader2023,
       author = {{Strader}, J. and {Aydi}, E. and {Chomiuk}, L. and {Kyer}, R. and {Urquhart}, R. and {Sokolovsky}, K.~V. and {Stanek}, K.~Z. and {Kochanek}, C.~S. and {Shappee}, B.~J.},
        title = "{SOAR spectroscopic follow-up of ASASSN-23hd, a nova in the LMC}",
      journal = {The Astronomer's Telegram},
         year = 2023,
        month = oct,
       volume = {16294},
        pages = {1},
       adsurl = {https://ui.adsabs.harvard.edu/abs/2023ATel16294....1S}
}

@ARTICLE{Aydi2022,
       author = {{Aydi}, E. and {Strader}, J. and {Chomiuk}, L. and {Sokolovsky}, K.~V. and {Kawash}, A.},
        title = "{SOAR spectroscopic classification of PNV J05222788-6936333 as a classical nova in the LMC}",
      journal = {The Astronomer's Telegram},
         year = 2022,
        month = may,
       volume = {15392},
        pages = {1},
       adsurl = {https://ui.adsabs.harvard.edu/abs/2022ATel15392....1A}
}

@ARTICLE{Stanek2018,
       author = {{Stanek}, K.~Z. and {Kochanek}, C.~S. and {Shields}, J.~V. and {Thompson}, T.~A. and {Chomiuk}, L. and {Strader}, J. and {Shappee}, B.~J. and {Holoien}, T.~W.-S. and {Prieto}, J.~L. and {Dong}, Subo and {Stritzinger}, M.},
        title = "{ASAS-SN Discovery of a Possible LMC Nova ASASSN-18pf}",
      journal = {The Astronomer's Telegram},
         year = 2018,
        month = jul,
       volume = {11857},
        pages = {1},
       adsurl = {https://ui.adsabs.harvard.edu/abs/2018ATel11857....1S}
}

@ARTICLE{Chomiuk2018c,
       author = {{Chomiuk}, L. and {Strader}, J. and {Shishkovsky}, L. and {Swihart}, S. and {Stanek}, K.~Z. and {Kochanek}, C.~S. and {Shields}, J.~V. and {Thompson}, T.~A. and {Shappee}, B.~J. and {Holoien}, T.~W.-S. and {Prieto}, J.~L. and {Dong}, Subo and {Stritzinger}, M.},
        title = "{ASAS-SN 18jj is a classical nova in the LMC}",
      journal = {The Astronomer's Telegram},
         year = {2018c},
        month = may,
       volume = {11610},
        pages = {1},
       adsurl = {https://ui.adsabs.harvard.edu/abs/2018ATel11610....1C}
}

@ARTICLE{Walter2018,
       author = {{Walter}, Frederick M.},
        title = "{OGLE-2018-NOVA-01 (N LMC 1996) is a He-N Nova}",
      journal = {The Astronomer's Telegram},
         year = 2018,
        month = mar,
       volume = {11390},
        pages = {1},
       adsurl = {https://ui.adsabs.harvard.edu/abs/2018ATel11390....1W}
}

@ARTICLE{Chomiuk2018b,
       author = {{Chomiuk}, L. and {Strader}, J. and {Stanek}, K.~Z. and {Kochanek}, C.~S. and {Shields}, J.~V. and {Thompson}, T.~A. and {Shappee}, B.~J. and {Holoien}, T.~W.-S. and {Prieto}, J.~L. and {Dong}, Subo and {Stritzinger}, M.},
        title = "{Pre-Discovery Light Curve of OGLE-2018-NOVA-01 from ASAS-SN}",
      journal = {The Astronomer's Telegram},
         year = {2018b},
        month = mar,
       volume = {11392},
        pages = {1},
       adsurl = {https://ui.adsabs.harvard.edu/abs/2018ATel11392....1C}
}

@ARTICLE{DiMille2016,
       author = {{Di Mille}, Francesco and {Angeloni}, Rodolfo and {Morrell}, Nidia},
        title = "{Spectroscopic follow-up of the recurrent nova LMCN 1968-12a}",
      journal = {The Astronomer's Telegram},
         year = 2016,
        month = jan,
       volume = {8586},
        pages = {1},
       adsurl = {https://ui.adsabs.harvard.edu/abs/2016ATel.8586....1D}
}

@ARTICLE{Aydi2024a,
       author = {{Aydi}, E. and {Chomiuk}, L. and {Strader}, J. and {Sokolovsky}, K.~V. and {Williams}, R.~E. and {Buckley}, D.~A.~H. and {Ederoclite}, A. and {Izzo}, L. and {Kyer}, R. and {Linford}, J.~D. and {Kniazev}, A. and {Metzger}, B.~D. and {Miko{\l}ajewska}, J. and {Molaro}, P. and {Molina}, I. and {Mukai}, K. and {Munari}, U. and {Orio}, M. and {Panurach}, T. and {Shappee}, B.~J. and {Shen}, K.~J. and {Sokoloski}, J.~L. and {Urquhart}, R. and {Walter}, F.~M.},
        title = "{Revisiting the classics: on the evolutionary origin of the 'Fe II' and 'He/N' spectral classes of novae}",
      journal = {\mnras},
         year = {2024a},
        month = jan,
       volume = {527},
       number = {3},
        pages = {9303-9321},
          doi = {10.1093/mnras/stad3342},
archivePrefix = {arXiv},
       eprint = {2309.07097},
 primaryClass = {astro-ph.SR},
       adsurl = {https://ui.adsabs.harvard.edu/abs/2024MNRAS.527.9303A}
}

@ARTICLE{Aydi2024b,
       author = {{Aydi}, E. and {Strader}, J. and {Molina}, I. and {Kyer}, R. and {Chomiuk}, L. and {Urquhart}, R. and {Dage}, K.~C. and {Sokolovsky}, K.~V. and {Stanek}, K.~Z. and {Kochanek}, C.~S. and {Shappee}, B.~J.},
        title = "{SOAR spectroscopic classification of ASASSN-24ck (AT 2024fjh) as a nova in the LMC}",
      journal = {The Astronomer's Telegram},
         year = {2024b},
        month = apr,
       volume = {16583},
        pages = {1},
       adsurl = {https://ui.adsabs.harvard.edu/abs/2024ATel16583....1A}
}

@ARTICLE{Munari2016,
       author = {{Munari}, U. and {Walter}, F.~M. and {Hambsch}, F.-J. and {Frigo}, A.},
        title = "{Photometric evolution of the 2016 outburst of recurrent Nova LMC 1968: the first three weeks}",
      journal = {Information Bulletin on Variable Stars},
         year = 2016,
        month = mar,
       volume = {6162},
        pages = {1},
          doi = {10.48550/arXiv.1603.02492},
archivePrefix = {arXiv},
       eprint = {1603.02492},
 primaryClass = {astro-ph.SR},
       adsurl = {https://ui.adsabs.harvard.edu/abs/2016IBVS.6162....1M}
}

@ARTICLE{Aydi2019a,
       author = {{Aydi}, E. and {Chomiuk}, L. and {Strader}, J. and {Swihart}, S.~J. and {Bahramian}, A. and {Harvey}, E.~J. and {Britt}, C.~T. and {Buckley}, D.~A.~H. and {Chen}, P. and {Dage}, K. and {Darnley}, M.~J. and {Dong}, S. and {Hambsch}, F-J. and {Holoien}, T.~W. -S. and {Jha}, S.~W. and {Kochanek}, C.~S. and {Kuin}, N.~P.~M. and {Li}, K.~L. and {Monard}, L.~A.~G. and {Mukai}, K. and {Page}, K.~L. and {Prieto}, J.~L. and {Richardson}, N.~D. and {Shappee}, B.~J. and {Shishkovsky}, L. and {Sokolovsky}, K.~V. and {Stanek}, K.~Z. and {Thompson}, T.},
        title = "{Flaring, Dust Formation, And Shocks In The Very Slow Nova ASASSN-17pf (LMCN 2017-11a)}",
      journal = {arXiv e-prints},
         year = 2019,
        month = mar,
          eid = {arXiv:1903.09232},
        pages = {arXiv:1903.09232},
          doi = {10.48550/arXiv.1903.09232},
archivePrefix = {arXiv},
       eprint = {1903.09232},
 primaryClass = {astro-ph.HE},
       adsurl = {https://ui.adsabs.harvard.edu/abs/2019arXiv190309232A}
}

@INPROCEEDINGS{Shida2004,
       author = {{Shida}, Raquel Yumi and {Liller}, William},
        title = "{The distribution of novae in the Magellanic Clouds}",
    booktitle = {IAU Colloquium 193: Variable Stars in the Local Group},
         year = 2004,
       editor = {{Kurtz}, Donald W. and {Pollard}, Karen R.},
       series = {Astronomical Society of the Pacific Conference Series},
       volume = {310},
        month = may,
        pages = {184},
       adsurl = {https://ui.adsabs.harvard.edu/abs/2004ASPC..310..184S}
}

@article{Danilet2015,
  author = {Danilet, B. and others},
  title = {ASASSN-15fd: a classical nova in the LMC},
  journal = {The Astronomer's Telegram},
  volume = {7260},
  year = {2015},
  url = {https://www.astronomerstelegram.org/?read=7260}
}

@article{Mroz2016b,
  author = {Mr\'oz, P. and Udalski, A.},
  title = {OGLE-2016-NOVA-01 = LMC 1968-12a 2016 eruption},
  journal = {The Astronomer's Telegram},
  volume = {8578},
  year = {2016b},
  url = {https://www.astronomerstelegram.org/?read=8578}
}

@article{Gorbovskoy2016,
  author = {Gorbovskoy, E. and others},
  title = {MASTER OT J051032.58-692130.4 = LMCN 2016-04a},
  journal = {The Astronomer's Telegram},
  year = {2016},
  url = {https://www.astronomerstelegram.org/?read=9039}
}

@article{Mroz2018,
  author = {Mr\'oz, P. and Udalski, A.},
  title = {OGLE-2018-NOVA-01 in the LMC},
  journal = {The Astronomer's Telegram},
  volume = {11384},
  year = {2018},
  url = {https://www.astronomerstelegram.org/?read=11384}
}

@misc{Jacques2019,
  author = {Jacques, C. and others},
  title = {AT 2019lvm in the LMC},
  year = {2019},
  howpublished = {\url{https://wis-tns.org/object/2019lvm}}
}

@misc{Pimentel2019,
  author = {Pimentel, E. and others},
  title = {AT 2019uni in the LMC},
  year = {2019},
  howpublished = {\url{https://wis-tns.org/object/2019uni}}
}

@misc{Schmeer2020,
  author = {Schmeer, P.},
  title = {LMC V1341 = LMC 1968-12a new recurrent nova eruption},
  year = {2020},
  howpublished = {\url{https://www.aavso.org/lmc-v1341-new-recurrent-nova-eruption}}
}

@misc{Bohlsen2016,
  author = {Bohlsen, T.},
  title = {``MASTER OT J051032.58-692130.4: Possible Nova (12.0 mag) in the Large Magellanic Cloud (LMC)". AAVSO Forums},
  year = {2016},
  howpublished = {https://www.aavso.org/master-ot-j05103258-6921304-possible-nova-120-mag-large-magellanic-cloud-lmc}
}
\bibliographystyle{aasjournalv7}

\end{document}